\documentclass[aps,showpacs,preprintnumbers,twocolumn,superscriptaddress,prl,floatfix,nofootinbib]{revtex4-2}

\usepackage{amsfonts}
\usepackage{amsmath}
\usepackage{amssymb}
\usepackage{graphicx}
\usepackage{dcolumn}
\usepackage{bm,bbm}
\usepackage{tabularx}
\usepackage{array}
\usepackage{epstopdf}
\usepackage{xkcdcolors}
\usepackage{xcolor}
\usepackage{mathrsfs}
\usepackage{subfigure}
\usepackage{mathtools}
\usepackage{float}
\usepackage[linesnumbered,lined,boxed,commentsnumbered,ruled,vlined]{algorithm2e}
\usepackage{algpseudocode}
\usepackage{verbatim}
\usepackage{comment}
\usepackage{ragged2e}
\usepackage{physics}
\usepackage{hyperref}
\usepackage{cleveref}
\usepackage{xr}
\usepackage{url}
\usepackage{tikz}
\usepackage{caption}
\usepackage{lineno}
\DeclareCaptionJustification{justified}{\justifying}
\hypersetup{colorlinks=true, citecolor=orange, urlcolor=blue, linkcolor=magenta}

\usepackage{stackengine}

\makeatletter
\renewcommand\@biblabel[1]{#1.}

\makeatother

\newcommand{\mytriangle}{\tikz\draw[black] (0,0)--(0.16,0)--(0.08,0.12)--(0,0);}

\definecolor{colSRGM}{RGB}{51,117,56}
\definecolor{colSRGMFT}{RGB}{220,205,125}
\definecolor{colSCM}{RGB}{194,106,119}
\definecolor{colSCMFT}{RGB}{46,37,133}

\begin{document}

\title{Testing structural balance theories in heterogeneous signed networks}

\author{Anna Gallo}
\affiliation{IMT School for Advanced Studies, Piazza San Francesco 19, 55100 Lucca (Italy)}
\affiliation{INdAM-GNAMPA Istituto Nazionale di Alta Matematica (Italy)}
\author{Diego Garlaschelli}
\affiliation{IMT School for Advanced Studies, Piazza San Francesco 19, 55100 Lucca (Italy)}
\affiliation{INdAM-GNAMPA Istituto Nazionale di Alta Matematica (Italy)}
\affiliation{Lorentz Institute for Theoretical Physics, University of Leiden, Niels Bohrweg 2, 2333 CA Leiden (The Netherlands)}
\author{Renaud Lambiotte}
\affiliation{Mathematical Institute, University of Oxford, Woodstock Road, OX2 6GG Oxford (United Kingdom)}
\author{Fabio Saracco}
\affiliation{`Enrico Fermi' Research Center (CREF), Via Panisperna 89A, 00184 Rome (Italy)}
\affiliation{Institute for Applied Computing `Mauro Picone' (IAC), National Research Council, Via dei Taurini 19, 00185 Rome (Italy)}
\affiliation{IMT School for Advanced Studies, Piazza San Francesco 19, 55100 Lucca (Italy)}
\author{Tiziano Squartini}
\email{tiziano.squartini@imtlucca.it}
\affiliation{IMT School for Advanced Studies, Piazza San Francesco 19, 55100 Lucca (Italy)}
\affiliation{INdAM-GNAMPA Istituto Nazionale di Alta Matematica (Italy)}
\affiliation{Institute for Advanced Study, University of Amsterdam, Oude Turfmarkt 145, 1012 GC Amsterdam (The Netherlands)}

\date{\today}

\begin{abstract}
\noindent The abundance of data about social relationships allows the human behavior to be analyzed as any other natural phenomenon. Here we focus on balance theory, stating that social actors tend to avoid establishing cycles with an odd number of negative links. This statement, however, can be supported only after a comparison with a benchmark. Since the existing ones disregard actors' heterogeneity, we extend Exponential Random Graphs to signed networks with both global and local constraints and employ them to assess the significance of empirical unbalanced patterns. We find that the nature of balance crucially depends on the null model: while homogeneous benchmarks favor the weak balance theory, according to which only triangles with one negative link should be under-represented, heterogeneous benchmarks favor the strong balance theory, according to which also triangles with all negative links should be under-represented. Biological networks, instead, display strong frustration under any benchmark, confirming that structural balance inherently characterizes social networks.
\end{abstract}

\maketitle

\section{Introduction}

Network theory has emerged as a powerful framework in many disciplines to model different kinds of real-world systems, by representing their units as nodes and the interactions between them as links. In social science, the study of networks with signed edges has recently seen its popularity revived \cite{antal2006social,leskovec2010signed,zaslavsky2012mathematical,tang2016survey}, because the signed character of links can be used to represent the positive as well as the negative social interactions that are currently identifiable in empirical data.

From a historical perspective, the interest towards signed networks is rooted into the psychological theory named \emph{balance theory} (BT), firstly proposed by Heider~\cite{heider1946attitudes}. The choice of adopting signed graphs to model it has, then, led Cartwright and Harary~\cite{cartwright1956structural} to introduce its structural version (SBT), which has found application not only in the study of human relationships, but also in that of biological, ecological and economic systems \cite{harary2002signed,ou2015detecting,iorio2016efficient,saiz2017evidence}.

\begin{figure}[t!]
\centering
\subfigure[]{\includegraphics[width=0.46\textwidth]{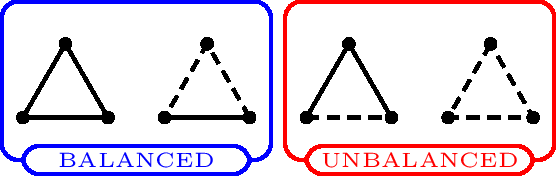}}
\subfigure[]{\includegraphics[width=0.46\textwidth]{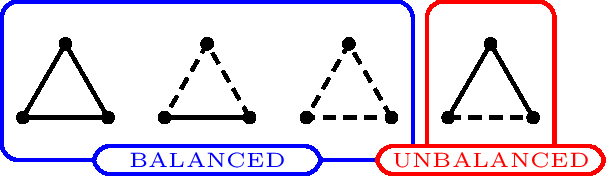}}
\caption{\textbf{Balanced and unbalanced motifs.} Fundamental triadic patterns, or motifs, considered as balanced (blue) and unbalanced (red) by the strong \textbf{(a)} and weak \textbf{(b)} versions of the balance theory.}
\label{fig:1}
\end{figure}

BT deals with the concept of \emph{balance}: a complete, signed graph is said to be balanced if all its triads have an even number of negative edges, i.e. either zero (in this case, the three edges are all positive) or two (see Fig. \ref{fig:1}). Informally speaking, BT formalizes the principles `the friend of my friend is my friend' and `the enemy of my enemy is my friend'. The so-called \emph{structure theorem} states that a complete, signed graph is balanced if and only if its set of nodes can be partitioned into two, disjoint subsets whose intra-modular links are all positive and whose inter-modular links are all negative. Cartwright and Harary extended the definition of balance to incomplete graphs~\cite{cartwright1956structural} by including cycles of length larger than three: a (connected) network is said to be balanced when all cycles are positive, i.e. they contain an even number of negative edges. Taken together, the criteria above form the so-called \emph{structural strong balance theory} (SSBT).

The framework of SSBT has been extended by Davis \cite{davis1967clustering} by introducing the concept of \emph{$k$-balanced} networks, according which signed graphs are balanced if their set of nodes can be partitioned into $k$ disjoint subsets with positive intra-modular links and negative inter-modular links. This generalized definition of balance leads to the formulation of \emph{structural weak balance theory} (SWBT), according to which triads with all negative edges are balanced, since each of their nodes can be thought of as a group on its own if necessary (see Fig. \ref{fig:1}).

Several metrics to decide whether signed networks are strongly or weakly balanced have been proposed. For instance, the level of balance of a signed network has been quantified as the number of edges that need be removed, or whose sign need be reversed, in order to obtain a network where each cycle has an even number of negative links \cite{akiyama1981balancing,harary1959measurement}. Alternatively, it has been defined as the number of balanced, closed walks (i.e. closed walks with an even number of negative links) that are present in the network \cite{estrada2014walk,singh2017measuring,estrada2019rethinking,kirkley2019balance}. In \cite{easley2012networks} an incomplete, signed network is considered balanced if it is possible to fill in all its missing links to obtain a complete, balanced graph according to SSBT. In \cite{aref2020multilevel} the authors define three different levels of balance: at the micro-scale, involving triads; at the meso-scale, involving larger subgraphs; at the macro-scale, involving the entire network. Still, as firstly noticed in \cite{cartwright1956structural}, `\emph{it may happen that only cycles of length 3 and 4 are important for the purpose of determining balance}'; this is further stressed in \cite{talaga2023polarization}, where it can be read that `\emph{this intuition has been later justified empirically by demonstrating that it is easier for people to memorize the valences of ties in shorter cycles}', and confirmed in \cite{giscard2017evaluating}, where it is noticed that `\emph{analyses based on counting simple cycles demonstrated that real networks often have a relatively low cycle length threshold after which the degree of balance measures quickly decrease}'.\\

Other approaches have been adopted in \cite{kunegis2010spectral,terzi2011spectral,anchuri2012communities}, where the problem is studied from a spectral perspective, and in \cite{belaza2017statistical}, where the problem is studied by employing concepts borrowed from statistical physics (each signed triad is assigned an energy and the networks at the `lowest temperature' have triangles without negative edges).

Other authors, instead, have focused on the complementary notion of \emph{frustration}, trying to quantify the extent to which signed networks are far from balanced \cite{zaslavsky1987balanced,aref2019balance,aref2020multilevel,aref2020modeling}. In \cite{zaslavsky1987balanced}, the authors define the so-called \emph{balanced decomposition number}, i.e. the (minimum) number of balanced groups into which nodes can be partitioned, and evaluate it by counting the (minimum) number of edges whose removal increases a network balance. In \cite{traag2019partitioning}, instead, the same index is evaluated by adopting the so-called \emph{switching signs method} introduced in \cite{abelson1958symbolic} and prescribing to count the (minimum) number of signs that must be reversed to balance a network. In \cite{kunegis2010spectral}, the level of (im)balance of a network is proxied by the magnitude of the smallest eigenvalue of the Laplacian matrix.

Empirical observations seem to point out that real-world, signed networks tend to be $k$-balanced, i.e. to avoid establishing the patterns that are considered as frustrated by SWBT: as an example, in \cite{anchuri2012communities} the authors study a pair of online, social networks induced by the relationships between users, showing that balance increases as the number of clusters into which nodes are partitioned is larger than two. In \cite{kirkley2019balance}, the authors notice that the weak formulation of SBT allows a better performance in predicting signs to be achieved.

In the present paper, we approach the concept of balance (or frustration) from a statistical perspective, comparing the empirical value of a chosen metric with the outcome of a properly defined benchmark model, i.e. a reference model preserving some of the network properties while randomizing the rest. The most common null model for signed graphs is perhaps the one obtained by keeping the positions of edges fixed while shuffling their signs \cite{leskovec2010signed,kirkley2019balance}. Reference \cite{facchetti2011computing} implements what we may call (for reasons that will be clear later) the canonical variant of the aforementioned exercise, assigning signs by means of a Bernoulli distribution. Reference \cite{singh2017measuring} introduces a null model for randomizing both the presence and the sign of links. In \cite{saiz2017evidence}, the signed version of the Local Rewiring Algorithm is implemented (at each step, two edges with the same sign are selected and rewired, to preserve the total number of signed links incident to each node). The canonical variant of this model is implemented in \cite{derr2018signed}, where the Balanced Signed Chung-Lu model (BSCL) is proposed (although it additionally constrains also the average number of signed triangles each edge is part of). Finally, refs. \cite{huitsing2012univariate,lerner2016structural,becatti2019,fritz2022exponential} define models constraining the structural properties of signed networks within the framework of Exponential Random Graphs (ERG).

Our contribution here focuses on binary, undirected signed networks and is motivated by two key considerations. First, real-world social networks have different levels of sparsity and we therefore aim at extending the ERG framework to include null models suitable for the analysis of signed graphs with plus (positive), minus (negative) and additionally zero (missing) edges. 
Second, as in the analysis of most other networks, we recognize the importance of preserving the inherent heterogeneity of different nodes and we therefore define new null models that can constrain the number of plus, minus and zero edges of each node separately. As we shall see, controlling for the different tendencies of actors of establishing friendly and unfriendly relationships
can change the estimated statistical significance of balance quite dramatically.
After defining a suite of such null models, we will use them to inspect the statistical significance of the most commonly studied (un)balanced patterns at both local and global levels, i.e. signed triangles and signed communities.

\section{Results}

\subsection*{Datasets description}

We now employ the benchmarks introduced and discussed in `Materials and Methods' and summed up in Table \ref{tab1} to analyze various real-world networks. Although most of them represent social relationships, we have also considered biological data as a comparison to check for specific patterns characterizing social structures.

The first dataset is the \emph{Correlates of Wars} (CoW) dataset \cite{doreian2015structural}. It provides a picture of the international political relationships over the years 1946-1997 and consists of 13 snapshots of 4 years each. A positive edge between any two countries indicates an alliance, a political agreement or the membership to the same governmental organization. Conversely, a negative edge indicates that the two countries are enemies, have a political disagreement or are part of different, governmental organizations.

The second dataset collects information about the relationships among the $\simeq300.000$ players of a massive multiplayer online game (MMOG) \cite{szell2010multirelational}. A positive edge between two players indicates a friendship, an alliance, or an economic relation. Conversely, a negative edge indicates the existence of an enmity, a conflict, or a fight. Since the network is directed, we have made it undirected by applying the following rules: if any two agents have the same opinion about the other, the undirected connection preserve the sign (i.e. $+1\cdot+1=+1$ and $-1\cdot-1=-1$); if any two agents have opposite opinions, we assume their undirected connection to have a negative sign (i.e. $+1\cdot-1=-1\cdot+1=-1$). Furthermore, in order to preserve the total number of nodes, we treat non-reciprocal connections as reciprocal, by preserving the original sign (i.e. $+1\cdot0=0\cdot+1=+1$ and $-1\cdot0=0\cdot-1=-1$).

The remaining datasets we consider are those collected in \cite{aref2020dataset} and analyzed in \cite{aref2019balance}. These include three socio-political networks (SPNs): \emph{N.G.H. Tribes}, \emph{Senate US}, \emph{Monastery}; two financial networks (FNs): \emph{Bitcoin Alpha} and \emph{Bitcoin OTC}; and three gene-regulatory networks (GRNs): \emph{E. Coli}, \emph{Macrophage}, \emph{Epidermal Growth Factor Receptor}.

In the SPNs, N.G.H. Tribes collects data about New Guinean Highland Tribes (here, a positive/negative link denotes alliance/rivalry), Monastery corresponds to the last frame of Sampson's data about the relationships between novices in a monastery \cite{sampson1968novitiate} (here, a positive/negative link indicates a positive/negative interaction), and Senate US collects data about the members of the 108th US Senate Congress (here, a positive/negative link indicates trust/distrust or similar/dissimilar political opinions).

The FNs are `who-trust-whom' networks of Bitcoin traders on an online platform: a positive/negative link indicates trust/distrust between users \cite{kumar2016edge}. The networks representing the FNs are weighted, directed ones: hence, after having binarized them by replacing each positive (negative) weight with a $+1$ ($-1$), we have made them undirected by applying the same rules adopted for the MMOG dataset.

\begin{table*}
\begin{center}
\begin{tabular}{p{.11\textwidth}|p{.43\textwidth}|p{.43\textwidth}}
Null model & Topology: free & Topology: fixed\\
\hline
Homogenous & SRGM: each pair of nodes is assigned a plus, a minus or a zero edge with a probability that is pair-independent; all nodes are statistically equivalent. Differently from the recipe adopted in \cite{singh2017measuring,el2012balance}, the parameters defining our SRGM can be unambiguously tuned to reproduce the empirical number of plus and minus edges of any (binary, undirected, signed) network.  & SRGM-FT: the topology is the same as in the real network and the connected pairs of nodes are assigned either a plus one or a minus one, with a probability that is pair-independent. Differently from the recipe adopted in \cite{facchetti2011computing,leskovec2010signed}, the parameters defining our SRGM-FT can be unambiguously tuned to reproduce the empirical number of plus and minus edges of any (binary, undirected, signed) network. The SRGM-FT is the conditional version of the SRGM.\\
\hline
Heterogenous & SCM: each pair of nodes is assigned a plus, a minus or a zero edge, with a probability that is pair-dependent and determined by the different tendencies of nodes to establish positive and negative interactions. This model represents the canonical variant of the one employed in \cite{saiz2017evidence}. & SCM-FT: the topology is the same as in the real network and the connected pairs of nodes are assigned either a plus one or a minus one, with a probability that is pair-dependent and determined by the different tendencies of nodes to establish positive and negative interactions. The SCM-FT is the conditional version of the SCM.\\
\hline
\end{tabular}
\end{center}
\caption{\textbf{Descriptive summary of signed benchmarks.} Table summarizing the properties of the four, signed null models introduced in this article, i.e. the Signed Random Graph Model (SRGM), the Signed Random Graph Model with Fixed Topology (SRGM-FT), the Signed Configuration Model (SCM), the Signed Configuration Model with Fixed Topology (SCM-FT).}
\label{tab1}
\end{table*}

Lastly, in the GRNs each node represents a gene, with positive links indicating activating connections and negative links indicating inhibiting connections. Specifically, \emph{E. Coli} collects data about a transcriptional network of the bacterium \emph{Escherichia Coli}; \emph{Macrophage} collects data about a blood cell that eliminates substances such as cancer cells, cellular debris and microbes; \emph{Epidermal Growth Factor Receptor} collects data about the protein that is responsible for cell division and survival in epidermal tissues.

The vast majority of the networks considered here is characterized by a small link density $c=2L/N(N-1)$ but a large fraction $L^+/L$ of positive links. The density of the CoW network decreases over time from $\simeq0.2$ to $\simeq0.1$ and the percentage of positive links is roughly stationary around $\simeq88\%$; on the other hand, the link density of the MMOG network is stationary around $0.003$ and the percentage of positive links decreases from $\simeq98\%$ to $\simeq60\%$. The SPNs have the largest values of link density among the configurations in our basket, ranging from $\simeq0.3$ to $\simeq0.5$, and percentages of positive links ranging from $\simeq50\%$ to $\simeq75\%$.  \emph{Bitcoin Alpha} has a link density of $\simeq0.002$ and a percentage of positive links of $\simeq90\%$, while \emph{Bitcoin OTC} has a link density of $\simeq0.001$ and a percentage of positive links of $\simeq85\%$. Lastly, the GRNs have a link density ranging from $\simeq10^{-3}$ to $\simeq10^{-2}$ and a percentage of positive links ranging from $\simeq58\%$ to $\simeq66\%$. For more details on the basic descriptive statistics of the networks considered in the present work, see the Supplementary Note 4.

\begin{figure*}[t!]
\centering
\includegraphics[width=0.3\textwidth]{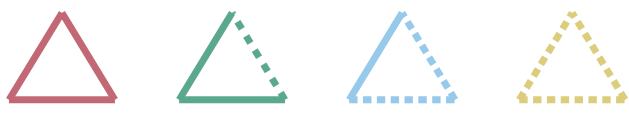}\\
\subfigure[]{\includegraphics[width=0.49\textwidth]{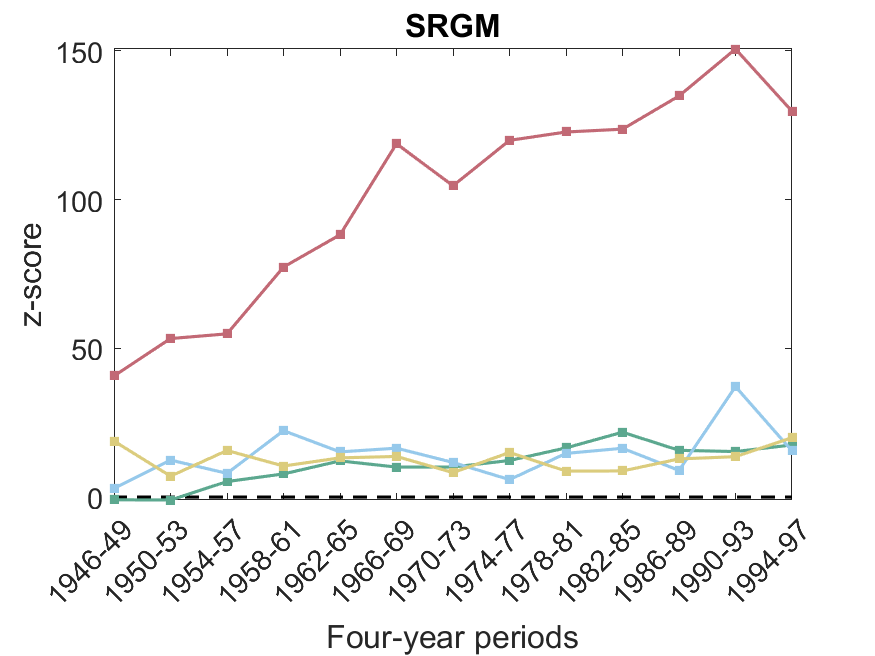}}
\subfigure[]{\includegraphics[width=0.49\textwidth]{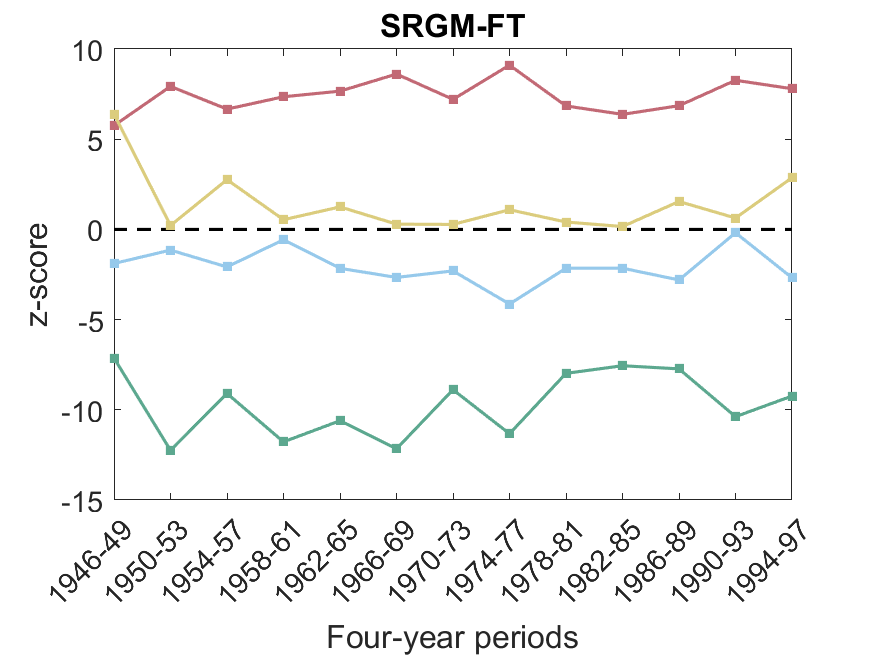}}\\
\subfigure[]{\includegraphics[width=0.49\textwidth]{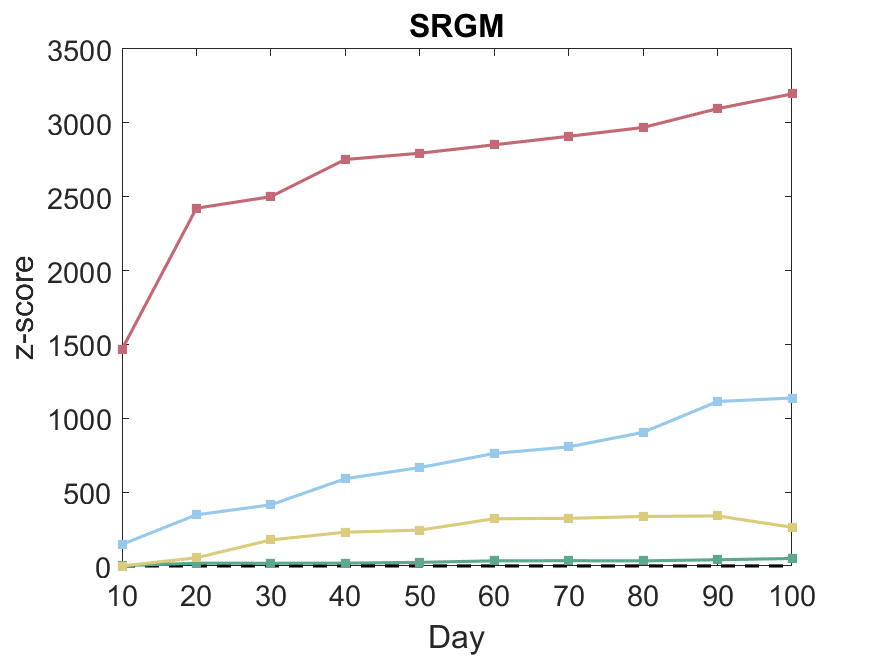}}
\subfigure[]{\includegraphics[width=0.49\textwidth]{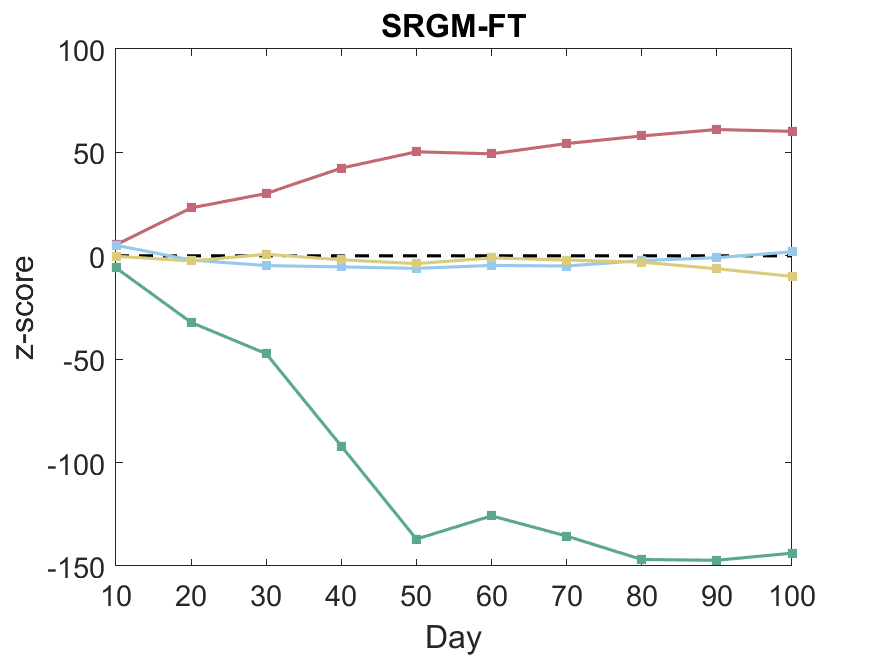}}
\caption{\textbf{Structural (im)balance in the Correlates of Wars and MMOG datasets under homogeneous benchmarks.} Structural (im)balance in the CoW and MMOG datasets: evolution of the $z$-scores of signed triangles under homogeneous benchmarks, i.e. the Signed Random Graph Model (SRGM) and the Signed Random Graph Model with Fixed Topology (SRGM-FT). \textbf{(a)$-$(b)} - 13 snapshots (of 4 years each) of the CoW dataset, covering the period 1946-1997. \textbf{(c)$-$(d)} - 10 snapshots of the MMOG dataset. \textbf{(b)$-$(d)} - The SRGM-FT supports the structural weak balance theory (SWBT) because the only significantly under-represented pattern in the data is also the only one that SWBT considers frustrated (triangle with only one negative link), while the $z$-score of the triangle with all negative edges (which the structural strong balance theory would expect to be under-represented as well) is very low. In any case, the hypothesis that nodes tend to establish balanced triangles with all positive links is supported on both datasets. Results of this type constitute the backbone of the narrative according to which the weak version of the structural balance theory (SBT) is the one that is better supported by data. \textbf{(a)$-$(c)} - Note that the SRGM has all $z$-scores positive, thereby not supporting any version of SBT, a result due to the complete randomization of the topology along with the edge signs: the over-representation of all patterns in the data is merely due to the fact that triangles form with small probability at a purely topological level, given the low link density, irrespective of their signs.}
\label{fig:3}
\end{figure*}

\subsection*{Assessing balance}

In order to test the validity of the two formulations of SBT, at the local level, we need to compare the empirical abundance of the triadic motifs defined in the Methods section with the corresponding expected values calculated under the null models we have introduced. To this aim, a very useful indicator is represented by the $z$-score $z_m=[N_m(\mathbf A^*)-\langle N_m\rangle]/\sigma[N_m]$, where $N_m(\mathbf A^*)$ is the number of occurrences of pattern $m$ in the real network $\mathbf A^*$, $\langle N_m\rangle$ is the expected occurrence of the same pattern under the chosen null model and $\sigma[N_m]=\sqrt{\langle N_m^2\rangle-\langle N_m\rangle^2}$ is the standard deviation of $N_m$ under the same null model. $z_m$ quantifies the number of standard deviations by which the empirical abundance of pattern $m$ differs from the expected one. For instance, after checking for the Gaussianity of $N_m$ under the null model (since it is a sum of dependent random variables, this is ensured by the generalization of the Central Limit Theorem - see the Supplementary Note 6), a result $|z_m|\leq2$ ($|z_m|\leq3$) indicates that the empirical abundance of pattern $m$ is compatible with the one expected under the chosen null model at the $5\%$ ($1\%$) level of statistical significance. On the other hand, a value $|z_m|>2$ ($|z_m|>3$) indicates that the empirical abundance of pattern $m$ is not compatible with the null model at those significance levels. In the latter case, a value $z_m>0$ ($z_m<0$) indicates the tendency of the pattern to be over- (under-)represented in the data with respect to the null model.

\begin{figure*}[t!]
\centering
\includegraphics[width=0.3\textwidth]{fig2.png}\\
\subfigure[]{\includegraphics[width=0.49\textwidth]{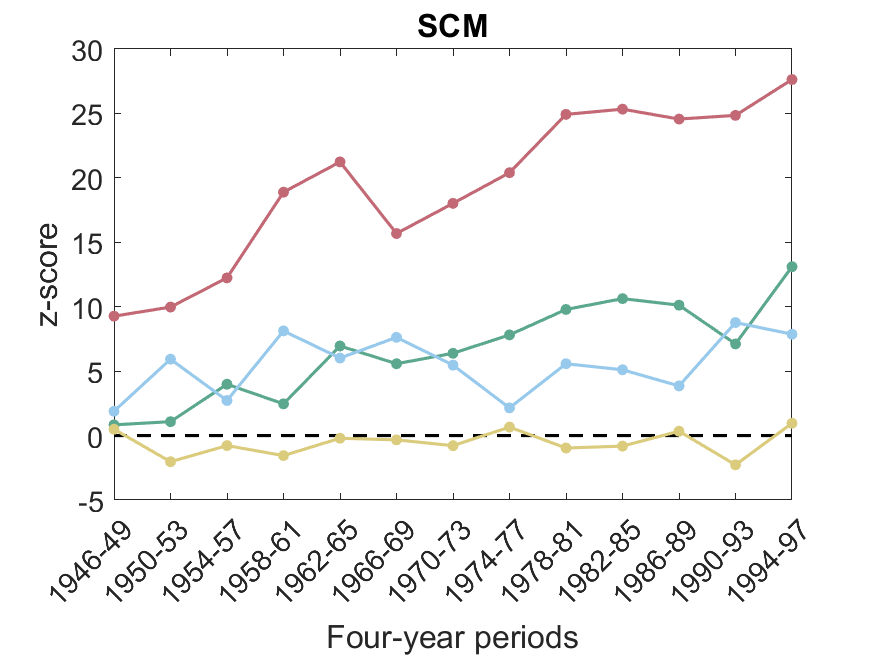}}
\subfigure[]{\includegraphics[width=0.49\textwidth]{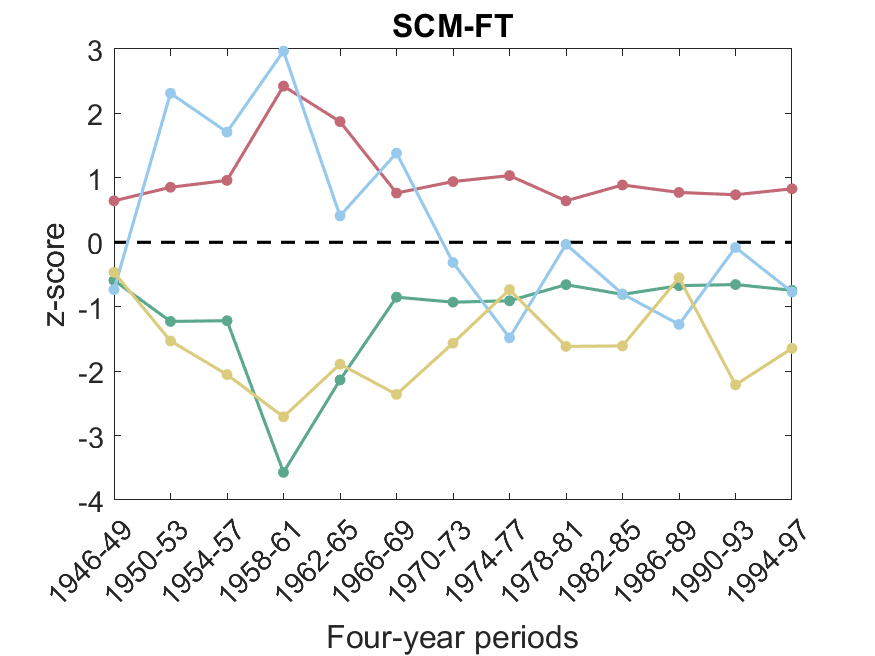}}\\
\subfigure[]{\includegraphics[width=0.49\textwidth]{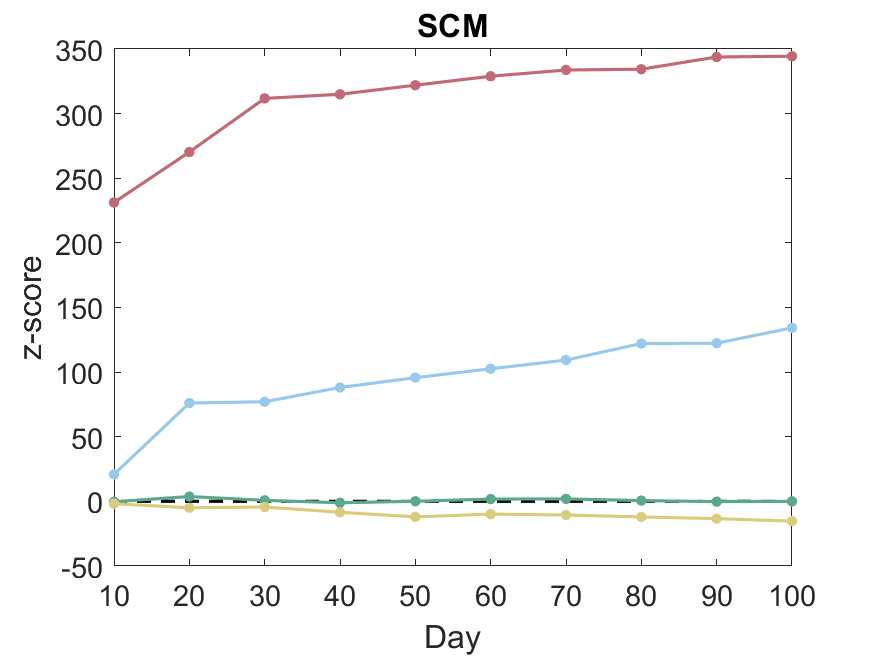}}
\subfigure[]{\includegraphics[width=0.49\textwidth]{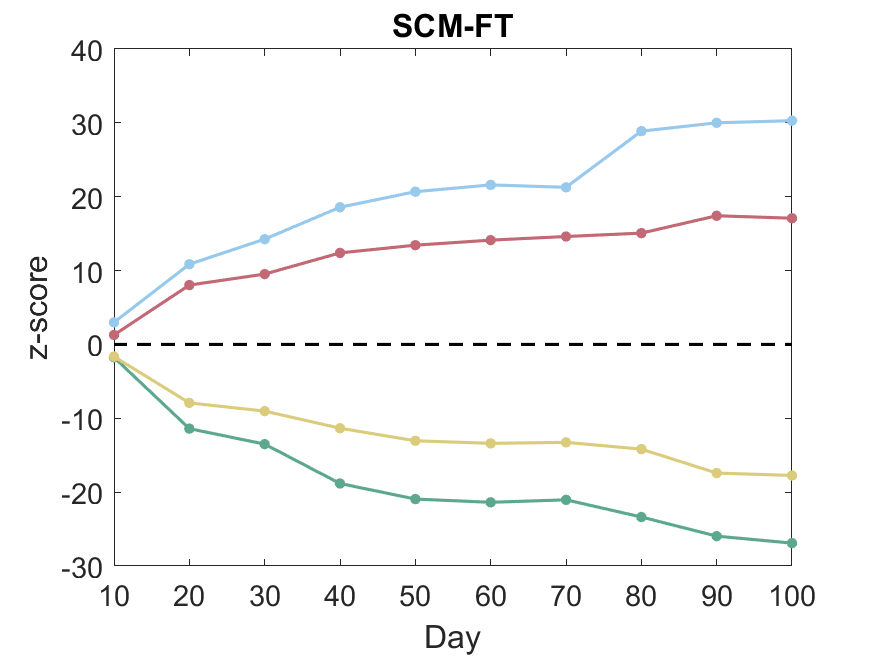}}
\caption{\textbf{Structural (im)balance in the Correlates of Wars and MMOG datasets under heterogeneous benchmarks.} Structural (im)balance in the CoW and MMOG datasets: evolution of the $z$-scores of signed triangles under heterogeneous benchmarks, i.e. the Signed Configuration Model (SCM) and the Signed Configuration Model with Fixed Topology (SCM-FT). \textbf{(a)$-$(b)} - 13 snapshots (of 4 years each) of the CoW dataset, covering the period 1946-1997. \textbf{(c)$-$(d)} - 10 snapshots of the MMOG dataset. The $z$-scores produced by the SCM \textbf{(a)$-$(c)} and the SCM-FT \textbf{(b)$-$(d)} are much smaller, in absolute value, than the corresponding ones produced by the Signed Random Graph Model and the Signed Random Graph Model with Fixed Topology (see Fig.~\ref{fig:3}), showing that node heterogeneity contributes significantly to the overall abundance of signed triangles. The all-positive (balanced) triangle is still strongly over-represented in all cases, but additionally the all-negative (frustrated) triangle is now always under-represented. Under the SCM-FT, the other frustrated triangle (the one with a single negative link) is also systematically under-represented, and these combined results provide support for the structural strong balance theory (SSBT) (particularly evidently for the MMOG data). By contrast, the structural weak balance theory (SWBT) (according to which one would expect the under-representation of only the triangle with a single negative link) is no longer supported. These results provide an alternative narrative w.r.t. the usual one: when the heterogeneity of the signed degrees of nodes is accounted for, statistical evidence supports SSBT rather than SWBT.}
\label{fig:4}
\end{figure*}

$z$-scores can be evaluated either analytically or numerically: implementing the first alternative requires employing the formulas provided in the Supplementary Note 6; implementing the second alternative requires numerically sampling the ensembles of graphs defined by our null models. Since the entries of the adjacency matrix are independent random variables, the unbiased generation of a random matrix $\mathbf{A}\in\mathbb{A}$ can be carried out by drawing a real number $u_{ij}\in U[0,1]$ and posing: for models with varying topology, $a_{ij}=-1$ if $0\leq u_{ij}\leq p_{ij}^-$, $a_{ij}=+1$ if $p_{ij}^-<u_{ij}<p_{ij}^-+p_{ij}^+$ and $a_{ij}=0$ if $p_{ij}^-+p_{ij}^+\leq u_{ij}\leq 1$, for all pairs $i<j$; for models with fixed topology, $a_{ij}=-1$ if $0\leq u_{ij}\leq p_{ij}^-$ and $a_{ij}=+1$ if $p_{ij}^-<u_{ij}\leq 1$, for all pairs ${i<j}$ such that $|a^*_{ij}|=1$ (see the Supplementary Note 5 for an estimation of the time required to sample the ensemble induced by each of our models, for each of our datasets).

\begin{figure*}[t!]
\centering
\includegraphics[width=0.3\textwidth]{fig2.png}\\
\subfigure[]{\includegraphics[width=0.49\textwidth]{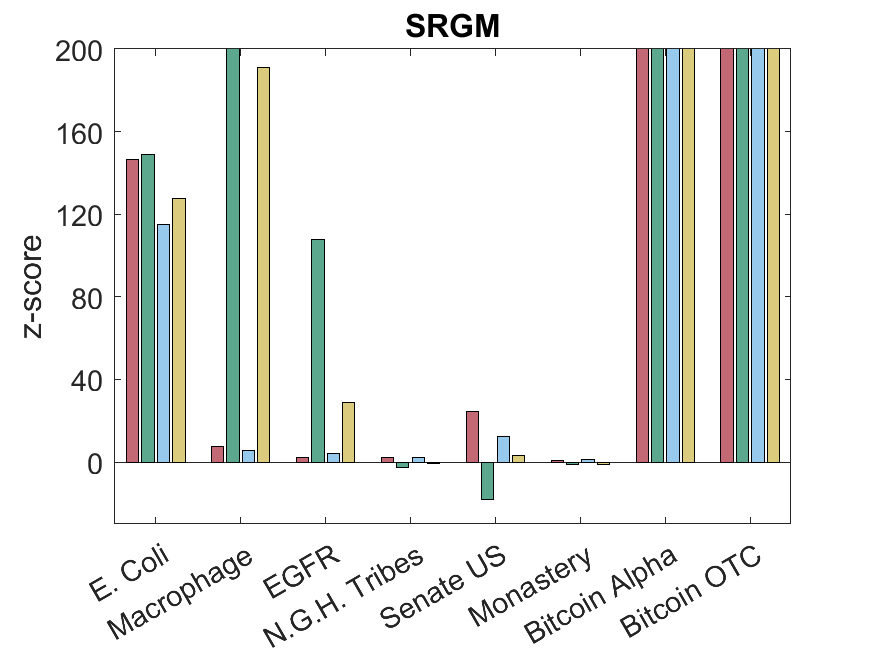}}
\subfigure[]{\includegraphics[width=0.49\textwidth]{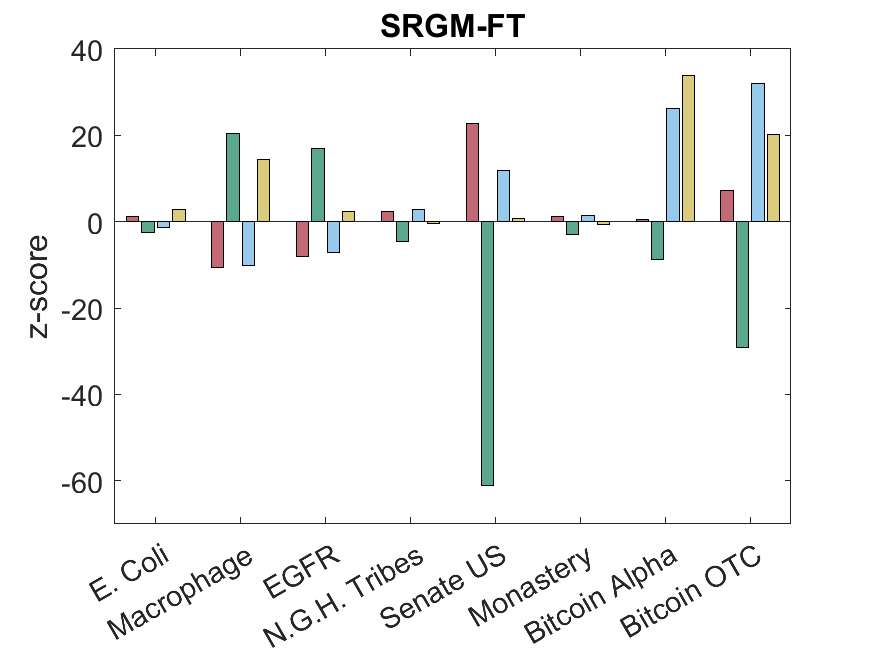}}\\
\subfigure[]{\includegraphics[width=0.49\textwidth]{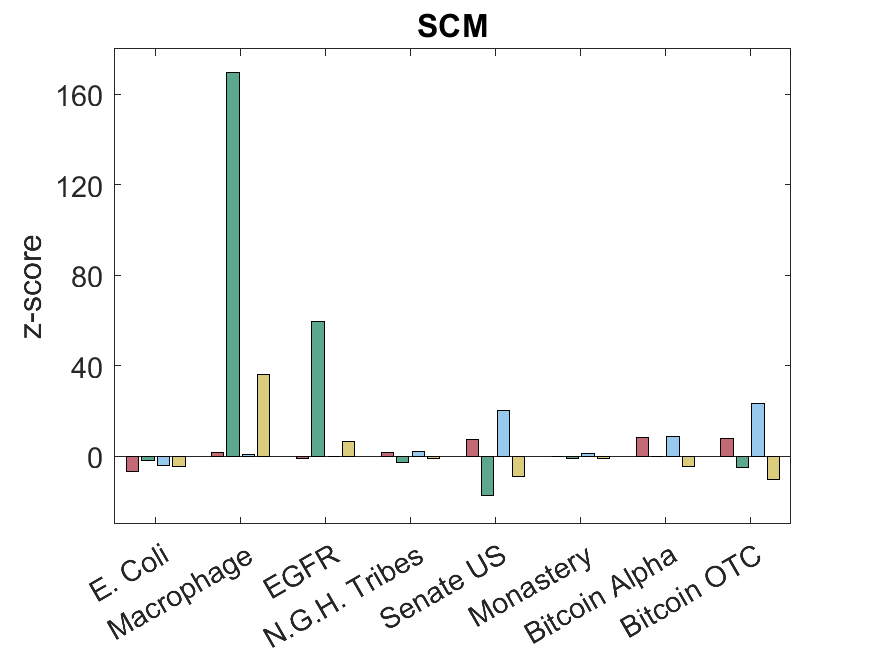}}
\subfigure[]{\includegraphics[width=0.49\textwidth]{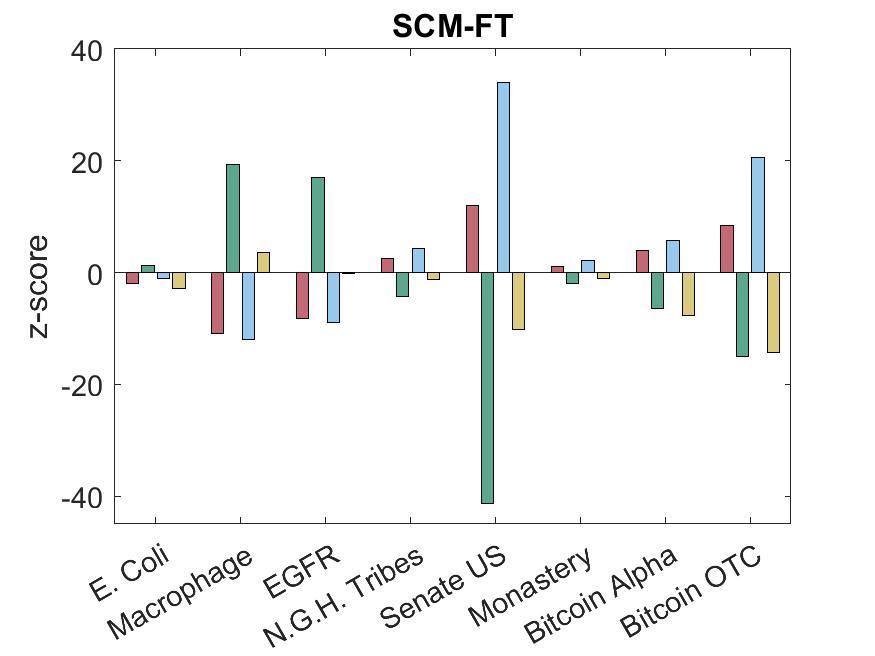}}
\caption{\textbf{Structural (im)balance in social and biological networks under homogeneous and heterogeneous benchmarks.} Structural (im)balance in social and biological networks under homogeneous \textbf{(a)$-$(b)} and heterogeneous \textbf{(c)$-$(d)} null models: $z$-scores of signed triangles for three, socio-political networks (N.G.H. Tribes, Senate US, Monastery), two, financial networks (Bitcoin Alpha, Bitcoin OTC) and, as a comparison, three, biological networks (E. Coli, Macrophage, EGFR). The Signed Random Graph Model (SRGM) produces $z$-scores that are almost always positive and very large for all triangles (balanced and unbalanced) and all signed networks (social and biological), a result confirming that this null model is completely uninformative about structural (im)balance, as it merely highlights that the formation of any triangle, irrespective of its signs, is highly unlikely if the topology is randomized completely. By contrast, the Signed Random Graph Model with Fixed Topology (SRGM-FT) largely supports the structural weak balance theory on social networks, as the only pattern under-represented in the data is the frustrated triangle with a single negative link. Heterogeneous null models, i.e. the Signed Configuration Model (SCM) and the Signed Configuration Model with Fixed Topology (SCM-FT)}, instead, systematically support the structural strong balance theory because they assign positive $z$-scores to the two balanced triangles (all-positive and with two negative links) and negative $z$-scores to the two frustrated triangles (all-negative and with one negative link). Additionally, in biological networks they tend to assign opposite signs (w.r.t. social networks) to most $z$-scores, highlight a strong tendency towards imbalance. These results are fully in line with what already observed with the Correlates of Wars and MMOG datasets in Figs.~\ref{fig:3} and~\ref{fig:4}.
\label{fig:5}
\end{figure*}

\subsection*{Testing structural balance at the microscopic scale}

We report our results starting from the network datasets that have several temporal snapshots (CoW and MMOG). Fig. \ref{fig:3} shows the temporal trends of the $z$-scores for the two networks under the homogeneous null models (SRGM and SRGM-FT). Panels \textbf{(a)$-$(c)} refer to the SRGM and show that the $z$-scores for all triangles, irrespective of their signs, are strong and positive. This means that all triangles are over-represented in the data, with respect to a null model that completely randomizes the topology. This result is not unexpected, as it merely indicates that, given the empirical density of links, it is very unlikely to form triangles completely by chance. These results simply tell us that the SRGM is uninformative about the (im)balance in the data, as it is entirely biased by a purely topological effect. This conclusion is in line with the results in \cite{kirkley2019balance}, which suggested that the SRGM-FT is to be preferred over the SRGM as it provides a better explanation of empirical network structures.

\begin{figure*}[t!]
\centering
\subfigure[]{\includegraphics[width=0.43\textwidth]{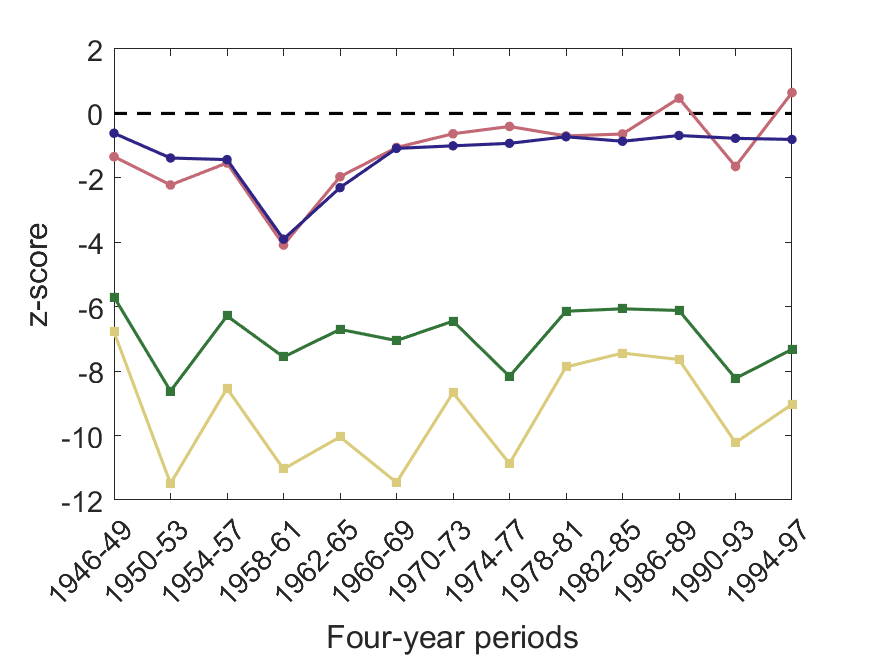}}
\subfigure[]{\includegraphics[width=0.43\textwidth]{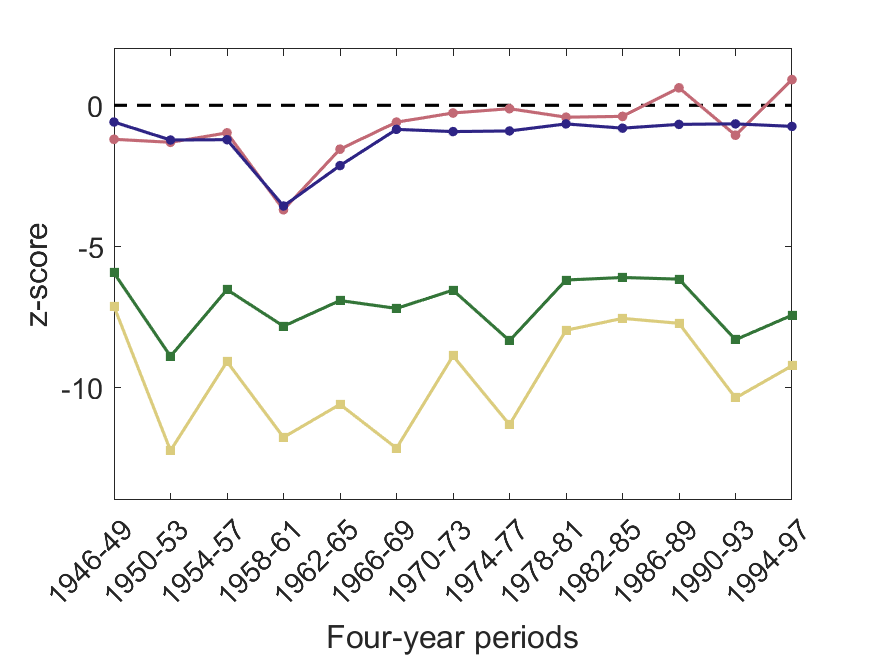}}\\
\subfigure[]{\includegraphics[width=0.43\textwidth]{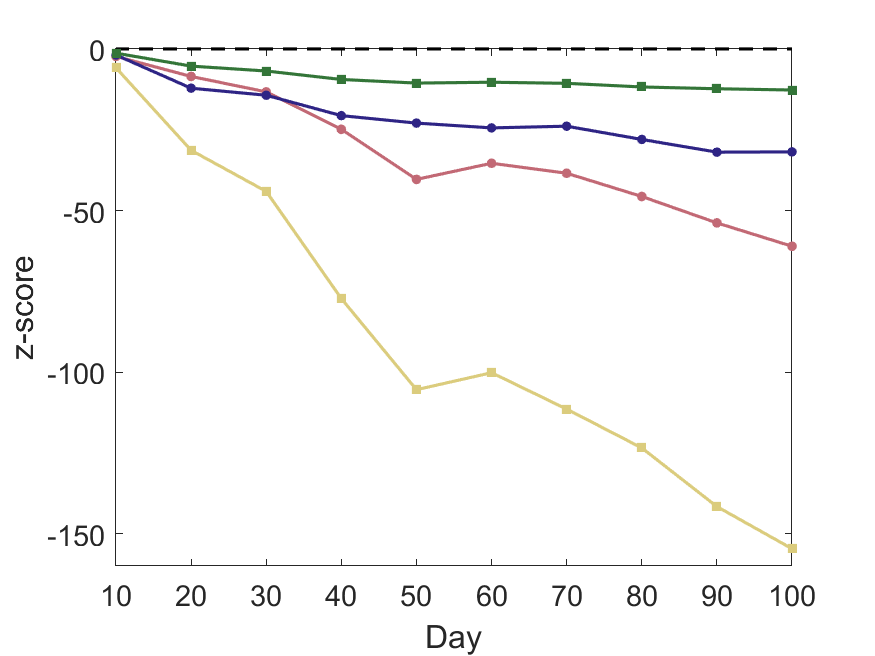}}
\subfigure[]{\includegraphics[width=0.43\textwidth]{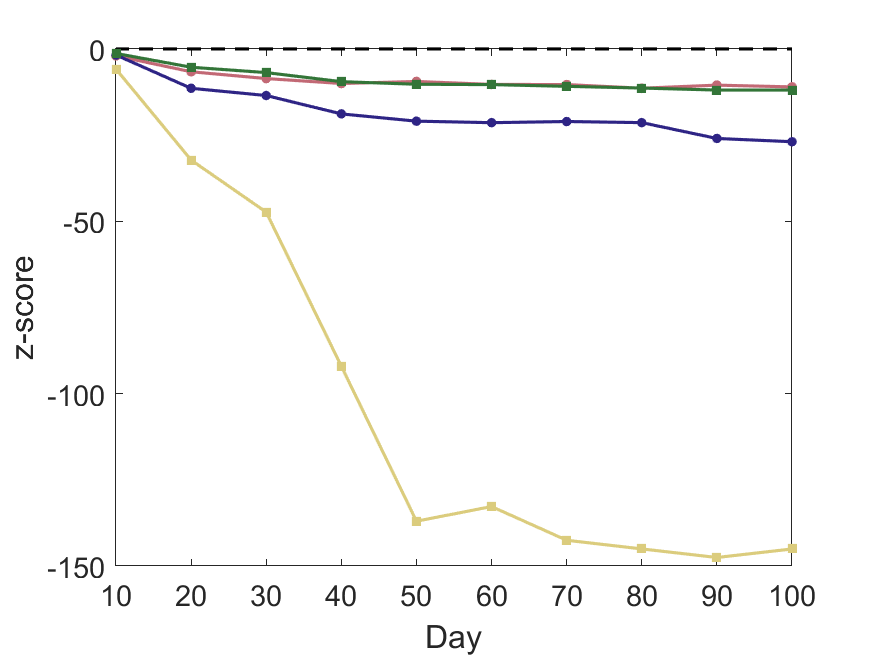}}\\
\subfigure[]{\includegraphics[width=0.43\textwidth]{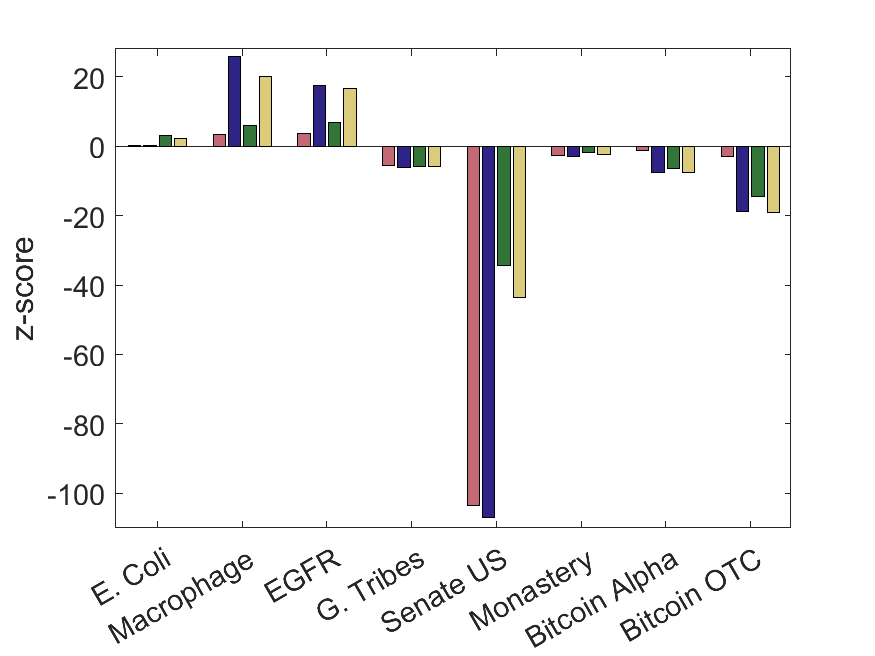}}
\subfigure[]{\includegraphics[width=0.43\textwidth]{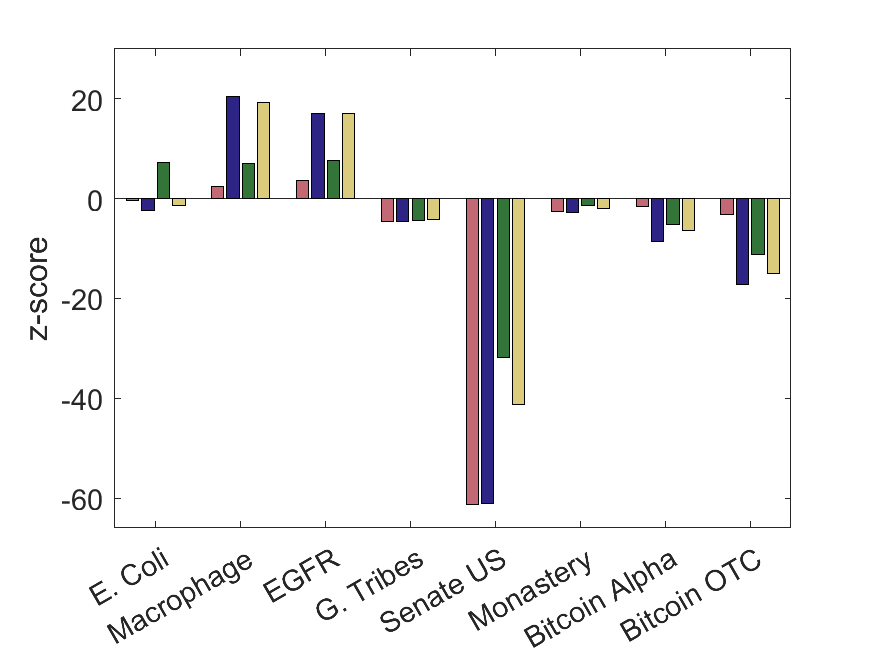}}
\caption{\textbf{Analysis of the $z$-scores of the degree of frustration indices.} Analysis of the $z$-scores of the strong degree of frustration defined in Eq.~\ref{SFI} and the weak degree of frustration defined in Eq.~\ref{WFI}. \textbf{(a)$-$(b)} - 13 snapshots of 4 years each of the Correlates of Wars dataset (covering the period 1946-1997). \textbf{(c)$-$(d)} - 10 snapshots of the MMOG dataset. \textbf{(e)$-$(f)} - Set of social and biological networks. $z$-scores are computed under the Signed Random Graph Model (SRGM) (\textcolor{colSRGM}{$\bullet$}), the Signed Random Graph Model with Fixed Topology (SRGM-FT) (\textcolor{colSRGMFT}{$\bullet$}), the Signed Configuration Model (SCM) (\textcolor{colSCM}{$\bullet$}) and the Signed Configuration Model with Fixed Topology (SCM-FT) (\textcolor{colSCMFT}{$\bullet$}). We see that, with respect to all null models, frustration is under-represented in all social network data and over-represented in all biological data.}
\label{fig:6}
\end{figure*}

By contrast, the results generated under the SRGM-FT clearly support SWBT (see panels \textbf{(b)$-$(d)}). Indeed, the only significantly over-represented pattern in the data is precisely the only one that SWBT considers frustrated (the triangle with a single negative link), whereas the empirical abundance of the triangle with all negative edges (which SSBT would predict to be over-represented as well) remains largely compatible with the null model. Notice that also the empirical abundance of the balanced triangle with two negative edges is close to the one expected under the SRGM-FT, although its $z$-score is typically smaller than the $z$-score of the all-negative triangle. In any case, the abundance of the balanced triangle with three positive edges is significantly over-represented on both datasets. This type of results constitute the backbone of the narrative according to which the weak version of SBT is the one that is better supported by data \cite{kirkley2019balance,leskovec2010signed}.

However, since both the SRGM and the SRGM-FT do not constrain the local (node-specific) signed properties (i.e. the signed degrees of nodes), they cannot disentangle the effects of node heterogeneity from the revealed overall structural (im)balance. For this reason, In Fig. \ref{fig:4} we repeat the analysis of the CoW and MMOG datasets using the SCM and SCM-FT null models. As expected, the resulting $z$-scores are much smaller in absolute value, showing that node heterogeneity in the real networks is in general strong and is responsible for a significant part of the overall measured (im)balance. Therefore, controlling for the local signed degrees is a way to filter out the effects of node heterogeneity in the statistical analysis of structural balance. In general, we see that the triangle will all negative links has now negative $z$-scores in both datasets, under both null models. Similarly, the all-positive triangle remains with positive $z$-scores in all cases. The level of statistical significance (i.e. the absolute value of the $z$-score) is however quite different in the various cases: in general we see an overwhelming over-representation of the two balanced triangles (the all-positive one and the one with two negative links) in the MMOG data under both null models, while for the CoW data the only clearly significant patter is the over-representation of the all-positive triangle in the SCM. Nicely, the SCM-FT gives always negative $z$-scores to both the frustrated triangles (the all-negative one and the one with only one negative link), and most of the time positive $z$-scores to the two balanced triangles. Although the statistical evidence is much stronger for the MMOG data, this result indicates that, if any, the version of SBT supported by the data is SSBT, rather than SWBT. Therefore, as soon as the heterogeneity of the signed degrees of nodes is accounted for, SWBT loses its statistical support, and SSBT is favoured by the data.

We now move to the results obtained on datasets which include other social networks as well as various biological networks, providing a different real-world benchmark where structural balance theory is not expected to apply. From Fig. \ref{fig:5} we confirm that the SRGM is completely uninformative about structural (im)balance, as it produces $z$-scores that are typically positive and very large for all triangles (balanced and unbalanced) and all networks (social and biological). This result simply means that the formation of any triangle, irrespective of its defining signs, is highly unlikely if the topology is completely randomized. By contrast, under the SRGM-FT the only pattern that is under-represented in social network data is the frustrated triangle with a single negative link, a result that largely supports SWBT (on biological data, this pattern is instead either not significant or over-represented). Heterogeneous null models (SCM and SCM-FT), instead, assign positive $z$-scores to the two balanced triangles (all-positive and with two negative links) and negative $z$-scores to the two frustrated triangles (all-negative and with one negative link), thereby systematically supporting SSBT. When used on biological networks, they instead highlight a strong tendency towards imbalance, as they tend to assign opposite signs (w.r.t. social networks) to most $z$-scores. These results confirm and extend what discussed above for the CoW and MMOG datasets, and additionally show that biological networks behave very differently from social networks, somehow favouring frustration. This is an indication that structural balance is indeed an inherent property of social networks.

\begin{figure*}[t!]
\centering
\includegraphics[width=0.8\textwidth]{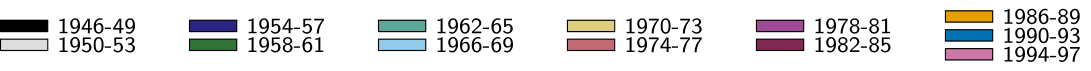}\\
\subfigure[]{\includegraphics[width=0.49\textwidth]{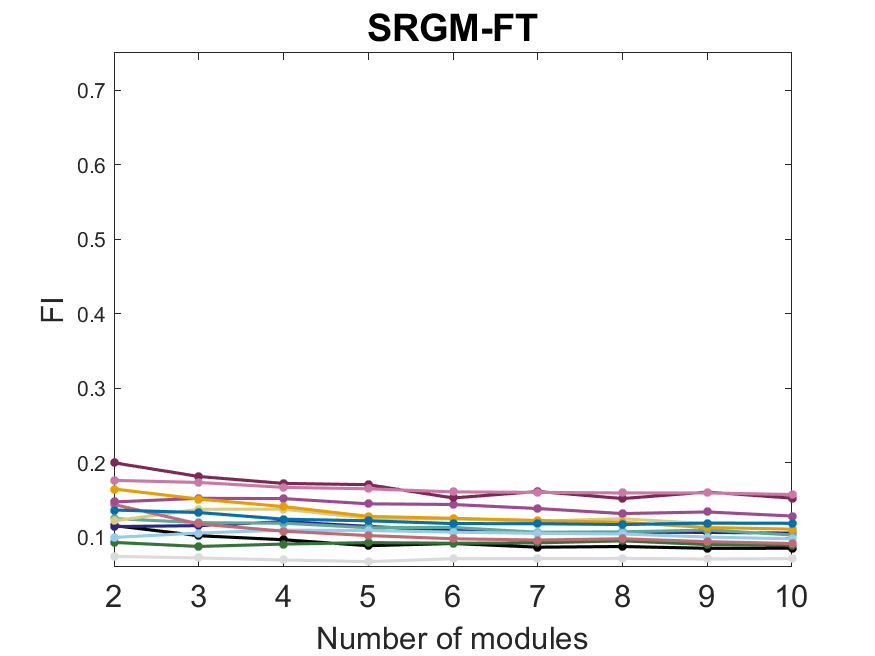}}
\subfigure[]{\includegraphics[width=0.49\textwidth]{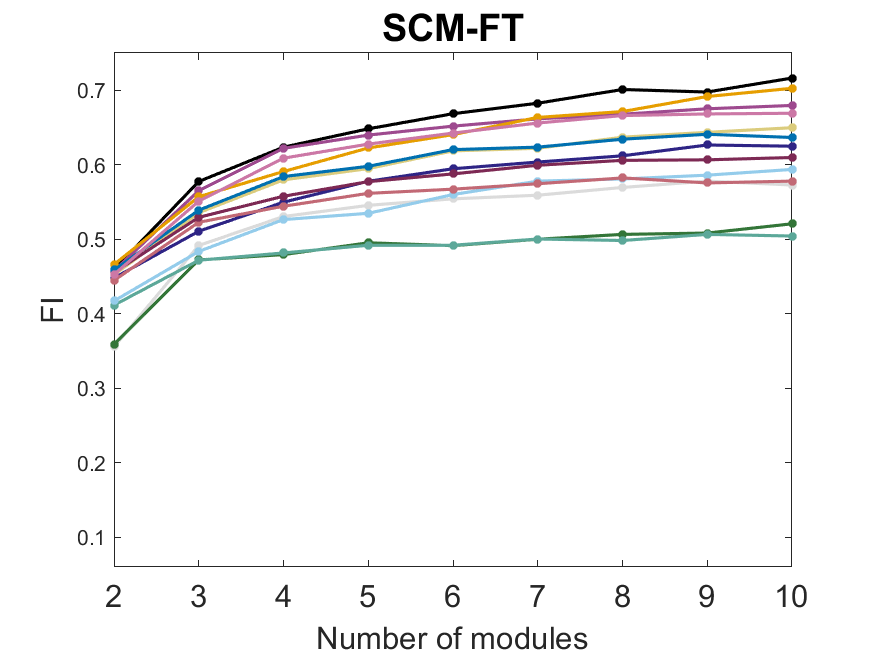}}
\caption{\textbf{Values of the frustration index on several, optimal partitions of the Correlates of Wars dataset.} Value of the frustration index (FI) on several, optimal partitions of the 13 snapshots (of 4 years each) of the CoW dataset, each obtained by maximizing the modularity $Q=-L\cdot(\text{FI}-\langle\text{FI}\rangle)$ for a given number $K$ of modules (communities), using as null models the Signed Random Graph Model with Fixed Topology (SRGM-FT) \textbf{(a)} and the Signed Configuration Model with Fixed Topology (SCM-FT) \textbf{(b)}. While the SRGM-FT reveals a rather flat profile of FI as a function of $K$, with the minimum obtained for a number of groups which is larger than two, the SCM-FT reveals that FI is always clearly minimized for a number of groups $K=2$. Taken together, these results extend our findings at the mesoscale level.}
\label{fig:7}
\end{figure*}

As further evidence supporting the above conclusion, in Fig. \ref{fig:6} we show, for all networks and under all null models, the $z$-scores of the frustration indices SDoF and WDoF defined in Eqs.~\ref{SFI} and~\ref{WFI} respectively. Note that, while the raw values of SDoF and WDoF would not discount the effects of the imposed structural constraints on the raw values of frustration, the $z$-scores measure the level of statistical significance of the `residual' frustration, after the structural constraints are accounted for. We see that, under all null models, the $z$-scores (when significant) are always negative for the social networks (signalling under-representation of the frustration indices in the data) and always positive for the biological networks (signalling over-representation of frustration in the data). Moreover, for the models with fixed topology, the $z$-scores for the heterogeneous null model (SCM-FT) are systematically smaller (in absolute value) than the ones for the corresponding homogeneous model (SRG-FT), indicating that, compared with the latter, the former model `explains more' of the level of empirical frustration observed in the data. The same relation does not apply systematically between the models with varying topology (SCM and SRG), suggesting that models with fixed topology lead to more robust conclusions, as already observed in terms of their support for SWBT or SSBT.

\subsection*{Testing structural balance at the mesoscopic scale}

Motivated by the last observation, we now use the null models with fixed topology to probe the patterns of structural (im)balance at a larger, mesoscopic level, i.e. as portrayed by the community structure deriving from optimally partitioning the nodes into communities with positive internal links and negative external ones. As anticipated, SSBT predicts that the overall level of intra-community frustration, as measured by the FI defined in Eq.~\ref{HOFI}, should be observed after optimally partitioning the nodes into two communities, dominated by positive signs internally and negative signs across. By contrast, SWBT allows for potentially any number of communities, because it bases the idea of balance precisely at the level of communities, so that all-negative triangles (and in principle all-negative cycles of any length) can be explained by placing the constituent nodes across distinct communities. To extract information about the signed community structure from our data, given a null model $\langle a_{ij}\rangle$ for the signed adjacency matrix entry $a_{ij}$, we look for the partition that maximizes the signed modularity, as defined in \cite{gomez2009analysis}:

\begin{align}
Q&=\sum_{i=1}^N\sum_{j>i}[a_{ij}^*-\langle a_{ij}\rangle]\delta_{c_ic_j}\\
&=\sum_{i=1}^N\sum_{j>i}[(a_{ij}^+)^*-(a_{ij}^-)^*-(p_{ij}^+-p_{ij}^-)]\delta_{c_ic_j}\nonumber\\
&=L_\bullet^+-L_\bullet^--\langle L_\bullet^+-L_\bullet^-\rangle\nonumber\\
&=-[(L_\circ^++L_\bullet^-)-\langle L_\circ^++L_\bullet^-\rangle],\nonumber
\end{align}
where $\bullet$ indicates quantities inside and $\circ$ outside the communities (note that $L_\bullet^+=L^+-L_\circ^+$ and that the total number of positive links is preserved under any null model considered here). For null models with fixed topology, a stronger result holds true, i.e. $Q=-L\cdot(\text{FI}-\langle\text{FI}\rangle)$ so that, since $L>0$, maximizing $Q$ becomes equivalent to minimizing  the difference between FI and its expected value (see the Supplementary Note 7). The minimization of FI is another popular approach to finding the optimal partition \cite{doreian1996partitioning} which, however, neglects the information embodied in a null model. Here, we consider a varying number $K=2\dots 10$ of communities and, for each value of $K$, look for the partition maximizing $Q$, using as null model both the SRGM-FT and the SCM-FT.
We then compute the value of FI as a function of $K$, as plotted in Fig.~\ref{fig:7} for the CoW dataset. We find that the trends produced under the SRGM-FT are quite flat, and in no case the minimum of FI is achieved by $K=2$. This result is in line with SWBT, under whose assumptions there is no specific characteristic number of communities that would characterize real networks. By contrast, the SCM-FT produces clearly increasing trends, all starting from a minimum of FI at $K=2$. This result strongly supports SSBT, according to which structural balance can be achieved by placing negative links between two communities, and positive links inside them. Taken together, these results extend our finding that SWBT (SSBT) is supported by homogeneous (heterogeneous) null models.\\

\section{Discussion}

Motivated by the widespread observation that actors in real social networks are characterized by a strong heterogeneity (typically signalled by broad distributions of node-specific topological properties), we have introduced a class of null models for signed networks characterized by either global or local constraints and with either fixed or varying topology. Our formalism provides the equivalent of various important ERGs to the domain of signed graphs. We have used our null models to address the problem of structural balance in real social networks. Our results show that the nature (weak or strong) and statistical strength of evidence of structural balance strongly depends on the null model adopted. In particular, we have shown that the occurrences of signed triangles favour SWBT when a homogeneous, global null model is considered. By contrast, SSBT is favoured by heterogeneous models with local constraints.

Generally speaking, adopting fixed-topology benchmarks seems to enhance the detection of frustration with the corresponding, homogeneous (heterogeneous) variant favouring SWBT (SSBT). As a possible behavioural explanation, we may advance the following one. Social agents are characterized by a certain level of tolerance. Such a level can be set by choosing the proper benchmark: null models constraining global quantities assume agents to be characterized on average by the same expected level of tolerance; by contrast, null models constraining local quantities account for the different levels of tolerance characterizing different agents. Let us imagine that relationships were established according to a random mechanism that preserves the total number of friends and enemies: should this be the case, our results indicate that equally tolerant agents would establish many more $(+,+,-)$ motifs than observed; instead, real-world agents are found to avoid engaging in relationships that lead to the formation of the $(+,+,-)$ pattern. Let us, now, refine the aforementioned picture and imagine that relationships were established according to a random mechanism that preserves the local number of friends and enemies: in this case, diversely tolerant agents would establish many more $(+,+,-)$ and $(-,-,-)$ motifs than observed; instead, real-world agents are found to avoid engaging in relationships that lead to the formation of both the $(+,+,-)$ and the $(-,-,-)$ patterns. Overall, then, agents that cannot choose with whom to interact, but only how, adopt a behaviour strongly avoiding engagement in frustrated relationships.

The same results have been extended to the mesoscale structural level, by finding that the optimal number of communities minimizing the overall level of frustration is $K=2$ with respect to a heterogeneous null model (strongly supporting SSBT), while there is no characteristic optimal number with respect to a homogeneous null model (in line with SWBT). Importantly, we have considered a set of biological networks as a benchmark of real-world systems for which structural balance theory is not expected to apply. We have found a strong level of frustration in biological systems, indicating that structural balance (in either strong or weak form) indeed characterizes social networks.\\

Future directions along which the present analysis could be extended concern the possibility of defining ERGs for directed, as well as weighted, signed networks - the main technical difficulty lying in the proper definition of (binary, directed; weighted, both undirected and directed) constraints. The most natural application of such a formalism would be represented by the statistical validation of the so-called \emph{status theory}, describing social interactions when hierarchies play a role \cite{leskovec2010signed}.

\section{Methods}

\subsection{Formalism and basic quantities}

A \emph{signed} graph is a graph where each edge can be \emph{positive}, \emph{negative} or \emph{missing}. In what follows, we will focus on binary, undirected, signed networks: hence, each edge will be `plus one', `minus one' or `zero'. More formally, for any two nodes $i$ and $j$, the corresponding entry of the signed adjacency matrix $\mathbf{A}$ will be assumed to be $a_{ij}=-1,0,+1$ (with $a_{ij}=a_{ji}$, $\forall\:i<j$). Since the total number of node pairs is $\frac{N(N-1)}{2}=\binom{N}{2}$ and any node pair can be positively connected, negatively connected or disconnected, the total number of possible graph configurations is $|\mathbb{A}|=3^{\binom{N}{2}}$. To ease mathematical manipulations, let us define the following three quantities:

\begin{align}
a_{ij}^-=[a_{ij}=-1],\quad a_{ij}^0=[a_{ij}=0],\quad a_{ij}^+=[a_{ij}=+1]
\end{align}
where we have employed Iverson's brackets notation (see the Supplementary Note 1). These new variables are mutually exclusive, i.e. $\{a_{ij}^-,a_{ij}^0,a_{ij}^+\}=\{(1,0,0),(0,1,0),(0,0,1)\}$, sum to 1, i.e. $a_{ij}^-+a_{ij}^0+a_{ij}^+=1$, and induce two non-negative matrices $\mathbf{A}^+, \mathbf{A}^-$ such that $\mathbf{A}=\mathbf{A}^+-\mathbf{A}^-$ and $|\mathbf{A}|=\mathbf{A}^++\mathbf{A}^-$.

The numbers of positive and negative links are defined as

\begin{equation}
L^+=\sum_{i=1}^N\sum_{j>i}a_{ij}^+\quad\text{and}\quad L^-=\sum_{i=1}^N\sum_{j>i}a_{ij}^-.
\end{equation}

Analogously, the positive and negative degrees of node $i$ are 

\begin{equation}
k^+_i=\sum_{j\neq i}a_{ij}^+\quad\text{and}\quad k^-_i=\sum_{j\neq i}a_{ij}^-
\end{equation}
(naturally, $2L^+=\sum_{i=1}^Nk_i^+$ and $2L^-=\sum_{i=1}^Nk_i^-$). The advantage of adopting Iverson's brackets is that each quantity is now computed from a matrix with positive entries, so that all quantities of interest are positive as well.

\begin{figure}[t!]
\centering
\includegraphics[width=0.11\textwidth]{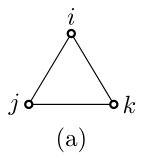}
\includegraphics[width=0.11\textwidth]{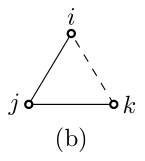}
\includegraphics[width=0.11\textwidth]{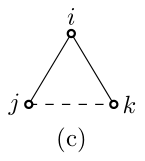}
\includegraphics[width=0.11\textwidth]{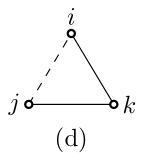}\\
\includegraphics[width=0.11\textwidth]{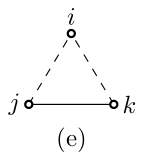}
\includegraphics[width=0.11\textwidth]{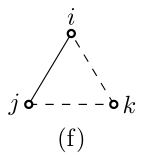}
\includegraphics[width=0.11\textwidth]{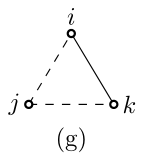}
\includegraphics[width=0.11\textwidth]{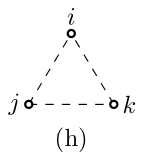}
\caption{\textbf{Signed triadic motifs.} Signed triangles involving three representative nodes $i,j,k$. Solid lines denote positive edges while dashed lines denote negative edges. According to the strong version of balance theory, triangles (a), (e), (f) and (g) are balanced, while triangles (b), (c), (d) and (h) are unbalanced. According to the weak version, triangles (a), (e), (f), (g) and (h) are balanced, while triangles (b), (c) and (d) are unbalanced.}
\label{fig:2}
\end{figure}

Let us now follow \cite{giscard2017evaluating}, according to which `\emph{local measures attain efficiency by focusing only on cycles of particular, usually short, length, such as 3-cycles (triads)}', and consider the signed triads depicted in Fig. \ref{fig:2}. As mentioned above, according to BT social systems tend to arrange themselves into configurations satisfying the principles `the friend of my friend is my friend', `the friend of my enemy is my enemy', `the enemy of my friend is my enemy', `the enemy of my enemy is my friend' \cite{heider1946attitudes}. SSBT formalizes this concept by stating that the overall network balance increases with the fraction of triangles having an even number of negative edges (said to be balanced or `positive' since the product of the edge sings is a `plus') and decreases with the fraction of triangles having an odd number of negative edges (said to be unbalanced or `negative' since the product of the edge sings is a `minus'). SWBT, on the other hand, considers the triangle with all negative edges balanced as well.

Notice that the product of an arbitrary number of matrices of type $\mathbf{A}^+$ and $\mathbf{A}^-$ allows us to count the abundance of closed walks whose signature matches the sequence of signs of the matrices. For example, the expression $[\mathbf{A}^+\mathbf{A}^-\mathbf{A}^+]_{ii}$ counts the number of closed walks, starting from and ending at $i$,  of length 3 and signature $(+-+)$. Similarly, the expression $[\mathbf{A}^+\mathbf{A}^+\mathbf{A}^-\mathbf{A}^+]_{ii}=[(\mathbf{A}^+)^2\mathbf{A}^-\mathbf{A}^+]_{ii}$ counts the number of closed walks, starting from and ending at $i$, of length 4 and signature $(++-+)$. Therefore, the level of balance of a network can be quantified by the abundance of (non-degenerate) triangles with an even number of negative links, i.e.

\begin{align}
T^{(+++)}&=\frac{1}{3}\sum_{i=1}^NT^{(+++)}_i=\frac{\text{Tr}[(\mathbf{A}^+)^3]}{6},\\
T^{(+--)}&=\frac{1}{2}\sum_{i=1}^NT^{(+--)}_i=\frac{\text{Tr}[\mathbf{A}^+(\mathbf{A}^-)^2]}{2}.
\end{align}

Similarly, the level of frustration of a network can be quantified by the abundance of (non-degenerate) triangles with an odd number of negative links, i.e.

\begin{align}
T^{(---)}&=\frac{1}{3}\sum_{i=1}^NT^{(---)}_i=\frac{\text{Tr}[(\mathbf{A}^-)^3]}{6},\\
T^{(++-)}&=\frac{1}{2}\sum_{i=1}^NT^{(++-)}_i=\frac{\text{Tr}[(\mathbf{A}^+)^2\mathbf{A}^-]}{2}
\end{align}
(see 
the Supplementary Note 2 for more details).

The above expressions form the basis for the definition of several indices quantifying the level of balance of a network. For instance, the total number of balanced patterns according to SSBT is $\#_{\mytriangle}^{sb}=T^{(+++)}+T^{(+--)}$, while the total number of unbalanced patterns is $\#_{\mytriangle}^{su}=T^{(---)}+T^{(++-)}$. Hence, we may naturally define a `strong degree of balance' index (SDoB) and a corresponding `strong degree of frustration' index (SDoF) as

\begin{equation}
\text{SDoB}=\frac{\#_{\mytriangle}^{sb}}{\#_{\mytriangle}^{sb}+\#_{\mytriangle}^{su}},\quad \text{SDoF}=1-\text{SDoB}.\label{SFI}
\end{equation}

On the other hand, the total number of balanced patterns according to SWBT is $\#_{\mytriangle}^{wb}=T^{(+++)}+T^{(+--)}+T^{(---)}$, while the total number of unbalanced patterns is $\#_{\mytriangle}^{wu}=T^{(++-)}$. Hence, we can introduce a `weak degree of balance' index (WDoB) and a corresponding `weak degree of frustration' index (WDoF) as

\begin{equation}
\text{WDoB}=\frac{\#_{\mytriangle}^{wb}}{\#_{\mytriangle}^{wb}+\#_{\mytriangle}^{wu}},\quad \text{WDoF}=1-\text{WDoB}.\label{WFI}
\end{equation}

The indices defined above quantify imbalance by counting the abundance of locally frustrated, short cycles. Other indices of frustration account for the effect of structural (im)balance at larger scales. In particular, at the mescoscopic level, the effect of structural balance would result in a signed network being partitioned  into communities of nodes, where intra-community links would be preferentially positive and inter-community links would be preferentially negative. Correspondingly, one can define the \emph{frustration index}

\begin{equation}
\text{FI}=\frac{L^+_\circ+L^-_\bullet}{L}
\label{HOFI}
\end{equation}
measuring the percentage of misplaced links, i.e. the total number $L^+_\circ$ of positive links between communities, plus the total number $L^-_\bullet$ of negative links within communities, divided by the total number $L$ of links (the formalism is adapted from the one in \cite{marchese2022detecting}). According to SSBT, the node partition minimizing frustration (and, correspondingly, the FI) should be the one corresponding to only two communities, because such bipartition can be realized without creating all-negative triangles. By contrast, SWBT allows for a larger number of communities, because the theory justifies the presence of all-negative triangles precisely by assuming that the three participating nodes are all placed in different communities.

\subsection{Null models of binary, undirected, signed graphs}

Here we generalize the ERG framework to account for models of binary, undirected, signed graphs. We will follow the analytical approach introduced in \cite{park2004statistical}, and further developed in \cite{Squartinia}, aimed at identifying the functional form of the maximum-entropy probability distribution (over all graphs of a chosen type) that preserves a desired set of empirical constraints on average. Specifically, this approach looks for the graph probability $P(\mathbf{A})$ that maximizes Shannon entropy

\begin{equation}
S=-\sum_{\mathbf{A}\in\mathbb{A}}P(\mathbf{A})\ln P(\mathbf{A})
\end{equation}
(where the sum runs over the set $\mathbb{A}$, of cardinality $|\mathbb{A}|=3^{\binom{N}{2}}$, of all binary, undirected, signed graphs) under a set of constraints enforcing the expected value of a chosen set of properties. The formal solution to this problem is the exponential probability $P(\mathbf{A})=e^{-H(\mathbf{A})}/Z$ where $H(\mathbf{A})$ (the \emph{Hamiltonian}) is a linear combination of the constrained properties, each multiplied by a corresponding Lagrange multiplier, and $Z=\sum_{\mathbf{A}\in\mathbb{A}}e^{-H(\mathbf{A})}$ is the normalizing constant (or \emph{partition function}).

In what follows, we will consider two classes of models, i.e. those keeping the network topology fixed and those letting the topology vary along with the edge signs. The first class is better suited for studying systems where actors cannot choose `with whom' to interact, but only `how' (e.g. because workers necessarily interact with colleagues at the same workplace or because countries necessarily interact with each other). On the other hand, the second class is better suited for studying systems where actors can choose their neighbours as well \cite{kirkley2019balance}. Whatever the situation, comparing the two types of models for the same network is in any case instructive, as it allows the role played by signed constraints to be disentangled from the one played by non-signed (purely topological) constraints.

\subsubsection{1. Signed Random Graph Model}

As the simplest example, the Signed Random Graph Model (SRGM) is defined by two, global constraints: $L^+(\mathbf{A})$ and $L^-(\mathbf{A})$. The Hamiltonian

\begin{equation}
H(\mathbf{A})=\alpha L^+(\mathbf{A})+\beta L^-(\mathbf{A})
\end{equation}
leads to a graph probability $P_\text{SRGM}(\mathbf{A})$ that factorizes over the individual entries of the matrix $\mathbf{A}$, which are i.i.d. random variables described by the finite scheme

\begin{equation}
a_{ij}\sim
\begin{pmatrix}
-1 & 0 & +1\\
p^- & p^0 & p^+
\end{pmatrix}\quad\forall\:i<j
\end{equation}
with $p^0\equiv1-p^--p^+$ and

\begin{align}
p^-&\equiv\frac{e^{-\beta}}{1+e^{-\alpha}+e^{-\beta}}\equiv\frac{y}{1+x+y},\\
p^+&\equiv\frac{e^{-\alpha}}{1+e^{-\alpha}+e^{-\beta}}\equiv\frac{x}{1+x+y},
\end{align}
where $x\equiv e^{-\alpha}$ and $y\equiv e^{-\beta}$ are transformed Lagrange multipliers (see the Supplementary Note 3 for more details). In other words, positive, negative and missing links appear with probability $p^+$, $p^-$ and $p^0$ respectively. The parameters $(x,y)$ determining these probabilities are tuned by maximizing the log-likelihood function $\mathcal{L}_\text{SRGM}(x,y)\equiv\ln P_\text{SRGM}(\mathbf{A}^*|x,y)$ where $\mathbf{A}^*$ denotes the specific, empirical network under analysis. This maximization, according to a general result~\cite{Garlaschelli2008}, leads to an equality between the expected and the empirical values of the constraints, i.e. $\langle L^+\rangle_\text{SRGM}=L^+(\mathbf{A}^*)$ and $\langle L^-\rangle_\text{SRGM}=L^-(\mathbf{A}^*)$ . This leads to $p^0\equiv1-p^--p^+$ and

\begin{align}
p^+=\frac{2L^+(\mathbf{A}^*)}{N(N-1)},\quad p^-=\frac{2L^-(\mathbf{A}^*)}{N(N-1)}.
\end{align}

\subsubsection{2. Signed Random Graph Model with Fixed Topology}

We can also consider a variant of the SRGM that keeps the topology of the network under analysis fixed while (solely) randomizing the edge signs. The Hamiltonian is again $H(\mathbf{A})=\alpha L^+(\mathbf{A})+\beta L^-(\mathbf{A})$, but the random variables are now only the entries of the adjacency matrix corresponding to the connected pairs of nodes in the original network $\mathbf{A}^*$, i.e. the ones for which $|a^*_{ij}|=1$. These entries obey the finite scheme

\begin{equation}
a_{ij}\sim
\begin{pmatrix}
-1 & +1\\
p^- & p^+
\end{pmatrix}\quad\forall\:i<j\:|\:|a^*_{ij}|=1
\end{equation}
with

\begin{align}
p^-&\equiv\frac{e^{-\beta}}{e^{-\alpha}+e^{-\beta}}\equiv\frac{y}{x+y},\\
p^+&\equiv\frac{e^{-\alpha}}{e^{-\alpha}+e^{-\beta}}\equiv\frac{x}{x+y}.
\end{align}

In other words, each entry for which $|a^*_{ij}|=1$ obeys a Bernoulli distribution with probabilities determined by the (Lagrange multipliers of the) imposed constraints (see the Supplementary Note 3 for more details). 
The maximization of the likelihood function $\mathcal{L}_\text{SRGM-FT}(x,y)\equiv\ln P_\text{SRGM-FT}(\mathbf{A}^*|x,y)$ (where FT stands for `fixed topology') leads to

\begin{align}
p^+=\frac{L^+(\mathbf{A}^*)}{L(\mathbf{A}^*)},\quad p^-=\frac{L^-(\mathbf{A}^*)}{L(\mathbf{A}^*)}
\end{align}
with $L(\mathbf{A}^*)$ representing the (empirical) number of links.

The SRGM and the SRGM-FT are related via the simple expression

\begin{equation}
P_\text{SRGM}(\mathbf{A})=P_\text{RGM}(\mathbf{A})\cdot P_\text{SRGM-FT}(\mathbf{A})
\end{equation}
involving the probability of the usual `unsigned' (Erd\H{o}s-R\'enyi) Random Graph Model (RGM) and stating that the probability of connecting any two nodes with, say, a positive link can be rewritten as the probability of connecting them with an unsigned link times the probability of assigning the latter a `plus one': in formulas, $p^+_\text{SRGM}/p_\text{RGM}^{\textcolor{white}{+}}=p^+_\text{SRGM-FT}$ (see the Supplementary Note 3 for more details). Notice that if the network under analysis is completely connected, the SRGM and the SRGM-FT coincide.

Although the recipes implemented in \cite{singh2017measuring} and \cite{facchetti2011computing,leskovec2010signed} are similar in spirit to the SRGM and the SRGM-FT, we provide the rigorous derivation of both models, together with the proof that the latter is nothing but the conditional version of the former.

\subsubsection{3. Signed Configuration Model}

The two aforementioned versions of the SRGM are defined by constraints which are global in nature. However, real social networks are characterized by an inherent heterogeneity of actors, which results in broad distributions of the number of connections of actors. To avoid statistical conclusions about structural balance that are biased by the application of homogeneous null models to intrinsically heterogeneous networks, it is therefore important to introduce models with local (node-specific) constraints. 

We, therefore, introduce the Signed Configuration Model (SCM) via the Hamiltonian

\begin{equation}
H(\mathbf{A})=\sum_{i=1}^N[\alpha_i k_i^+(\mathbf{A})+\beta_i k_i^-(\mathbf{A})]
\end{equation}
which constraints the expected value of the signed degrees $\{k_i^+(\mathbf{A})\}_{i=1}^N$ and $\{k_i^-(\mathbf{A})\}_{i=1}^N$ of all nodes. The resulting graph probability $P_\text{SCM}(\mathbf{A})$ is still factorized over independent entries of the matrix $\mathbf{A}$, however these entries are no longer identically distributed. Rather, they obey the finite scheme

\begin{equation}
a_{ij}\sim
\begin{pmatrix}
-1 & 0 & +1\\
p_{ij}^- & p_{ij}^0 & p_{ij}^+
\end{pmatrix}\quad\forall\:i<j
\end{equation}
with
\begin{align}
p_{ij}^-&\equiv\frac{e^{-(\beta_i+\beta_j)}}{1+e^{-(\alpha_i+\alpha_j)}+e^{-(\beta_i+\beta_j)}}\equiv\frac{y_iy_j}{1+x_ix_j+y_iy_j},\\
p_{ij}^+&\equiv\frac{e^{-(\alpha_i+\alpha_j)}}{1+e^{-(\alpha_i+\alpha_j)}+e^{-(\beta_i+\beta_j)}}\equiv\frac{x_ix_j}{1+x_ix_j+y_iy_j}
\end{align}
and $p_{ij}^0\equiv1-p_{ij}^--p_{ij}^+$ (see the Supplementary Note 3 for more details). In other words, the two nodes $i$ and $j$ are connected by a positive, negative or missing link with probability $p_{ij}^+$, $p_{ij}^-$ or $p_{ij}^0$ respectively. The parameters of the SCM are found by maximizing the log-likelihood $\mathcal{L}_\text{SCM}(\{x_i\}_{i=1}^N,\{y_i\}_{i=1}^N)\equiv\ln P_\text{SCM}(\mathbf{A}^*|\{x_i\}_{i=1}^N,\{y_i\}_{i=1}^N)$, and the result ensures that $\langle k_i^+\rangle_\text{SCM}=k_i^+(\mathbf{A}^*)$ and $\langle k_i^-\rangle_\text{SCM}=k_i^-(\mathbf{A}^*)$, $\forall\:i$. Explicitly,

\begin{align}
k_i^+(\mathbf{A}^*)&=\sum_{j\neq i}\frac{x_ix_j}{1+x_ix_j+y_iy_j}=\langle k_i^+\rangle\quad\forall\:i,\\
k_i^-(\mathbf{A}^*)&=\sum_{j\neq i}\frac{y_iy_j}{1+x_ix_j+y_iy_j}=\langle k_i^-\rangle\quad\forall\:i,
\end{align}
which is a system of $2N$ coupled non-linear equations that have a unique solution to be found numerically, e.g. following the guidelines provided in \cite{vallarano2021} (see the Supplementary Note 4). If $x_i\ll1$ and $y_i\ll1$ $\forall\:i$, a `sparse' approximation of the SCM holds true and one can factorize the probabilities as $p_{ij}^+\simeq x_ix_j$ and $p_{ij}^-\simeq y_iy_j$, $\forall\:i<j$. Such a manipulation leads us to

\begin{align}
p_{ij}^+\simeq\frac{k_i^+(\mathbf{A}^*)k_j^+(\mathbf{A}^*)}{2L^+(\mathbf{A}^*)},\quad p_{ij}^-\simeq\frac{k_i^-(\mathbf{A}^*)k_j^-(\mathbf{A}^*)}{2L^-(\mathbf{A}^*)},
\end{align}
a result that we may call the Signed Chung-Lu Model (SCLM).

To the best of our knowledge, the canonical SCM described here has no precedents in the literature: Ref. \cite{saiz2017evidence} provides a microcanonical version of the model, while the variant considered in \cite{derr2018signed} is just an approximation of the full canonical model derived here. Notice that the bipartite version of the SCM can be recovered as a special case of the Bipartite Score Configuration Model, proposed in \cite{becatti2019}.

\subsubsection{4. Signed Configuration Model with Fixed Topology}

As for the SRGM, a variant of the SCM that keeps the topology of the network under analysis fixed while (solely) randomizing the signs of the edges can be defined. Again, the Hamiltonian reads $H(\mathbf{A})=\sum_{i=1}^N[\alpha_i k_i^+(\mathbf{A})+\beta_i k_i^-(\mathbf{A})]$ but the only random variables are those corresponding to the connected pairs of nodes in the empirical graph, i.e. the ones for which $|a^*_{ij}|=1$. Each of them obeys the finite scheme

\begin{equation}
a_{ij}\sim
\begin{pmatrix}
-1 & +1\\
p_{ij}^- & p_{ij}^+
\end{pmatrix}\quad\forall\:i<j\:|\:|a^*_{ij}|=1
\end{equation}
with
\begin{align}
p_{ij}^-&\equiv\frac{e^{-(\beta_i+\beta_j)}}{e^{-(\alpha_i+\alpha_j)}+e^{-(\beta_i+\beta_j)}}\equiv\frac{y_iy_j}{x_ix_j+y_iy_j},\\
p_{ij}^+&\equiv\frac{e^{-(\alpha_i+\alpha_j)}}{e^{-(\alpha_i+\alpha_j)}+e^{-(\beta_i+\beta_j)}}\equiv\frac{x_ix_j}{x_ix_j+y_iy_j}.
\end{align}

Maximizing the log-likelihood $\mathcal{L}_\text{SCM-FT}(\{x_i\}_{i=1}^N,\{y_i\}_{i=1}^N)\equiv\ln P_\text{SCM-FT}(\mathbf{A}^*|\{x_i\}_{i=1}^N,\{y_i\}_{i=1}^N)$ leads to the equations

\begin{align}
k_i^+(\mathbf{A}^*)&=\sum_{j\neq i}|a^*_{ij}|\frac{x_ix_j}{x_ix_j+y_iy_j}=\langle k_i^+\rangle\quad\forall\:i,\\
k_i^-(\mathbf{A}^*)&=\sum_{j\neq i}|a^*_{ij}|\frac{y_iy_j}{x_ix_j+y_iy_j}=\langle k_i^-\rangle\quad\forall\:i,
\end{align}
which can be solved numerically - again, along the guidelines provided in \cite{vallarano2021} (see the Supplementary Note 4 for more details).

Similarly to what has been observed for the SRGM and the SRGM-FT, the SCM and the SCM-FT are related via

\begin{equation}
P_\text{SCM}(\mathbf{A})=P_\text{ICM}(\mathbf{A})\cdot P_\text{SCM-FT}(\mathbf{A}),
\end{equation}
an expression involving the probability of an ordinary (unsigned) `induced' Configuration Model (ICM) with probabilities such that $(p_{ij}^+)_\text{SCM}/(p_{ij}^{\textcolor{white}{+}})_\text{ICM}=(p_{ij}^+)_\text{SCM}/[(p_{ij}^+)_\text{SCM}+(p_{ij}^-)_\text{SCM}]=(p_{ij}^+)_\text{SCM-FT}$, for any pair of nodes (see the Supplementary Note 3). Notice that, if the network under consideration is completely connected, then the SCM and the SCM-FT coincide.

\section{Acknowledgments}

We thank Michael Szell for sharing the Pardus dataset employed for the present analysis.

This work is supported by the European Union - NextGenerationEU - National Recovery and Resilience Plan (Piano Nazionale di Ripresa e Resilienza, PNRR), project `SoBigData.it - Strengthening the Italian RI for Social Mining and Big Data Analytics' - Grant IR0000013 (n. 3264, 28/12/2021). This work is also supported by the project NetRes - `Network analysis of economic and financial resilience', Italian DM n. 289, 25-03-2021 (PRO3 Scuole), CUP D67G22000130001 (\url{https://netres.imtlucca.it}). DG acknowledges support from the Dutch Econophysics Foundation (Stichting Econophysics, Leiden, the Netherlands) and the Netherlands Organization for Scientific Research (NWO/OCW). RL acknowledges support from the EPSRC grants n. EP/V013068/1 and EP/V03474X/1.

\section{Author contributions statement}

Study conception and design: DG, RL, FS, TS. Data collection: AG. Analysis and interpretation of results: AG, DG, RL, FS, TS. Draft manuscript preparation: DG, TS.

\section{Data availability}

Data concerning CoW are described in \cite{doreian2015structural} and can be found at the address \url{http://mrvar.fdv.uni-lj.si/pajek/SVG/CoW/}. Data concerning E. coli, Macrophage, EGFR, N.G.H. Tribes and Monastery are are described in \cite{aref2019balance} and can be found at the address \url{https://figshare.com/articles/dataset/Signed_networks_from_sociology_and_political_science_biology_international_relations_finance_and_computational_chemistry/5700832}. Data concerning Bitcoin Alpha and Bitcoin OTC are described in \cite{aref2020multilevel} and can be found at the address \url{https://figshare.com/articles/dataset/Dataset_of_directed_signed_networks_from_social_domain/12152628}. Data concerning MMOG, described in \cite{szell2010multirelational} are subject to proprietary restrictions and cannot be shared. 

\section{Code availability} The codes implementing the null models employed for the present analysis are available upon request.\\

\section{Competing Interests} The authors declare no competing financial interests.

\bibliography{bibmain.bib}

\begin{thebibliography}{49}%
\makeatletter
\providecommand \@ifxundefined [1]{%
 \@ifx{#1\undefined}
}%
\providecommand \@ifnum [1]{%
 \ifnum #1\expandafter \@firstoftwo
 \else \expandafter \@secondoftwo
 \fi
}%
\providecommand \@ifx [1]{%
 \ifx #1\expandafter \@firstoftwo
 \else \expandafter \@secondoftwo
 \fi
}%
\providecommand \natexlab [1]{#1}%
\providecommand \enquote  [1]{``#1''}%
\providecommand \bibnamefont  [1]{#1}%
\providecommand \bibfnamefont [1]{#1}%
\providecommand \citenamefont [1]{#1}%
\providecommand \href@noop [0]{\@secondoftwo}%
\providecommand \href [0]{\begingroup \@sanitize@url \@href}%
\providecommand \@href[1]{\@@startlink{#1}\@@href}%
\providecommand \@@href[1]{\endgroup#1\@@endlink}%
\providecommand \@sanitize@url [0]{\catcode `\\12\catcode `\$12\catcode
  `\&12\catcode `\#12\catcode `\^12\catcode `\_12\catcode `\%12\relax}%
\providecommand \@@startlink[1]{}%
\providecommand \@@endlink[0]{}%
\providecommand \url  [0]{\begingroup\@sanitize@url \@url }%
\providecommand \@url [1]{\endgroup\@href {#1}{\urlprefix }}%
\providecommand \urlprefix  [0]{URL }%
\providecommand \Eprint [0]{\href }%
\providecommand \doibase [0]{https://doi.org/}%
\providecommand \selectlanguage [0]{\@gobble}%
\providecommand \bibinfo  [0]{\@secondoftwo}%
\providecommand \bibfield  [0]{\@secondoftwo}%
\providecommand \translation [1]{[#1]}%
\providecommand \BibitemOpen [0]{}%
\providecommand \bibitemStop [0]{}%
\providecommand \bibitemNoStop [0]{.\EOS\space}%
\providecommand \EOS [0]{\spacefactor3000\relax}%
\providecommand \BibitemShut  [1]{\csname bibitem#1\endcsname}%
\let\auto@bib@innerbib\@empty
\bibitem [{\citenamefont {Antal}\ \emph {et~al.}(2006)\citenamefont {Antal},
  \citenamefont {Krapivsky},\ and\ \citenamefont {Redner}}]{antal2006social}%
  \BibitemOpen
  \bibfield  {author} {\bibinfo {author} {\bibfnamefont {T.}~\bibnamefont
  {Antal}}, \bibinfo {author} {\bibfnamefont {P.~L.}\ \bibnamefont
  {Krapivsky}},\ and\ \bibinfo {author} {\bibfnamefont {S.}~\bibnamefont
  {Redner}},\ }\bibfield  {title} {\bibinfo {title} {Social balance on
  networks: {T}he dynamics of friendship and enmity},\ }\href
  {https://doi.org/10.1016/j.physd.2006.09.028} {\bibfield  {journal} {\bibinfo
   {journal} {Physica D: Nonlinear Phenomena}\ }\textbf {\bibinfo {volume}
  {224}},\ \bibinfo {pages} {130} (\bibinfo {year} {2006})}\BibitemShut
  {NoStop}%
\bibitem [{\citenamefont {Leskovec}\ \emph {et~al.}(2010)\citenamefont
  {Leskovec}, \citenamefont {Huttenlocher},\ and\ \citenamefont
  {Kleinberg}}]{leskovec2010signed}%
  \BibitemOpen
  \bibfield  {author} {\bibinfo {author} {\bibfnamefont {J.}~\bibnamefont
  {Leskovec}}, \bibinfo {author} {\bibfnamefont {D.}~\bibnamefont
  {Huttenlocher}},\ and\ \bibinfo {author} {\bibfnamefont {J.}~\bibnamefont
  {Kleinberg}},\ }\bibfield  {title} {\bibinfo {title} {Signed networks in
  social media},\ }in\ \href {https://doi.org/10.1145/1753326.1753532} {\emph
  {\bibinfo {booktitle} {Proceedings of the SIGCHI Conference on Human Factors
  in Computing Systems}}}\ (\bibinfo {year} {2010})\ pp.\ \bibinfo {pages}
  {1361--1370}\BibitemShut {NoStop}%
\bibitem [{\citenamefont {Zaslavsky}(2012)}]{zaslavsky2012mathematical}%
  \BibitemOpen
  \bibfield  {author} {\bibinfo {author} {\bibfnamefont {T.}~\bibnamefont
  {Zaslavsky}},\ }\bibfield  {title} {\bibinfo {title} {A {M}athematical
  {B}ibliography of {S}igned and {G}ain {G}raphs and {A}llied {A}reas},\ }\href
  {https://doi.org/10.37236/29} {\bibfield  {journal} {\bibinfo  {journal} {The
  Electronic Journal of Combinatorics}\ ,\ \bibinfo {pages} {DS8}} (\bibinfo
  {year} {2012})}\BibitemShut {NoStop}%
\bibitem [{\citenamefont {Tang}\ \emph {et~al.}(2016)\citenamefont {Tang},
  \citenamefont {Chang}, \citenamefont {Aggarwal},\ and\ \citenamefont
  {Liu}}]{tang2016survey}%
  \BibitemOpen
  \bibfield  {author} {\bibinfo {author} {\bibfnamefont {J.}~\bibnamefont
  {Tang}}, \bibinfo {author} {\bibfnamefont {Y.}~\bibnamefont {Chang}},
  \bibinfo {author} {\bibfnamefont {C.}~\bibnamefont {Aggarwal}},\ and\
  \bibinfo {author} {\bibfnamefont {H.}~\bibnamefont {Liu}},\ }\bibfield
  {title} {\bibinfo {title} {A {S}urvey of {S}igned {N}etwork {M}ining in
  {S}ocial {M}edia},\ }\href {https://doi.org/10.1145/2956185} {\bibfield
  {journal} {\bibinfo  {journal} {ACM Computing Surveys (CSUR)}\ }\textbf
  {\bibinfo {volume} {49}},\ \bibinfo {pages} {1} (\bibinfo {year}
  {2016})}\BibitemShut {NoStop}%
\bibitem [{\citenamefont {Heider}(1946)}]{heider1946attitudes}%
  \BibitemOpen
  \bibfield  {author} {\bibinfo {author} {\bibfnamefont {F.}~\bibnamefont
  {Heider}},\ }\bibfield  {title} {\bibinfo {title} {Attitudes and {C}ognitive
  {O}rganization},\ }\href {https://doi.org/10.1080/00223980.1946.9917275}
  {\bibfield  {journal} {\bibinfo  {journal} {The Journal of Psychology}\
  }\textbf {\bibinfo {volume} {21}},\ \bibinfo {pages} {107} (\bibinfo {year}
  {1946})}\BibitemShut {NoStop}%
\bibitem [{\citenamefont {Cartwright}\ and\ \citenamefont
  {Harary}(1956)}]{cartwright1956structural}%
  \BibitemOpen
  \bibfield  {author} {\bibinfo {author} {\bibfnamefont {D.}~\bibnamefont
  {Cartwright}}\ and\ \bibinfo {author} {\bibfnamefont {F.}~\bibnamefont
  {Harary}},\ }\bibfield  {title} {\bibinfo {title} {Structural balance: a
  generalization of {H}eider's theory.},\ }\href
  {https://psycnet.apa.org/doi/10.1037/h0046049} {\bibfield  {journal}
  {\bibinfo  {journal} {Psychological Review}\ }\textbf {\bibinfo {volume}
  {63}},\ \bibinfo {pages} {277} (\bibinfo {year} {1956})}\BibitemShut
  {NoStop}%
\bibitem [{\citenamefont {Harary}\ \emph {et~al.}(2002)\citenamefont {Harary},
  \citenamefont {Lim},\ and\ \citenamefont {Wunsch}}]{harary2002signed}%
  \BibitemOpen
  \bibfield  {author} {\bibinfo {author} {\bibfnamefont {F.}~\bibnamefont
  {Harary}}, \bibinfo {author} {\bibfnamefont {M.-H.}\ \bibnamefont {Lim}},\
  and\ \bibinfo {author} {\bibfnamefont {D.~C.}\ \bibnamefont {Wunsch}},\
  }\bibfield  {title} {\bibinfo {title} {Signed graphs for portfolio analysis
  in risk management},\ }\href {https://doi.org/10.1093/imaman/13.3.201}
  {\bibfield  {journal} {\bibinfo  {journal} {IMA Journal of Management
  Mathematics}\ }\textbf {\bibinfo {volume} {13}},\ \bibinfo {pages} {201}
  (\bibinfo {year} {2002})}\BibitemShut {NoStop}%
\bibitem [{\citenamefont {Ou-Yang}\ \emph {et~al.}(2015)\citenamefont
  {Ou-Yang}, \citenamefont {Dai},\ and\ \citenamefont
  {Zhang}}]{ou2015detecting}%
  \BibitemOpen
  \bibfield  {author} {\bibinfo {author} {\bibfnamefont {L.}~\bibnamefont
  {Ou-Yang}}, \bibinfo {author} {\bibfnamefont {D.-Q.}\ \bibnamefont {Dai}},\
  and\ \bibinfo {author} {\bibfnamefont {X.-F.}\ \bibnamefont {Zhang}},\
  }\bibfield  {title} {\bibinfo {title} {Detecting {P}rotein {C}omplexes from
  {S}igned {P}rotein-{P}rotein {I}nteraction {N}etworks},\ }\href
  {https://doi.org/10.1109/tcbb.2015.2401014} {\bibfield  {journal} {\bibinfo
  {journal} {IEEE/ACM Transactions on Computational Biology and
  Bioinformatics}\ }\textbf {\bibinfo {volume} {12}},\ \bibinfo {pages} {1333}
  (\bibinfo {year} {2015})}\BibitemShut {NoStop}%
\bibitem [{\citenamefont {Iorio}\ \emph {et~al.}(2016)\citenamefont {Iorio},
  \citenamefont {Bernardo-Faura}, \citenamefont {Gobbi}, \citenamefont
  {Cokelaer}, \citenamefont {Jurman},\ and\ \citenamefont
  {Saez-Rodriguez}}]{iorio2016efficient}%
  \BibitemOpen
  \bibfield  {author} {\bibinfo {author} {\bibfnamefont {F.}~\bibnamefont
  {Iorio}}, \bibinfo {author} {\bibfnamefont {M.}~\bibnamefont
  {Bernardo-Faura}}, \bibinfo {author} {\bibfnamefont {A.}~\bibnamefont
  {Gobbi}}, \bibinfo {author} {\bibfnamefont {T.}~\bibnamefont {Cokelaer}},
  \bibinfo {author} {\bibfnamefont {G.}~\bibnamefont {Jurman}},\ and\ \bibinfo
  {author} {\bibfnamefont {J.}~\bibnamefont {Saez-Rodriguez}},\ }\bibfield
  {title} {\bibinfo {title} {Efficient randomization of biological networks
  while preserving functional characterization of individual nodes},\ }\href
  {https://doi.org/10.1186/s12859-016-1402-1} {\bibfield  {journal} {\bibinfo
  {journal} {BMC Bioinformatics}\ }\textbf {\bibinfo {volume} {17}},\ \bibinfo
  {pages} {1} (\bibinfo {year} {2016})}\BibitemShut {NoStop}%
\bibitem [{\citenamefont {Saiz}\ \emph {et~al.}(2017)\citenamefont {Saiz},
  \citenamefont {G{\'o}mez-Garde{\~n}es}, \citenamefont {Nuche}, \citenamefont
  {Gir{\'o}n}, \citenamefont {Pueyo},\ and\ \citenamefont
  {Alados}}]{saiz2017evidence}%
  \BibitemOpen
  \bibfield  {author} {\bibinfo {author} {\bibfnamefont {H.}~\bibnamefont
  {Saiz}}, \bibinfo {author} {\bibfnamefont {J.}~\bibnamefont
  {G{\'o}mez-Garde{\~n}es}}, \bibinfo {author} {\bibfnamefont {P.}~\bibnamefont
  {Nuche}}, \bibinfo {author} {\bibfnamefont {A.}~\bibnamefont {Gir{\'o}n}},
  \bibinfo {author} {\bibfnamefont {Y.}~\bibnamefont {Pueyo}},\ and\ \bibinfo
  {author} {\bibfnamefont {C.~L.}\ \bibnamefont {Alados}},\ }\bibfield  {title}
  {\bibinfo {title} {Evidence of structural balance in spatial ecological
  networks},\ }\href {https://doi.org/10.1111/ecog.02561} {\bibfield  {journal}
  {\bibinfo  {journal} {Ecography}\ }\textbf {\bibinfo {volume} {40}},\
  \bibinfo {pages} {733} (\bibinfo {year} {2017})}\BibitemShut {NoStop}%
\bibitem [{\citenamefont {Davis}(1967)}]{davis1967clustering}%
  \BibitemOpen
  \bibfield  {author} {\bibinfo {author} {\bibfnamefont {J.~A.}\ \bibnamefont
  {Davis}},\ }\bibfield  {title} {\bibinfo {title} {Clustering and {S}tructural
  {B}alance in {G}raphs},\ }\href {https://doi.org/10.1177/001872676702000206}
  {\bibfield  {journal} {\bibinfo  {journal} {Human Relations}\ }\textbf
  {\bibinfo {volume} {20}},\ \bibinfo {pages} {181} (\bibinfo {year}
  {1967})}\BibitemShut {NoStop}%
\bibitem [{\citenamefont {Akiyama}\ \emph {et~al.}(1981)\citenamefont
  {Akiyama}, \citenamefont {Avis}, \citenamefont {Chv{\'a}tal},\ and\
  \citenamefont {Era}}]{akiyama1981balancing}%
  \BibitemOpen
  \bibfield  {author} {\bibinfo {author} {\bibfnamefont {J.}~\bibnamefont
  {Akiyama}}, \bibinfo {author} {\bibfnamefont {D.}~\bibnamefont {Avis}},
  \bibinfo {author} {\bibfnamefont {V.}~\bibnamefont {Chv{\'a}tal}},\ and\
  \bibinfo {author} {\bibfnamefont {H.}~\bibnamefont {Era}},\ }\bibfield
  {title} {\bibinfo {title} {Balancing signed graphs},\ }\href
  {https://doi.org/10.1016/0166-218X(81)90001-9} {\bibfield  {journal}
  {\bibinfo  {journal} {Discrete Applied Mathematics}\ }\textbf {\bibinfo
  {volume} {3}},\ \bibinfo {pages} {227} (\bibinfo {year} {1981})}\BibitemShut
  {NoStop}%
\bibitem [{\citenamefont {Harary}(1959)}]{harary1959measurement}%
  \BibitemOpen
  \bibfield  {author} {\bibinfo {author} {\bibfnamefont {F.}~\bibnamefont
  {Harary}},\ }\bibfield  {title} {\bibinfo {title} {On the measurement of
  structural balance},\ }\href {https://doi.org/10.1002/bs.3830040405}
  {\bibfield  {journal} {\bibinfo  {journal} {Behavioral Science}\ }\textbf
  {\bibinfo {volume} {4}},\ \bibinfo {pages} {316} (\bibinfo {year}
  {1959})}\BibitemShut {NoStop}%
\bibitem [{\citenamefont {Estrada}\ and\ \citenamefont
  {Benzi}(2014)}]{estrada2014walk}%
  \BibitemOpen
  \bibfield  {author} {\bibinfo {author} {\bibfnamefont {E.}~\bibnamefont
  {Estrada}}\ and\ \bibinfo {author} {\bibfnamefont {M.}~\bibnamefont
  {Benzi}},\ }\bibfield  {title} {\bibinfo {title} {Walk-based measure of
  balance in signed networks: {D}etecting lack of balance in social networks},\
  }\href {https://link.aps.org/doi/10.1103/PhysRevE.90.042802} {\bibfield
  {journal} {\bibinfo  {journal} {Physical Review E}\ }\textbf {\bibinfo
  {volume} {90}},\ \bibinfo {pages} {042802} (\bibinfo {year}
  {2014})}\BibitemShut {NoStop}%
\bibitem [{\citenamefont {Singh}\ and\ \citenamefont
  {Adhikari}(2017)}]{singh2017measuring}%
  \BibitemOpen
  \bibfield  {author} {\bibinfo {author} {\bibfnamefont {R.}~\bibnamefont
  {Singh}}\ and\ \bibinfo {author} {\bibfnamefont {B.}~\bibnamefont
  {Adhikari}},\ }\bibfield  {title} {\bibinfo {title} {Measuring the balance of
  signed networks and its application to sign prediction},\ }\href
  {https://dx.doi.org/10.1088/1742-5468/aa73ef} {\bibfield  {journal} {\bibinfo
   {journal} {Journal of Statistical Mechanics: Theory and Experiment}\
  }\textbf {\bibinfo {volume} {2017}},\ \bibinfo {pages} {063302} (\bibinfo
  {year} {2017})}\BibitemShut {NoStop}%
\bibitem [{\citenamefont {Estrada}(2019)}]{estrada2019rethinking}%
  \BibitemOpen
  \bibfield  {author} {\bibinfo {author} {\bibfnamefont {E.}~\bibnamefont
  {Estrada}},\ }\bibfield  {title} {\bibinfo {title} {Rethinking structural
  balance in signed social networks},\ }\href
  {https://doi.org/10.1016/j.dam.2019.04.019} {\bibfield  {journal} {\bibinfo
  {journal} {Discrete Applied Mathematics}\ }\textbf {\bibinfo {volume}
  {268}},\ \bibinfo {pages} {70} (\bibinfo {year} {2019})}\BibitemShut
  {NoStop}%
\bibitem [{\citenamefont {Kirkley}\ \emph {et~al.}(2019)\citenamefont
  {Kirkley}, \citenamefont {Cantwell},\ and\ \citenamefont
  {Newman}}]{kirkley2019balance}%
  \BibitemOpen
  \bibfield  {author} {\bibinfo {author} {\bibfnamefont {A.}~\bibnamefont
  {Kirkley}}, \bibinfo {author} {\bibfnamefont {G.~T.}\ \bibnamefont
  {Cantwell}},\ and\ \bibinfo {author} {\bibfnamefont {M.~E.}\ \bibnamefont
  {Newman}},\ }\bibfield  {title} {\bibinfo {title} {Balance in signed
  networks},\ }\href {https://link.aps.org/doi/10.1103/PhysRevE.99.012320}
  {\bibfield  {journal} {\bibinfo  {journal} {Physical Review E}\ }\textbf
  {\bibinfo {volume} {99}},\ \bibinfo {pages} {012320} (\bibinfo {year}
  {2019})}\BibitemShut {NoStop}%
\bibitem [{\citenamefont {Easley}\ \emph {et~al.}(2012)\citenamefont {Easley},
  \citenamefont {Kleinberg} \emph {et~al.}}]{easley2012networks}%
  \BibitemOpen
  \bibfield  {author} {\bibinfo {author} {\bibfnamefont {D.}~\bibnamefont
  {Easley}}, \bibinfo {author} {\bibfnamefont {J.}~\bibnamefont {Kleinberg}},
  \emph {et~al.},\ }\bibfield  {title} {\bibinfo {title} {Networks, crowds, and
  markets},\ }\href {https://doi.org/10.1017/CBO9780511761942} {\bibfield
  {journal} {\bibinfo  {journal} {Cambridge Books}\ } (\bibinfo {year}
  {2012})}\BibitemShut {NoStop}%
\bibitem [{\citenamefont {Aref}\ \emph
  {et~al.}(2020{\natexlab{a}})\citenamefont {Aref}, \citenamefont {Dinh},
  \citenamefont {Rezapour},\ and\ \citenamefont
  {Diesner}}]{aref2020multilevel}%
  \BibitemOpen
  \bibfield  {author} {\bibinfo {author} {\bibfnamefont {S.}~\bibnamefont
  {Aref}}, \bibinfo {author} {\bibfnamefont {L.}~\bibnamefont {Dinh}}, \bibinfo
  {author} {\bibfnamefont {R.}~\bibnamefont {Rezapour}},\ and\ \bibinfo
  {author} {\bibfnamefont {J.}~\bibnamefont {Diesner}},\ }\bibfield  {title}
  {\bibinfo {title} {Multilevel structural evaluation of signed directed social
  networks based on balance theory},\ }\href
  {https://doi.org/10.1038/s41598-020-71838-6} {\bibfield  {journal} {\bibinfo
  {journal} {Scientific Reports}\ }\textbf {\bibinfo {volume} {10}},\ \bibinfo
  {pages} {1} (\bibinfo {year} {2020}{\natexlab{a}})}\BibitemShut {NoStop}%
\bibitem [{\citenamefont {Talaga}\ \emph {et~al.}(2023)\citenamefont {Talaga},
  \citenamefont {Stella}, \citenamefont {Swanson},\ and\ \citenamefont
  {Teixeira}}]{talaga2023polarization}%
  \BibitemOpen
  \bibfield  {author} {\bibinfo {author} {\bibfnamefont {S.}~\bibnamefont
  {Talaga}}, \bibinfo {author} {\bibfnamefont {M.}~\bibnamefont {Stella}},
  \bibinfo {author} {\bibfnamefont {T.~J.}\ \bibnamefont {Swanson}},\ and\
  \bibinfo {author} {\bibfnamefont {A.~S.}\ \bibnamefont {Teixeira}},\
  }\bibfield  {title} {\bibinfo {title} {Polarization and multiscale structural
  balance in signed networks},\ }\href
  {https://doi.org/10.1038/s42005-023-01467-8} {\bibfield  {journal} {\bibinfo
  {journal} {Communication Physics}\ }\textbf {\bibinfo {volume} {6}} (\bibinfo
  {year} {2023})}\BibitemShut {NoStop}%
\bibitem [{\citenamefont {Giscard}\ \emph {et~al.}(2017)\citenamefont
  {Giscard}, \citenamefont {Rochet},\ and\ \citenamefont
  {Wilson}}]{giscard2017evaluating}%
  \BibitemOpen
  \bibfield  {author} {\bibinfo {author} {\bibfnamefont {P.-L.}\ \bibnamefont
  {Giscard}}, \bibinfo {author} {\bibfnamefont {P.}~\bibnamefont {Rochet}},\
  and\ \bibinfo {author} {\bibfnamefont {R.~C.}\ \bibnamefont {Wilson}},\
  }\bibfield  {title} {\bibinfo {title} {Evaluating balance on social networks
  from their simple cycles},\ }\href {https://doi.org/10.1093/comnet/cnx005}
  {\bibfield  {journal} {\bibinfo  {journal} {Journal of Complex Networks}\
  }\textbf {\bibinfo {volume} {5}},\ \bibinfo {pages} {750} (\bibinfo {year}
  {2017})}\BibitemShut {NoStop}%
\bibitem [{\citenamefont {Kunegis}\ \emph {et~al.}(2010)\citenamefont
  {Kunegis}, \citenamefont {Schmidt}, \citenamefont {Lommatzsch}, \citenamefont
  {Lerner}, \citenamefont {De~Luca},\ and\ \citenamefont
  {Albayrak}}]{kunegis2010spectral}%
  \BibitemOpen
  \bibfield  {author} {\bibinfo {author} {\bibfnamefont {J.}~\bibnamefont
  {Kunegis}}, \bibinfo {author} {\bibfnamefont {S.}~\bibnamefont {Schmidt}},
  \bibinfo {author} {\bibfnamefont {A.}~\bibnamefont {Lommatzsch}}, \bibinfo
  {author} {\bibfnamefont {J.}~\bibnamefont {Lerner}}, \bibinfo {author}
  {\bibfnamefont {E.~W.}\ \bibnamefont {De~Luca}},\ and\ \bibinfo {author}
  {\bibfnamefont {S.}~\bibnamefont {Albayrak}},\ }\bibfield  {title} {\bibinfo
  {title} {Spectral {A}nalysis of {S}igned {G}raphs for {C}lustering,
  {P}rediction and {V}isualization},\ }in\ \href
  {https://doi.org/10.1137/1.9781611972801.49} {\emph {\bibinfo {booktitle}
  {Proceedings of the 2010 SIAM International Conference on Data Mining}}}\
  (\bibinfo {organization} {SIAM},\ \bibinfo {year} {2010})\ pp.\ \bibinfo
  {pages} {559--570}\BibitemShut {NoStop}%
\bibitem [{\citenamefont {Terzi}\ and\ \citenamefont
  {Winkler}(2011)}]{terzi2011spectral}%
  \BibitemOpen
  \bibfield  {author} {\bibinfo {author} {\bibfnamefont {E.}~\bibnamefont
  {Terzi}}\ and\ \bibinfo {author} {\bibfnamefont {M.}~\bibnamefont
  {Winkler}},\ }\bibfield  {title} {\bibinfo {title} {A {S}pectral {A}lgorithm
  for {C}omputing {S}ocial {B}alance},\ }in\ \href
  {https://doi.org/10.1007/978-3-642-21286-4_1} {\emph {\bibinfo {booktitle}
  {International Workshop on Algorithms and Models for the Web-Graph}}}\
  (\bibinfo {organization} {Springer},\ \bibinfo {year} {2011})\ pp.\ \bibinfo
  {pages} {1--13}\BibitemShut {NoStop}%
\bibitem [{\citenamefont {Anchuri}\ and\ \citenamefont
  {Magdon-Ismail}(2012)}]{anchuri2012communities}%
  \BibitemOpen
  \bibfield  {author} {\bibinfo {author} {\bibfnamefont {P.}~\bibnamefont
  {Anchuri}}\ and\ \bibinfo {author} {\bibfnamefont {M.}~\bibnamefont
  {Magdon-Ismail}},\ }\bibfield  {title} {\bibinfo {title} {Communities and
  {B}alance in {S}igned {N}etworks: {A} {S}pectral {A}pproach},\ }in\ \href
  {https://doi.org/10.1109/ASONAM.2012.48} {\emph {\bibinfo {booktitle} {2012
  IEEE/ACM International Conference on Advances in Social Networks Analysis and
  Mining}}}\ (\bibinfo {organization} {IEEE},\ \bibinfo {year} {2012})\ pp.\
  \bibinfo {pages} {235--242}\BibitemShut {NoStop}%
\bibitem [{\citenamefont {Belaza}\ \emph {et~al.}(2017)\citenamefont {Belaza},
  \citenamefont {Hoefman}, \citenamefont {Ryckebusch}, \citenamefont {Bramson},
  \citenamefont {van~den Heuvel},\ and\ \citenamefont
  {Schoors}}]{belaza2017statistical}%
  \BibitemOpen
  \bibfield  {author} {\bibinfo {author} {\bibfnamefont {A.~M.}\ \bibnamefont
  {Belaza}}, \bibinfo {author} {\bibfnamefont {K.}~\bibnamefont {Hoefman}},
  \bibinfo {author} {\bibfnamefont {J.}~\bibnamefont {Ryckebusch}}, \bibinfo
  {author} {\bibfnamefont {A.}~\bibnamefont {Bramson}}, \bibinfo {author}
  {\bibfnamefont {M.}~\bibnamefont {van~den Heuvel}},\ and\ \bibinfo {author}
  {\bibfnamefont {K.}~\bibnamefont {Schoors}},\ }\bibfield  {title} {\bibinfo
  {title} {Statistical physics of balance theory},\ }\href
  {https://doi.org/10.1371/journal.pone.0183696} {\bibfield  {journal}
  {\bibinfo  {journal} {PLoS One}\ }\textbf {\bibinfo {volume} {12}},\ \bibinfo
  {pages} {e0183696} (\bibinfo {year} {2017})}\BibitemShut {NoStop}%
\bibitem [{\citenamefont {Zas{\'l}avsky}(1987)}]{zaslavsky1987balanced}%
  \BibitemOpen
  \bibfield  {author} {\bibinfo {author} {\bibfnamefont {T.}~\bibnamefont
  {Zas{\'l}avsky}},\ }\bibfield  {title} {\bibinfo {title} {Balanced
  decompositions of a signed graph},\ }\href
  {https://doi.org/10.1016/0095-8956(87)90026-8} {\bibfield  {journal}
  {\bibinfo  {journal} {Journal of Combinatorial Theory, Series B}\ }\textbf
  {\bibinfo {volume} {43}},\ \bibinfo {pages} {1} (\bibinfo {year}
  {1987})}\BibitemShut {NoStop}%
\bibitem [{\citenamefont {Aref}\ and\ \citenamefont
  {Wilson}(2019)}]{aref2019balance}%
  \BibitemOpen
  \bibfield  {author} {\bibinfo {author} {\bibfnamefont {S.}~\bibnamefont
  {Aref}}\ and\ \bibinfo {author} {\bibfnamefont {M.~C.}\ \bibnamefont
  {Wilson}},\ }\bibfield  {title} {\bibinfo {title} {Balance and {F}rustration
  in {S}igned {N}etworks},\ }\href {https://doi.org/10.1093/comnet/cny015}
  {\bibfield  {journal} {\bibinfo  {journal} {Journal of Complex Networks}\
  }\textbf {\bibinfo {volume} {7}},\ \bibinfo {pages} {163} (\bibinfo {year}
  {2019})}\BibitemShut {NoStop}%
\bibitem [{\citenamefont {Aref}\ \emph
  {et~al.}(2020{\natexlab{b}})\citenamefont {Aref}, \citenamefont {Mason},\
  and\ \citenamefont {Wilson}}]{aref2020modeling}%
  \BibitemOpen
  \bibfield  {author} {\bibinfo {author} {\bibfnamefont {S.}~\bibnamefont
  {Aref}}, \bibinfo {author} {\bibfnamefont {A.~J.}\ \bibnamefont {Mason}},\
  and\ \bibinfo {author} {\bibfnamefont {M.~C.}\ \bibnamefont {Wilson}},\
  }\bibfield  {title} {\bibinfo {title} {A modeling and computational study of
  the frustration index in signed networks},\ }\href
  {https://doi.org/10.1002/net.21907} {\bibfield  {journal} {\bibinfo
  {journal} {Networks}\ }\textbf {\bibinfo {volume} {75}},\ \bibinfo {pages}
  {95} (\bibinfo {year} {2020}{\natexlab{b}})}\BibitemShut {NoStop}%
\bibitem [{\citenamefont {Traag}\ \emph {et~al.}(2019)\citenamefont {Traag},
  \citenamefont {Doreian},\ and\ \citenamefont
  {Mrvar}}]{traag2019partitioning}%
  \BibitemOpen
  \bibfield  {author} {\bibinfo {author} {\bibfnamefont {V.}~\bibnamefont
  {Traag}}, \bibinfo {author} {\bibfnamefont {P.}~\bibnamefont {Doreian}},\
  and\ \bibinfo {author} {\bibfnamefont {A.}~\bibnamefont {Mrvar}},\ }\bibfield
   {title} {\bibinfo {title} {Partitioning {S}igned {N}etworks},\ }\href
  {https://doi.org/10.1002/9781119483298.ch8} {\bibfield  {journal} {\bibinfo
  {journal} {Advances in Network Clustering and Blockmodeling}\ ,\ \bibinfo
  {pages} {225}} (\bibinfo {year} {2019})}\BibitemShut {NoStop}%
\bibitem [{\citenamefont {Abelson}\ and\ \citenamefont
  {Rosenberg}(1958)}]{abelson1958symbolic}%
  \BibitemOpen
  \bibfield  {author} {\bibinfo {author} {\bibfnamefont {R.~P.}\ \bibnamefont
  {Abelson}}\ and\ \bibinfo {author} {\bibfnamefont {M.~J.}\ \bibnamefont
  {Rosenberg}},\ }\bibfield  {title} {\bibinfo {title} {Symbolic psycho-logic:
  {A} model of attitudinal cognition},\ }\href
  {https://psycnet.apa.org/doi/10.1002/bs.3830030102} {\bibfield  {journal}
  {\bibinfo  {journal} {Behavioral Science}\ }\textbf {\bibinfo {volume} {3}},\
  \bibinfo {pages} {1} (\bibinfo {year} {1958})}\BibitemShut {NoStop}%
\bibitem [{\citenamefont {Facchetti}\ \emph {et~al.}(2011)\citenamefont
  {Facchetti}, \citenamefont {Iacono},\ and\ \citenamefont
  {Altafini}}]{facchetti2011computing}%
  \BibitemOpen
  \bibfield  {author} {\bibinfo {author} {\bibfnamefont {G.}~\bibnamefont
  {Facchetti}}, \bibinfo {author} {\bibfnamefont {G.}~\bibnamefont {Iacono}},\
  and\ \bibinfo {author} {\bibfnamefont {C.}~\bibnamefont {Altafini}},\
  }\bibfield  {title} {\bibinfo {title} {Computing global structural balance in
  large-scale signed social networks},\ }\href
  {https://doi.org/10.1073/pnas.1109521108} {\bibfield  {journal} {\bibinfo
  {journal} {Proceedings of the National Academy of Sciences}\ }\textbf
  {\bibinfo {volume} {108}},\ \bibinfo {pages} {20953} (\bibinfo {year}
  {2011})}\BibitemShut {NoStop}%
\bibitem [{\citenamefont {Derr}\ \emph {et~al.}(2018)\citenamefont {Derr},
  \citenamefont {Aggarwal},\ and\ \citenamefont {Tang}}]{derr2018signed}%
  \BibitemOpen
  \bibfield  {author} {\bibinfo {author} {\bibfnamefont {T.}~\bibnamefont
  {Derr}}, \bibinfo {author} {\bibfnamefont {C.}~\bibnamefont {Aggarwal}},\
  and\ \bibinfo {author} {\bibfnamefont {J.}~\bibnamefont {Tang}},\ }\bibfield
  {title} {\bibinfo {title} {Signed {N}etwork {M}odeling {B}ased on
  {S}tructural {B}alance {T}heory},\ }in\ \href
  {https://doi.org/10.1145/3269206.3271746} {\emph {\bibinfo {booktitle}
  {Proceedings of the 27th ACM International Conference on Information and
  Knowledge Management}}}\ (\bibinfo {year} {2018})\ pp.\ \bibinfo {pages}
  {557--566}\BibitemShut {NoStop}%
\bibitem [{\citenamefont {Huitsing}\ \emph {et~al.}(2012)\citenamefont
  {Huitsing}, \citenamefont {Van~Duijn}, \citenamefont {Snijders},
  \citenamefont {Wang}, \citenamefont {Sainio}, \citenamefont {Salmivalli},\
  and\ \citenamefont {Veenstra}}]{huitsing2012univariate}%
  \BibitemOpen
  \bibfield  {author} {\bibinfo {author} {\bibfnamefont {G.}~\bibnamefont
  {Huitsing}}, \bibinfo {author} {\bibfnamefont {M.~A.}\ \bibnamefont
  {Van~Duijn}}, \bibinfo {author} {\bibfnamefont {T.~A.}\ \bibnamefont
  {Snijders}}, \bibinfo {author} {\bibfnamefont {P.}~\bibnamefont {Wang}},
  \bibinfo {author} {\bibfnamefont {M.}~\bibnamefont {Sainio}}, \bibinfo
  {author} {\bibfnamefont {C.}~\bibnamefont {Salmivalli}},\ and\ \bibinfo
  {author} {\bibfnamefont {R.}~\bibnamefont {Veenstra}},\ }\bibfield  {title}
  {\bibinfo {title} {Univariate and multivariate models of positive and
  negative networks: Liking, disliking, and bully--victim relationships},\
  }\href {https://doi.org/10.1016/j.socnet.2012.08.001} {\bibfield  {journal}
  {\bibinfo  {journal} {Social Networks}\ }\textbf {\bibinfo {volume} {34}},\
  \bibinfo {pages} {645} (\bibinfo {year} {2012})}\BibitemShut {NoStop}%
\bibitem [{\citenamefont {Lerner}(2016)}]{lerner2016structural}%
  \BibitemOpen
  \bibfield  {author} {\bibinfo {author} {\bibfnamefont {J.}~\bibnamefont
  {Lerner}},\ }\bibfield  {title} {\bibinfo {title} {Structural balance in
  signed networks: {S}eparating the probability to interact from the tendency
  to fight},\ }\href {https://doi.org/10.1016/j.socnet.2015.12.002} {\bibfield
  {journal} {\bibinfo  {journal} {Social Networks}\ }\textbf {\bibinfo {volume}
  {45}},\ \bibinfo {pages} {66} (\bibinfo {year} {2016})}\BibitemShut {NoStop}%
\bibitem [{\citenamefont {Becatti}\ \emph {et~al.}(2019)\citenamefont
  {Becatti}, \citenamefont {Caldarelli},\ and\ \citenamefont
  {Saracco}}]{becatti2019}%
  \BibitemOpen
  \bibfield  {author} {\bibinfo {author} {\bibfnamefont {C.}~\bibnamefont
  {Becatti}}, \bibinfo {author} {\bibfnamefont {G.}~\bibnamefont
  {Caldarelli}},\ and\ \bibinfo {author} {\bibfnamefont {F.}~\bibnamefont
  {Saracco}},\ }\bibfield  {title} {\bibinfo {title} {Entropy-based
  randomization of rating networks},\ }\href
  {https://doi.org/10.1103/PhysRevE.99.022306} {\bibfield  {journal} {\bibinfo
  {journal} {Physical Review E}\ }\textbf {\bibinfo {volume} {99}},\ \bibinfo
  {pages} {022306} (\bibinfo {year} {2019})}\BibitemShut {NoStop}%
\bibitem [{\citenamefont {Fritz}\ \emph {et~al.}(2022)\citenamefont {Fritz},
  \citenamefont {Mehrl}, \citenamefont {Thurner} \emph
  {et~al.}}]{fritz2022exponential}%
  \BibitemOpen
  \bibfield  {author} {\bibinfo {author} {\bibfnamefont {C.}~\bibnamefont
  {Fritz}}, \bibinfo {author} {\bibfnamefont {M.}~\bibnamefont {Mehrl}},
  \bibinfo {author} {\bibfnamefont {P.~W.}\ \bibnamefont {Thurner}}, \emph
  {et~al.},\ }\bibfield  {title} {\bibinfo {title} {Exponential {R}andom
  {G}raph {M}odels for {D}ynamic {S}igned {N}etworks: {A}n {A}pplication to
  {I}nternational {R}elations},\ }\href
  {http://dx.doi.org/10.48550/arXiv.2205.13411} {\bibfield  {journal} {\bibinfo
   {journal} {arXiv preprint arXiv:2205.13411}\ } (\bibinfo {year}
  {2022})}\BibitemShut {NoStop}%
\bibitem [{\citenamefont {Doreian}\ and\ \citenamefont
  {Mrvar}(2015)}]{doreian2015structural}%
  \BibitemOpen
  \bibfield  {author} {\bibinfo {author} {\bibfnamefont {P.}~\bibnamefont
  {Doreian}}\ and\ \bibinfo {author} {\bibfnamefont {A.}~\bibnamefont
  {Mrvar}},\ }\bibfield  {title} {\bibinfo {title} {Structural {B}alance and
  {S}igned {I}nternational {R}elations},\ }\href
  {https://doi.org/10.21307/joss-2019-012} {\bibfield  {journal} {\bibinfo
  {journal} {Journal of Social Structure}\ }\textbf {\bibinfo {volume} {16}},\
  \bibinfo {pages} {1} (\bibinfo {year} {2015})}\BibitemShut {NoStop}%
\bibitem [{\citenamefont {Szell}\ \emph {et~al.}(2010)\citenamefont {Szell},
  \citenamefont {Lambiotte},\ and\ \citenamefont
  {Thurner}}]{szell2010multirelational}%
  \BibitemOpen
  \bibfield  {author} {\bibinfo {author} {\bibfnamefont {M.}~\bibnamefont
  {Szell}}, \bibinfo {author} {\bibfnamefont {R.}~\bibnamefont {Lambiotte}},\
  and\ \bibinfo {author} {\bibfnamefont {S.}~\bibnamefont {Thurner}},\
  }\bibfield  {title} {\bibinfo {title} {Multirelational organization of
  large-scale social networks in an online world},\ }\href
  {https://doi.org/10.1073/pnas.1004008107} {\bibfield  {journal} {\bibinfo
  {journal} {Proceedings of the National Academy of Sciences}\ }\textbf
  {\bibinfo {volume} {107}},\ \bibinfo {pages} {13636} (\bibinfo {year}
  {2010})}\BibitemShut {NoStop}%
\bibitem [{are(2018)}]{aref2020dataset}%
  \BibitemOpen
  \href
  {https://figshare.com/articles/dataset/Signed_networks_from_sociology_and_political_science_biology_international_relations_finance_and_computational_chemistry/5700832}
  {\bibinfo {title} {Signed networks from sociology and political science,
  systems biology, international relations, finance, and computational
  chemistry}} (\bibinfo {year} {2018})\BibitemShut {NoStop}%
\bibitem [{\citenamefont {Sampson}(1968)}]{sampson1968novitiate}%
  \BibitemOpen
  \bibfield  {author} {\bibinfo {author} {\bibfnamefont {S.~F.}\ \bibnamefont
  {Sampson}},\ }\href {https://doi.org/10.1016/0378-8733(95)00259-6} {\emph
  {\bibinfo {title} {A novitiate in a period of change: An experimental and
  case study of social relationships}}}\ (\bibinfo  {publisher} {Cornell
  University},\ \bibinfo {year} {1968})\BibitemShut {NoStop}%
\bibitem [{\citenamefont {Kumar}\ \emph {et~al.}(2016)\citenamefont {Kumar},
  \citenamefont {Spezzano}, \citenamefont {Subrahmanian},\ and\ \citenamefont
  {Faloutsos}}]{kumar2016edge}%
  \BibitemOpen
  \bibfield  {author} {\bibinfo {author} {\bibfnamefont {S.}~\bibnamefont
  {Kumar}}, \bibinfo {author} {\bibfnamefont {F.}~\bibnamefont {Spezzano}},
  \bibinfo {author} {\bibfnamefont {V.}~\bibnamefont {Subrahmanian}},\ and\
  \bibinfo {author} {\bibfnamefont {C.}~\bibnamefont {Faloutsos}},\ }\bibfield
  {title} {\bibinfo {title} {Edge {W}eight {P}rediction in {W}eighted {S}igned
  {N}etworks},\ }in\ \href {https://ieeexplore.ieee.org/document/7837846}
  {\emph {\bibinfo {booktitle} {2016 IEEE 16th International Conference on Data
  Mining (ICDM)}}}\ (\bibinfo {organization} {IEEE},\ \bibinfo {year} {2016})\
  pp.\ \bibinfo {pages} {221--230}\BibitemShut {NoStop}%
\bibitem [{\citenamefont {El~Maftouhi}\ \emph {et~al.}(2012)\citenamefont
  {El~Maftouhi}, \citenamefont {Manoussakis},\ and\ \citenamefont
  {Megalakaki}}]{el2012balance}%
  \BibitemOpen
  \bibfield  {author} {\bibinfo {author} {\bibfnamefont {A.}~\bibnamefont
  {El~Maftouhi}}, \bibinfo {author} {\bibfnamefont {Y.}~\bibnamefont
  {Manoussakis}},\ and\ \bibinfo {author} {\bibfnamefont {O.}~\bibnamefont
  {Megalakaki}},\ }\bibfield  {title} {\bibinfo {title} {{B}alance in {R}andom
  {S}igned {G}raphs},\ }\href {https://doi.org/10.1080/15427951.2012.675413}
  {\bibfield  {journal} {\bibinfo  {journal} {{I}nternet {M}athematics}\
  }\textbf {\bibinfo {volume} {8}},\ \bibinfo {pages} {364} (\bibinfo {year}
  {2012})}\BibitemShut {NoStop}%
\bibitem [{\citenamefont {G{\'o}mez}\ \emph {et~al.}(2009)\citenamefont
  {G{\'o}mez}, \citenamefont {Jensen},\ and\ \citenamefont
  {Arenas}}]{gomez2009analysis}%
  \BibitemOpen
  \bibfield  {author} {\bibinfo {author} {\bibfnamefont {S.}~\bibnamefont
  {G{\'o}mez}}, \bibinfo {author} {\bibfnamefont {P.}~\bibnamefont {Jensen}},\
  and\ \bibinfo {author} {\bibfnamefont {A.}~\bibnamefont {Arenas}},\
  }\bibfield  {title} {\bibinfo {title} {Analysis of community structure in
  networks of correlated data},\ }\href
  {https://doi.org/10.1103/PhysRevE.80.016114} {\bibfield  {journal} {\bibinfo
  {journal} {Physical {R}eview {E}}\ }\textbf {\bibinfo {volume} {80}},\
  \bibinfo {pages} {016114} (\bibinfo {year} {2009})}\BibitemShut {NoStop}%
\bibitem [{\citenamefont {Doreian}\ and\ \citenamefont
  {Mrvar}(1996)}]{doreian1996partitioning}%
  \BibitemOpen
  \bibfield  {author} {\bibinfo {author} {\bibfnamefont {P.}~\bibnamefont
  {Doreian}}\ and\ \bibinfo {author} {\bibfnamefont {A.}~\bibnamefont
  {Mrvar}},\ }\bibfield  {title} {\bibinfo {title} {A partitioning approach to
  structural balance},\ }\href@noop {} {\bibfield  {journal} {\bibinfo
  {journal} {Social Networks}\ }\textbf {\bibinfo {volume} {18}},\ \bibinfo
  {pages} {149} (\bibinfo {year} {1996})}\BibitemShut {NoStop}%
\bibitem [{\citenamefont {Marchese}\ \emph {et~al.}(2022)\citenamefont
  {Marchese}, \citenamefont {Caldarelli},\ and\ \citenamefont
  {Squartini}}]{marchese2022detecting}%
  \BibitemOpen
  \bibfield  {author} {\bibinfo {author} {\bibfnamefont {E.}~\bibnamefont
  {Marchese}}, \bibinfo {author} {\bibfnamefont {G.}~\bibnamefont
  {Caldarelli}},\ and\ \bibinfo {author} {\bibfnamefont {T.}~\bibnamefont
  {Squartini}},\ }\bibfield  {title} {\bibinfo {title} {Detecting mesoscale
  structures by surprise},\ }\href {https://doi.org/10.1038/s42005-022-00890-7}
  {\bibfield  {journal} {\bibinfo  {journal} {Communications Physics}\ }\textbf
  {\bibinfo {volume} {5}},\ \bibinfo {pages} {1} (\bibinfo {year}
  {2022})}\BibitemShut {NoStop}%
\bibitem [{\citenamefont {Park}\ and\ \citenamefont
  {Newman}(2004)}]{park2004statistical}%
  \BibitemOpen
  \bibfield  {author} {\bibinfo {author} {\bibfnamefont {J.}~\bibnamefont
  {Park}}\ and\ \bibinfo {author} {\bibfnamefont {M.~E.~J.}\ \bibnamefont
  {Newman}},\ }\bibfield  {title} {\bibinfo {title} {{Statistical mechanics of
  networks}},\ }\href {https://doi.org/10.1103/PhysRevE.70.066117} {\bibfield
  {journal} {\bibinfo  {journal} {Phys. Rev. E}\ }\textbf {\bibinfo {volume}
  {70}},\ \bibinfo {pages} {66117} (\bibinfo {year} {2004})}\BibitemShut
  {NoStop}%
\bibitem [{\citenamefont {Squartini}\ and\ \citenamefont
  {Garlaschelli}(2017)}]{Squartinia}%
  \BibitemOpen
  \bibfield  {author} {\bibinfo {author} {\bibfnamefont {T.}~\bibnamefont
  {Squartini}}\ and\ \bibinfo {author} {\bibfnamefont {D.}~\bibnamefont
  {Garlaschelli}},\ }\href {https://doi.org/10.1007/978-3-319-69438-2} {\emph
  {\bibinfo {title} {{Maximum-Entropy Networks. Pattern Detection, Network
  Reconstruction and Graph Combinatorics}}}}\ (\bibinfo  {publisher} {Springer
  International Publishing},\ \bibinfo {year} {2017})\ p.\ \bibinfo {pages}
  {116}\BibitemShut {NoStop}%
\bibitem [{\citenamefont {Garlaschelli}\ and\ \citenamefont
  {Loffredo}(2008)}]{Garlaschelli2008}%
  \BibitemOpen
  \bibfield  {author} {\bibinfo {author} {\bibfnamefont {D.}~\bibnamefont
  {Garlaschelli}}\ and\ \bibinfo {author} {\bibfnamefont {M.~I.}\ \bibnamefont
  {Loffredo}},\ }\bibfield  {title} {\bibinfo {title} {Maximum likelihood:
  {E}xtracting unbiased information from complex networks},\ }\href
  {https://doi.org/10.1103/PhysRevE.78.015101} {\bibfield  {journal} {\bibinfo
  {journal} {Physical Review E}\ }\textbf {\bibinfo {volume} {78}},\ \bibinfo
  {pages} {015101} (\bibinfo {year} {2008})}\BibitemShut {NoStop}%
\bibitem [{\citenamefont {Vallarano}\ \emph {et~al.}(2021)\citenamefont
  {Vallarano}, \citenamefont {Bruno}, \citenamefont {Marchese}, \citenamefont
  {Trapani}, \citenamefont {Saracco}, \citenamefont {Cimini}, \citenamefont
  {Zanon},\ and\ \citenamefont {Squartini}}]{vallarano2021}%
  \BibitemOpen
  \bibfield  {author} {\bibinfo {author} {\bibfnamefont {N.}~\bibnamefont
  {Vallarano}}, \bibinfo {author} {\bibfnamefont {M.}~\bibnamefont {Bruno}},
  \bibinfo {author} {\bibfnamefont {E.}~\bibnamefont {Marchese}}, \bibinfo
  {author} {\bibfnamefont {G.}~\bibnamefont {Trapani}}, \bibinfo {author}
  {\bibfnamefont {F.}~\bibnamefont {Saracco}}, \bibinfo {author} {\bibfnamefont
  {G.}~\bibnamefont {Cimini}}, \bibinfo {author} {\bibfnamefont
  {M.}~\bibnamefont {Zanon}},\ and\ \bibinfo {author} {\bibfnamefont
  {T.}~\bibnamefont {Squartini}},\ }\bibfield  {title} {\bibinfo {title} {Fast
  and scalable likelihood maximization for {E}xponential {R}andom {G}raph
  {M}odels with local constraints},\ }\href
  {https://doi.org/10.1038/s41598-021-93830-4} {\bibfield  {journal} {\bibinfo
  {journal} {Scientific Reports}\ }\textbf {\bibinfo {volume} {11}},\ \bibinfo
  {pages} {15227} (\bibinfo {year} {2021})}\BibitemShut {NoStop}%
\end{thebibliography}%

\clearpage

\onecolumngrid

\section*{Supplementary Note 1\\Representing binary, undirected, signed networks}\label{AppA}

The three functions $a_{ij}^-=[a_{ij}=-1]$, $a_{ij}^0=[a_{ij}=0]$ and $a_{ij}^+=[a_{ij}=+1]$ have been defined via the Iverson's brackets notation. Iverson's brackets work in a way that is reminiscent of the Heaviside step function, i.e. $\Theta[x]=[x>0]$; in fact,

\begin{equation}
a_{ij}^-=[a_{ij}=-1]=\begin{dcases}
1, & \text{if}\quad a_{ij}=-1\\
0, & \text{if}\quad a_{ij}=0,+1
\end{dcases}
\end{equation}
(i.e. $a_{ij}^-=1$ if $a_{ij}<0$ and zero otherwise),

\begin{equation}
a_{ij}^0=[a_{ij}=0]=\begin{dcases}
1, & \text{if}\quad a_{ij}=0\\
0, & \text{if}\quad a_{ij}=-1,+1
\end{dcases}
\end{equation}
(i.e. $a_{ij}^0=1$ if $a_{ij}=0$ and zero otherwise),

\begin{equation}
a_{ij}^+=[a_{ij}=+1]=\begin{dcases}
1, & \text{if}\quad a_{ij}=+1\\
0, & \text{if}\quad a_{ij}=-1,0
\end{dcases}
\end{equation}
(i.e. $a_{ij}^+=1$ if $a_{ij}>0$ and zero otherwise). The matrices $\mathbf{A}^+\equiv\{a_{ij}^+\}_{i,j=1}^N$ and $\mathbf{A}^-\equiv\{a_{ij}^-\}_{i,j=1}^N$, thus, remain naturally defined and induce the relationships $\mathbf{A}=\mathbf{A}^+-\mathbf{A}^-$, i.e. $a_{ij}=a_{ij}^+-a_{ij}^-$, $\forall\:i\neq j$ and $|\mathbf{A}|=\mathbf{A}^++\mathbf{A}^-$, i.e. $|a_{ij}|=a_{ij}^++a_{ij}^-$, $\forall\:i\neq j$.

\clearpage

\section*{Supplementary Note 2\\Counting triangles on binary, undirected, signed networks}\label{AppB}

A well-known result states that the abundance of node-specific, unsigned triangles reads

\begin{align}
2T_i&=\sum_{\substack{j=1\\(j\neq i)}}^N\sum_{\substack{k=1\\(k\neq i,j)}}^Na_{ij}a_{jk}a_{ki}=[\mathbf{A}\mathbf{A}\mathbf{A}]_{ii}=[\mathbf{A}]^3_{ii}\quad\forall\:i;
\end{align}
let us, now, consider signed networks: the abundances of node-specific, signed triangles with an even number of negative links read

\begin{align}
T_i^{(+++)}&=\sum_{\substack{j=1\\(j\neq i)}}^N\sum_{\substack{k=1\\(k\neq i,j)}}^N\frac{a_{ij}^+a_{jk}^+a_{ki}^+}{2}=\frac{[\mathbf{A}^+\mathbf{A}^+\mathbf{A}^+]_{ii}}{2}=\frac{[\mathbf{A}^+]^3_{ii}}{2}\quad\forall\:i,\\
T_i^{(-+-)}&=\sum_{\substack{j=1\\(j\neq i)}}^N\sum_{\substack{k=1\\(k\neq i,j)}}^N\frac{a_{ij}^-a_{jk}^+a_{ki}^-}{2}=\frac{[\mathbf{A}^-\mathbf{A}^+\mathbf{A}^-]_{ii}}{2}\quad\forall\:i,\\
T_i^{(+--)}&=\sum_{\substack{j=1\\(j\neq i)}}^N\sum_{\substack{k=1\\(k\neq i,j)}}^Na_{ij}^+a_{jk}^-a_{ki}^-=[\mathbf{A}^+\mathbf{A}^-\mathbf{A}^-]_{ii}=[\mathbf{A}^+(\mathbf{A}^-)^2]_{ii}\quad\forall\:i,\\
T_i^{(--+)}&=\sum_{\substack{j=1\\(j\neq i)}}^N\sum_{\substack{k=1\\(k\neq i,j)}}^Na_{ij}^-a_{jk}^-a_{ki}^+=[\mathbf{A}^-\mathbf{A}^-\mathbf{A}^+]_{ii}=[(\mathbf{A}^-)^2\mathbf{A}^+]_{ii}\quad\forall\:i
\end{align}
while the abundances of node-specific, signed triangles with an odd number of negative links read

\begin{align}
T_i^{(---)}&=\sum_{\substack{j=1\\(j\neq i)}}^N\sum_{\substack{k=1\\(k\neq i,j)}}^N\frac{a_{ij}^-a_{jk}^-a_{ki}^-}{2}=\frac{[\mathbf{A}^-\mathbf{A}^-\mathbf{A}^-]_{ii}}{2}=\frac{[\mathbf{A}^-]^3_{ii}}{2}\quad\forall\:i,\\
T_i^{(+-+)}&=\sum_{\substack{j=1\\(j\neq i)}}^N\sum_{\substack{k=1\\(k\neq i,j)}}^N\frac{a_{ij}^+a_{jk}^-a_{ki}^+}{2}=\frac{[\mathbf{A}^+\mathbf{A}^-\mathbf{A}^+]_{ii}}{2}\quad\forall\:i,\\
T_i^{(++-)}&=\sum_{\substack{j=1\\(j\neq i)}}^N\sum_{\substack{k=1\\(k\neq i,j)}}^Na_{ij}^+a_{jk}^+a_{ki}^-=[\mathbf{A}^+\mathbf{A}^+\mathbf{A}^-]_{ii}=[(\mathbf{A}^+)^2\mathbf{A}^-]_{ii}\quad\forall\:i,\\
T_i^{(-++)}&=\sum_{\substack{j=1\\(j\neq i)}}^N\sum_{\substack{k=1\\(k\neq i,j)}}^Na_{ij}^-a_{jk}^+a_{ki}^+=[\mathbf{A}^-\mathbf{A}^+\mathbf{A}^+]_{ii}=[\mathbf{A}^-(\mathbf{A}^+)^2]_{ii}\quad\forall\:i.
\end{align}

Let us, now, write the expressions for the total abundances of signed triangles with an even number of negative links:

\begin{align}
T^{(+++)}&=\frac{1}{3}\sum_{i=1}^NT^{(+++)}_i=\frac{1}{6}\sum_{i=1}^N[\mathbf{A}^+\mathbf{A}^+\mathbf{A}^+]_{ii}=\frac{\text{Tr}[\mathbf{A}^+\mathbf{A}^+\mathbf{A}^+]}{6}=\frac{\text{Tr}[(\mathbf{A}^+)^3]}{6},\\
T^{(-+-)}&=\frac{1}{2}\sum_{i=1}^NT^{(-+-)}_i=\frac{1}{2}\sum_{i=1}^N[\mathbf{A}^-\mathbf{A}^+\mathbf{A}^-]_{ii}=\frac{\text{Tr}[\mathbf{A}^-\mathbf{A}^+\mathbf{A}^-]}{2},\\
T^{(+--)}&=\frac{1}{2}\sum_{i=1}^NT^{(+--)}_i=\frac{1}{2}\sum_{i=1}^N[\mathbf{A}^+\mathbf A^-\mathbf{A}^-]_{ii}=\frac{\text{Tr}[\mathbf{A}^+\mathbf{A}^-\mathbf{A}^-]}{2}=\frac{\text{Tr}[\mathbf{A}^+(\mathbf{A}^-)^2]}{2},\\
T^{(--+)}&=\frac{1}{2}\sum_{i=1}^NT^{(--+)}_i=\frac{1}{2}\sum_{i=1}^N[\mathbf{A}^-\mathbf{A}^-\mathbf{A}^+]_{ii}=\frac{\text{Tr}[\mathbf{A}^-\mathbf A^-\mathbf{A}^+]}{2}=\frac{\text{Tr}[(\mathbf{A}^-)^2\mathbf{A}^+]}{2};
\end{align}
analogously, for the abundances of triangles with an odd number of negative links, reading

\begin{align}
T^{(---)}&=\frac{1}{3}\sum_{i=1}^NT^{(---)}_i=\frac{1}{6}\sum_{i=1}^N[\mathbf{A}^-\mathbf{A}^-\mathbf{A}^-]_{ii}=\frac{\text{Tr}[\mathbf{A}^-\mathbf{A}^-\mathbf{A}^-]}{6}=\frac{\text{Tr}[(\mathbf{A}^-)^3]}{6},\\
T^{(+-+)}&=\frac{1}{2}\sum_{i=1}^NT^{(+-+)}_i=\frac{1}{2}\sum_{i=1}^N[\mathbf{A}^+\mathbf{A}^-\mathbf{A}^+]_{ii}=\frac{\text{Tr}[\mathbf{A}^+\mathbf{A}^-\mathbf{A}^+]}{2},\\
T^{(++-)}&=\frac{1}{2}\sum_{i=1}^NT^{(++-)}_i=\frac{1}{2}\sum_{i=1}^N[\mathbf{A}^+\mathbf{A}^+\mathbf{A}^-]_{ii}=\frac{\text{Tr}[\mathbf{A}^+\mathbf{A}^+\mathbf{A}^-]}{2}=\frac{\text{Tr}[(\mathbf{A}^+)^2\mathbf{A}^-]}{2},\\
T^{(-++)}&=\frac{1}{2}\sum_{i=1}^NT^{(-++)}_i=\frac{1}{2}\sum_{i=1}^N[\mathbf{A}^-\mathbf{A}^+\mathbf{A}^+]_{ii}=\frac{\text{Tr}[\mathbf{A}^-\mathbf{A}^+\mathbf{A}^+]}{2}=\frac{\text{Tr}[\mathbf{A}^-(\mathbf{A}^+)^2]}{2}
\end{align}
(where each numeric factor avoids the corresponding pattern to be overcounted).\\

Since the trace of a matrix is invariant under a cyclic permutation of the members of its argument, the following result holds true

\begin{equation}
\frac{\text{Tr}[\mathbf{A}^+\mathbf A^-\mathbf{A}^-]}{2}=\frac{\text{Tr}[\mathbf{A}^-\mathbf A^+\mathbf{A}^-]}{2}=\frac{\text{Tr}[\mathbf{A}^-\mathbf A^-\mathbf{A}^+]}{2}
\end{equation}
further implying that $T^{(+--)}=T^{(-+-)}=T^{(--+)}$. As a consequence, the number of balanced patterns according to either variant of the SBT can be defined in several, equivalent ways, i.e. $\#_{\mytriangle}^{sb}=T^{(+++)}+T^{(+--)}=T^{(+++)}+T^{(-+-)}=T^{(+++)}+T^{(--+)}$; analogously, $\#_{\mytriangle}^{wb}=T^{(+++)}+T^{(+--)}+T^{(---)}=T^{(+++)}+T^{(-+-)}+T^{(---)}=T^{(+++)}+T^{(--+)}+T^{(---)}$.

\clearpage

\section*{Supplementary Note 3\\Probabilistic models for binary, undirected, signed networks}\label{AppC}

The generalization of the ERG formalism for the analysis of binary, undirected, signed graphs rests upon the constrained maximization of Shannon entropy, i.e.

\begin{equation}
\mathscr{L}=S[P]-\sum_{i=0}^M\theta_i[P(\mathbf{A})C_i(\mathbf{A})-\langle C_i\rangle]
\end{equation}
where $S=-\sum_{\mathbf{A}\in\mathbb{A}}P(\mathbf{A})\ln P(\mathbf{A})$, $C_0\equiv\langle C_0\rangle\equiv1$ sums up the normalization condition and the remaining $M-1$ constraints represent proper, topological properties. Such an optimization procedure defines the expression

\begin{equation}
P(\mathbf{A})=\frac{e^{-H(\mathbf{A})}}{Z}=\frac{e^{-H(\mathbf{A})}}{\sum_{\mathbf{A}\in\mathbb{A}}e^{-H(\mathbf{A})}}=\frac{e^{-\sum_{i=1}^M\theta_iC_i(\mathbf{A})}}{\sum_{\mathbf{A}\in\mathbb{A}}e^{-\sum_{i=1}^M\theta_iC_i(\mathbf{A})}}
\end{equation}
that can be made explicit only after a specific set of constraints has been chosen.\\

\subsection*{Signed Random Graph Model}

The first set of constraints we consider is represented by the properties $L^+(\mathbf{A})$ and $L^-(\mathbf{A})$. The Hamiltonian describing such a problem reads

\begin{equation}
H(\mathbf{A})=\alpha L^+(\mathbf{A})+\beta L^-(\mathbf{A});
\end{equation}
as a consequence, the partition function reads

\begin{align}
Z&=\sum_{\mathbf{A}\in\mathbb A}e^{-H(\mathbf A)}=\sum_{\mathbf{A}\in\mathbb A}e^{-\alpha L^+(\mathbf{A})-\beta L^-(\mathbf{A})}=\sum_{\mathbf A\in\mathbb A}e^{-\sum_{i=1}^N\sum_{j(>i)=1}^N(\alpha a_{ij}^++\beta a_{ij}^-)}=\sum_{\mathbf A\in\mathbb A}\prod_{i=1}^N\prod_{\substack{j=1\\(j>i)}}^Ne^{-\alpha a_{ij}^+-\beta a_{ij}^-}\nonumber\\
&=\prod_{i=1}^N\prod_{\substack{j=1\\(j>i)}}^N\sum_{a_{ij}=-1,0,1}e^{-\alpha a_{ij}^+-\beta a_{ij}^-}=\prod_{i=1}^N\prod_{\substack{j=1\\(j>i)}}^N(1+e^{-\alpha}+e^{-\beta})=(1+e^{-\alpha}+e^{-\beta})^{\binom{N}{2}}
\end{align}
and induces the expression

\begin{equation}
P_\text{SRGM}(\mathbf{A})=\frac{e^{-\alpha L^+(\mathbf{A})-\beta L^-(\mathbf{A})}}{(1+e^{-\alpha}+e^{-\beta})^{\binom N 2}}\equiv\frac{x^{L^+(\mathbf{A})}y^{L^-(\mathbf{A})}}{(1+x+y)^{\binom N 2}}\equiv(p^-)^{L^-}(p^0)^{L^0}(p^+)^{L^+}
\end{equation}
having posed $p^-\equiv\frac{e^{-\beta}}{1+e^{-\alpha}+e^{-\beta}}\equiv\frac{y}{1+x+y}$, $p^0\equiv\frac{1}{1+e^{-\alpha}+e^{-\beta}}\equiv\frac{1}{1+x+y}$ and $p^+\equiv\frac{e^{-\alpha}}{1+e^{-\alpha}+e^{-\beta}}\equiv\frac{x}{1+x+y}$, where $p^+$ is the probability that any two nodes are linked by a positive edge, $p^-$ is the probability that any two nodes are linked by a negative edge and $p^0$ is the probability that any two nodes are no linked at all. Hence, according to the SRGM, each entry of a signed network is a random variable following a generalized Bernoulli distribution, i.e. obeying the finite scheme

\begin{equation}
a_{ij}\sim
\begin{pmatrix}
-1 & 0 & +1\\
p^- & p^0 & p^+
\end{pmatrix}\quad\forall\:i<j;
\end{equation}
notice that while the expected value of the random variable $a_{ij}$ reads $\langle a_{ij}\rangle=p^+-p^-$, its variance reads $\text{Var}[a_{ij}]=p^-[1+(p^+-p^-)]+p^+[1-(p^+-p^-)]$ - such a notation, introduced by Khintchine in \emph{Mathematical Foundations of Information Theory}, compactly represent a discrete probability distribution, by listing its support on the first row and the probability of the elementary events constituting it on the second row. As a consequence, any network belonging to $\mathbb{A}$ is a collection of i.i.d. random variables and obeys the finite scheme

\begin{equation}
\mathbf{A}\sim\bigotimes
\begin{pmatrix}
-1 & 0 & +1\\
p^- & p^0 & p^+
\end{pmatrix}
\end{equation}
i.e. the direct product of the $\binom{N}{2}$ finite schemes above.\\

The probability, under the SRGM, that a graph has exactly $L^+$ positive links and $L^-$ negative links reads

\begin{equation}\label{multinomialedges}
P(L^-,L^+)=\binom{\binom{N}{2}}{L^-,L^0,L^+}(p^-)^{L^-}(p^0)^{L^0}(p^+)^{L^+};
\end{equation}
in other words, it is a multinomial distribution, i.e. a generalization of the binomial distribution in case there are more than two, possible outcomes for each trial. The combinatorial factor

\begin{equation}\label{multinomialcoeffbis}
\binom{\binom{N}{2}}{L^-,L^0,L^+}=\frac{\binom{N}{2}!}{L^-!L^0!L^+!}
\end{equation}
with $L^0=\binom{N}{2}-L=\binom{N}{2}-(L^-+L^+)$, is the (multinomial) coefficient counting the total number of ways $L$ links ($L^+$ of which are positive and $L^-$ of which are negative) can be placed among the node-pairs. Hence, Supplementary Equation \eqref{multinomialcoeffbis} also represents the total number of graphs with a given number of signed links.

Naturally, it is possible to define the marginal random variables $a_{ij}^+\sim\text{Ber}(p^+)$ and $a_{ij}^-\sim\text{Ber}(p^-)$ which, in turn, induce the marginal probability distributions $P(L^-)=\text{Bin}\left(\binom{N}{2},p^-\right)$, $P(L^0)=\text{Bin}\left(\binom{N}{2},p^0\right)$ and $P(L^+)=\text{Bin}\left(\binom{N}{2},p^+\right)$; from the latter ones, it follows that the total number of expected, positive links reads $\langle L^+\rangle=\binom{N}{2}p^+$ while the total number of expected, negative links reads $\langle L^-\rangle=\binom{N}{2}p^-$. Obviously, $\langle L\rangle=\langle L^+\rangle+\langle L^-\rangle=\binom{N}{2}(p^-+p^+)\equiv\binom{N}{2}p$. In other words, it is possible to define a `traditional' Random Graph Model whose parameter is $p\equiv p(a_{ij}=-1)+p(a_{ij}=+1)=p^-+p^+$.\\

\begin{figure}[t!]
\subfigure[]{\includegraphics[width=0.45\textwidth]{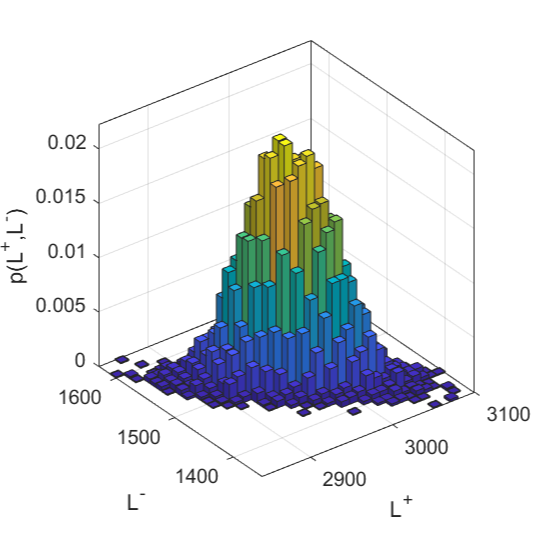}}
\subfigure[]{\includegraphics[width=0.45\textwidth]{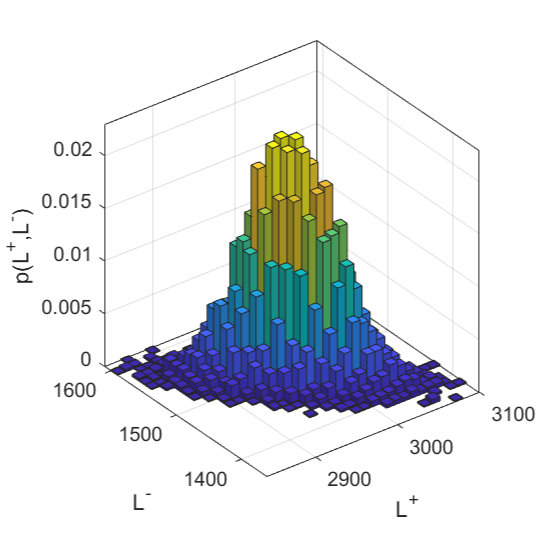}}\\
\subfigure[]{\includegraphics[width=0.49\textwidth]{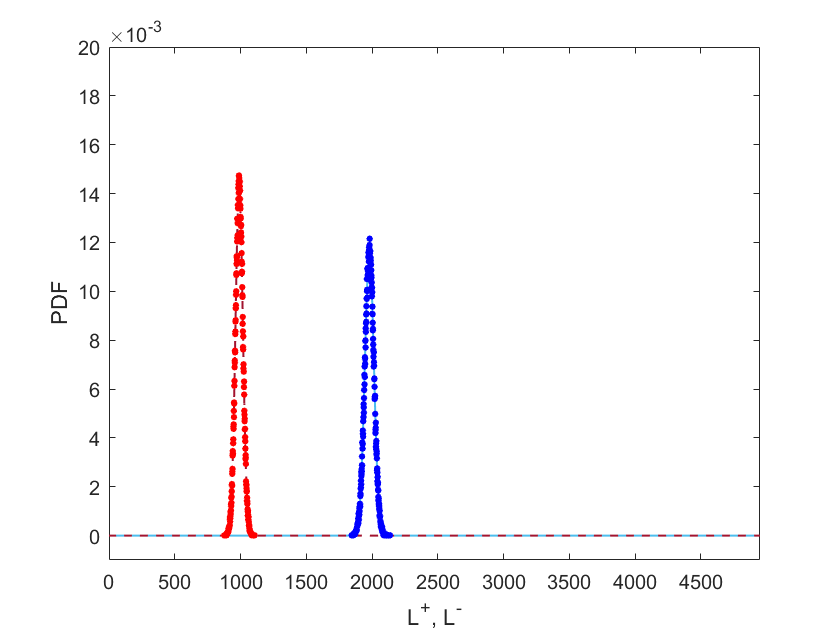}}
\subfigure[]{\includegraphics[width=0.49\textwidth]{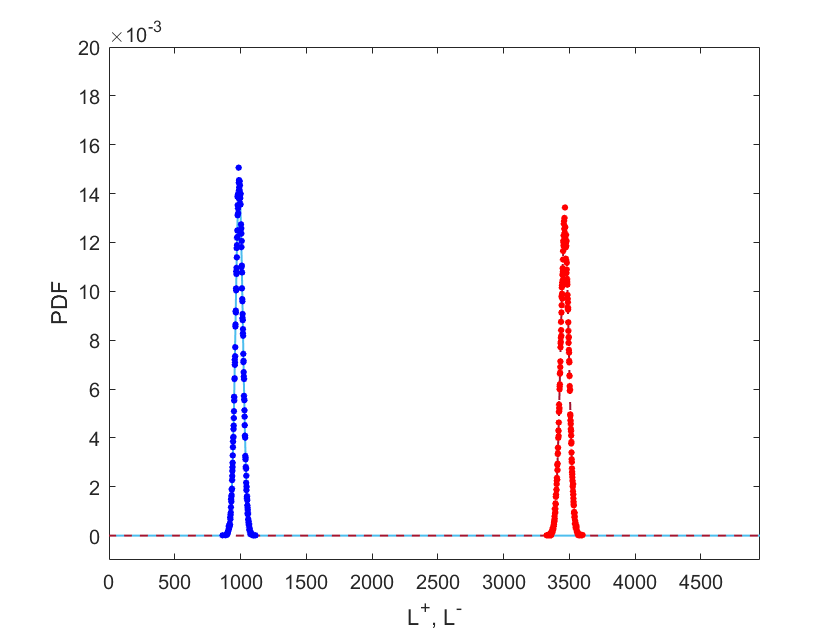}}
\caption{\textbf{Empirical distributions of $L^+$ and $L^-$ under the Signed Random Graph Model.} \textbf{(a)$-$(b)} - Empirical, joint distribution of $L^+$ and $L^-$ over an ensemble of $10.000$ configurations induced by the Signed Random Graph Model (SRGM) whose parameters have been tuned to $N=100$, $p^-=0.2$, $p^0=0.3$ and $p^+=0.5$ \textbf{(a)} and multinomial distribution $\text{Multi}\left(\binom{N}{2},\{p^-,p^0,p^+\}\right)$ \textbf{(b)}: the two have been sided for a visual comparison. \textbf{(c)$-$(d)} - Distributions of $L^+$ (blue dots) and $L^-$ (red dots) over an ensemble of $100.000$ configurations induced by the SRGM whose parameters have been tuned to $N=100$, $p^+=0.4$, $p^-=0.2$ \textbf{(c)} and  $N=100$, $p^+=0.2$, $p^-=0.7$ \textbf{(d)}. The red, dashed lines represent the binomial distributions $\text{Bin}\left(\binom{N}{2},p^-\right)$ while the blue, solid lines represent the binomial distributions $\text{Bin}\left(\binom{N}{2},p^+\right)$.}
\label{fig:1A}
\end{figure}

Let us, now, move to describe the behaviour of the degree. The probability, under the SRGM, that a node has exactly $k^+$ positive links and $k^-$ negative links reads

\begin{equation}\label{multinomialdeg}
P(k^-,k^+)=\binom{N-1}{k^-,k^0,k^+}(p^-)^{k^-}(p^0)^{k^0}(p^+)^{k^+};
\end{equation}
again, it obeys a multinomial distribution. The combinatorial factor

\begin{figure}[t!]
\subfigure[]{\includegraphics[width=0.45\textwidth]{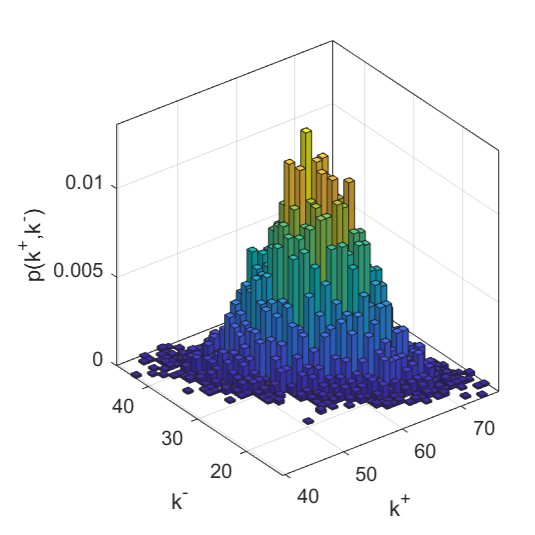}}
\subfigure[]{\includegraphics[width=0.46\textwidth]{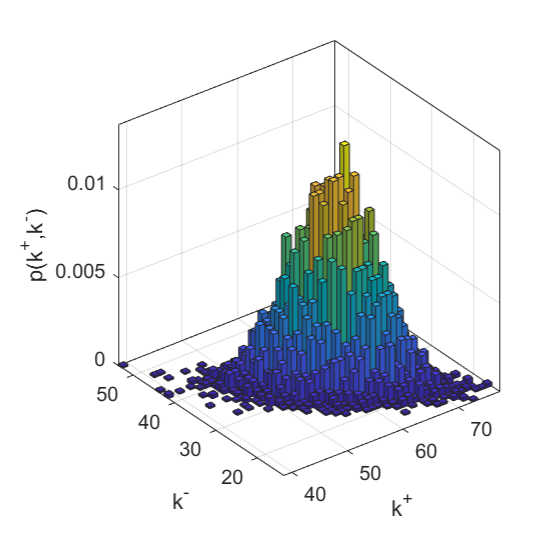}}\\
\subfigure[]{\includegraphics[width=0.49\textwidth]{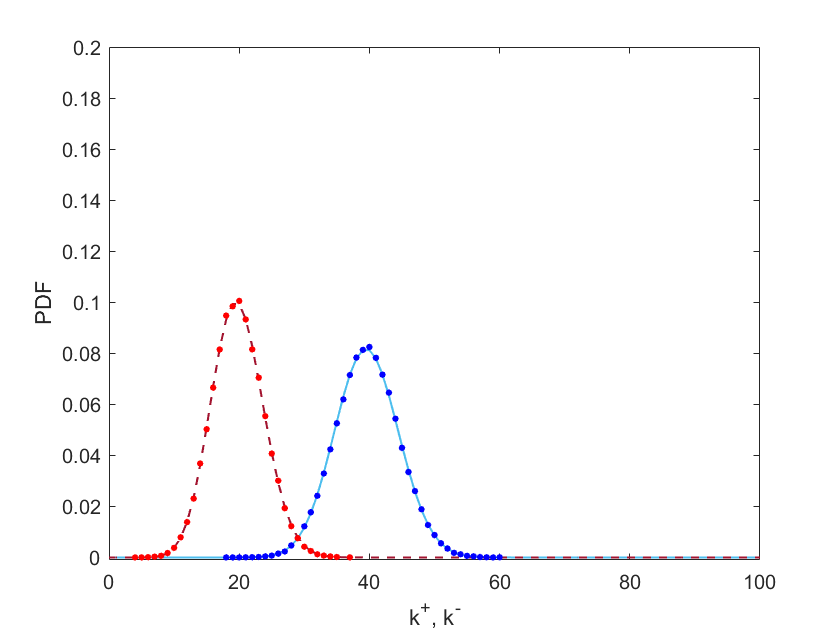}}
\subfigure[]{\includegraphics[width=0.49\textwidth]{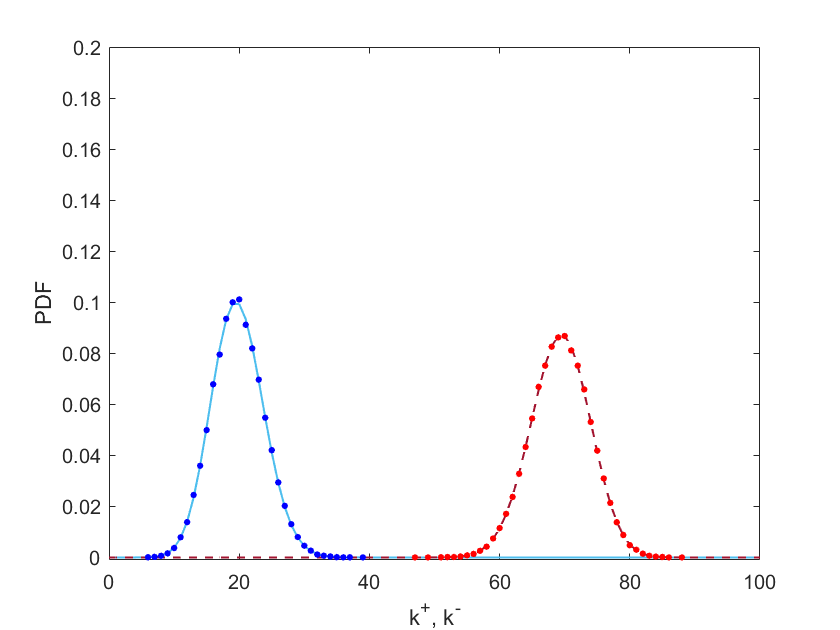}}
\caption{\textbf{Empirical distributions of $k^+$ and $k^-$ under the Signed Random Graph Model.} \textbf{(a)$-$(b)} - Empirical, joint distribution of $k^+$ and $k^-$ over an ensemble of $10.000$ configurations induced by the Signed Random Graph Model (SRGM) whose parameters have been tuned to $N=100$, $p^-=0.3$, $p^0=0.1$ and $p^+=0.6$ \textbf{(a)} and multinomial distribution $\text{Multi}\left(N-1,\{p^-,p^0,p^+\}\right)$ \textbf{(b)}: the two have been sided for a visual comparison. \textbf{(c)$-$(d)} - Distributions of $k^+$ (blue dots) and $k^-$ (red dots), for an arbitrarily chosen node, over an ensemble of $100.000$ configurations induced by the SRGM whose parameters have been tuned to $N=100$, $p^+=0.4$, $p^-=0.2$ \textbf{(c)} and $N=100$, $p^+=0.2$, $p^-=0.7$ \textbf{(d)}. The red, dashed lines represent the binomial distributions $\text{Bin}\left(N-1,p^-\right)$ while the blue, solid lines represent the binomial distributions $\text{Bin}\left(N-1,p^+\right)$.}
\label{fig:2A}
\end{figure}

\begin{equation}\label{multcoeff}
\binom{N-1}{k^-,k^0,k^+}=\frac{(N-1)!}{k^-!k^0!k^+!}
\end{equation}
with $k^0=(N-1)-k=(N-1)-(k^++k^-)$, is the (multinomial) coefficient counting the total number of ways $k$ links ($k^+$ of which are positive and $k^-$ of which are negative) can be placed among the $N-1$ node-pairs each node individuates. The marginal random variables $a_{ij}^+\sim\text{Ber}(p^+)$ and $a_{ij}^-\sim\text{Ber}(p^-)$ also induce the marginal probability distributions $P(k^-)=\text{Bin}\left(N-1,p^-\right)$, $P(k^0)=\text{Bin}\left(N-1,p^0\right)$ and $P(k^+)=\text{Bin}\left(N-1,p^+\right)$; from the latter ones, it follows that the expected, positive degree reads $\langle k^+\rangle=(N-1)p^+$ while the expected, negative degree reads $\langle k^-\rangle=(N-1)p^-$. Obviously, $\langle k\rangle=\langle k^+\rangle+\langle k^-\rangle=(N-1)(p^-+p^+)\equiv(N-1)p$.\\

In order to determine the parameters that define the SRGM, let us maximize the likelihood function

\begin{equation}
\mathcal{L}_\text{SRGM}(x,y)\equiv\ln P_\text{SRGM}(\mathbf{A}^*|x,y)=L^+(\mathbf{A}^*)\ln(x)+L^-(\mathbf{A}^*)\ln(y)-\binom{N}{2}\ln(1+x+y)
\end{equation}
with respect to $x$ and $y$. Upon doing so, we obtain the pair of equations

\begin{align}
\frac{\partial\mathcal{L}_\text{SRGM}(x,y)}{\partial x}=\frac{L^+(\mathbf{A}^*)}{x}-\binom{N}{2}\frac{1}{1+x+y},\quad\frac{\partial\mathcal{L}_\text{SRGM}(x,y)}{\partial y}=\frac{L^-(\mathbf{A}^*)}{y}-\binom{N}{2}\frac{1}{1+x+y};
\end{align}
equating them to zero leads us to find $L^+(\mathbf{A}^*)=\binom{N}{2}\frac{x}{1+x+y}=\binom{N}{2}p^+=\langle L^+\rangle$ and $L^-(\mathbf{A}^*)=\binom{N}{2}\frac{y}{1+x+y}=\binom{N}{2}p^-=\langle L^-\rangle$, i.e. 

\begin{align}
p^+=\frac{2L^+(\mathbf{A}^*)}{N(N-1)},\quad p^-=\frac{2L^-(\mathbf{A}^*)}{N(N-1)}.
\end{align}

Naturally, $p^0\equiv1-p^--p^+$.

\subsection*{Signed Random Graph Model with fixed topology}

Let us, again, consider the properties $L^+(\mathbf{A})$ and $L^-(\mathbf{A})$, to be satisfied by keeping a network topology fixed. In what follows, we will indicate the adopted topology as the one induced by the matrix $\mathbf{A}^*$. The Hamiltonian describing such a problem still reads

\begin{equation}
H(\mathbf{A})=\alpha L^+(\mathbf{A})+\beta L^-(\mathbf{A})
\end{equation}
but induces a partition function reading

\begin{align}
Z=&\sum_{\substack{\mathbf{A}\in\mathbb A\\(|\mathbf{A}|=|\mathbf{A}^*|)}}e^{-H(\mathbf{A})}=\sum_{\substack{\mathbf{A}\in\mathbb A\\(|\mathbf{A}|=|\mathbf{A}^*|)}}e^{-\alpha L^+(\mathbf{A})-\beta L^-(\mathbf{A})}=\sum_{\substack{\mathbf{A}\in\mathbb A\\(|\mathbf{A}|=|\mathbf{A}^*|)}}e^{-\sum_{i=1}^N\sum_{j(>i)=1}^N(\alpha a_{ij}^++\beta a_{ij}^-)}=\sum_{\substack{\mathbf{A}\in\mathbb A\\(|\mathbf{A}|=|\mathbf{A}^*|)}}\prod_{i=1}^N\prod_{\substack{j=1\\(j>i)}}^Ne^{-\alpha a_{ij}^+-\beta a_{ij}^-}\nonumber\\
=&\prod_{i=1}^N\prod_{\substack{j=1\\(j>i})}^N\left(\sum_{a_{ij}=-1,1}e^{-\alpha a_{ij}^+-\beta a_{ij}^-}\right)^{\left|a_{ij}^*\right|}=\prod_{i=1}^N\prod_{\substack{j=1\\(j>i)}}^N\left(e^{-\alpha}+e^{-\beta}\right)^{\left|a_{ij}^*\right|}=(e^{-\alpha}+e^{-\beta})^L;
\end{align}
in other words, the support of the distribution becomes the set of node pairs $i,j$, with $i<j$, such that $|a_{ij}^*|=1$, inducing a set of admissible configurations whose cardinality amounts at $2^L$. The expression above leads us to find

\begin{figure}[t!]
\subfigure[]{\includegraphics[width=0.49\textwidth]{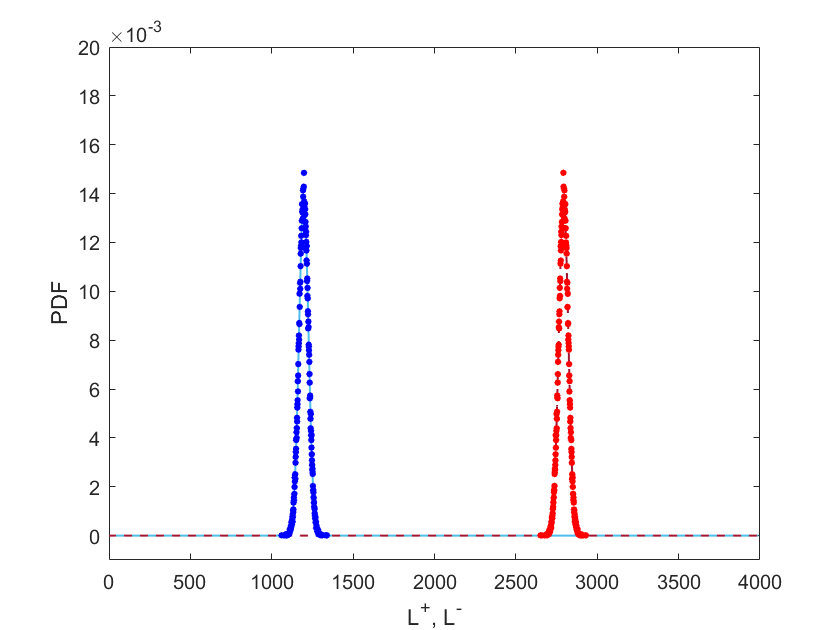}}
\subfigure[]{\includegraphics[width=0.49\textwidth]{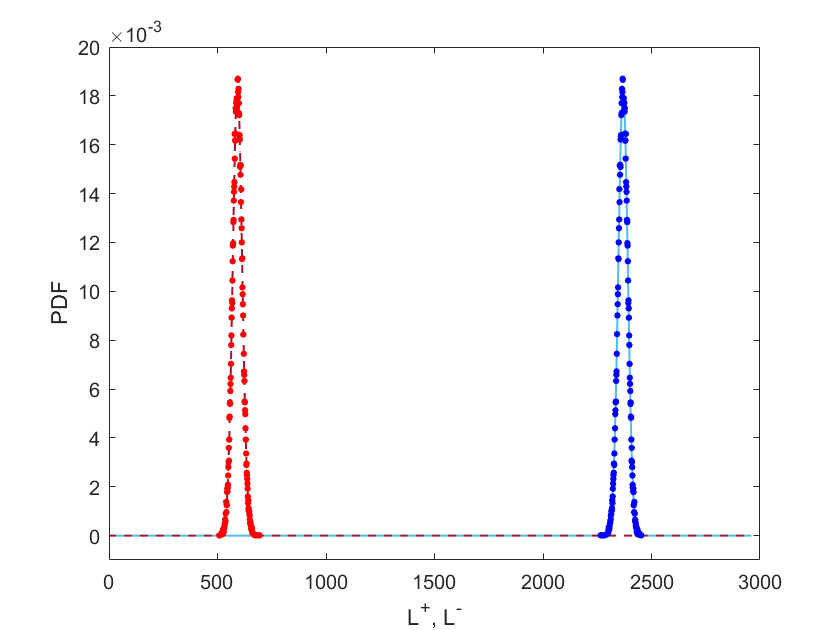}}
\caption{\textbf{Empirical distributions of $L^+$ and $L^-$ under the Signed Random Graph Model with Fixed Topology.}  Distributions of $L^+$ (blue dots) and $L^-$ (red dots) over an ensemble of $100.000$ configurations induced by the Signed Random Graph Model with Fixed Topology (SRGM-FT) whose parameters have been tuned to $N=100$, $p^+=0.3$, $p^-=0.7$ \textbf{(a)} and  $N=100$, $p^+=0.8$, $p^-=0.2$ \textbf{(b)}. The fixed topologies have been chosen by sampling a `traditional' Random Graph Model with $p=0.8$ \textbf{(a)} and $p=0.6$ \textbf{(b)}. The red, dashed lines represent the binomial distributions $\text{Bin}\left(L,p^-\right)$ while the blue, solid lines represent the binomial distributions $\text{Bin}\left(L,p^+\right)$.}
\label{fig:3A}
\end{figure}

\begin{align}\label{probgraph}
P_\text{SRGM-FT}(\mathbf{A})=\frac{e^{-\alpha L^+(\mathbf{A})-\beta L^-(\mathbf{A})}}{(e^{-\alpha}+e^{-\beta})^L}\equiv\frac{x^{L^+(\mathbf{A})}y^{L^-(\mathbf{A})}}{(x+y)^L}\equiv(p^-)^{L^-}(p^+)^{L^+}
\end{align}
having posed $p^-\equiv\frac{e^{-\beta}}{e^{-\alpha}+e^{-\beta}}\equiv\frac{y}{x+y}$ and $p^+\equiv\frac{e^{-\alpha}}{e^{-\alpha}+e^{-\beta}}\equiv\frac{x}{x+y}$ where $p^+$ is the probability that any two, connected nodes are linked by a positive edge and $p^-$ is the probability that any two, connected nodes are linked by a negative edge. Hence, according to the SRGM-FT, the generic entry of a signed network satisfying $|a_{ij}|=|a_{ij}^*|=1$ is a random variable following a Bernoulli distribution, i.e. obeying the finite scheme

\begin{equation}
a_{ij}\sim
\begin{pmatrix}
-1 & +1\\
p^- & p^+
\end{pmatrix}\quad\forall\:i<j\:|\:|a_{ij}^*|=1.
\end{equation}

\begin{figure}[t!]
\subfigure[]{\includegraphics[width=0.49\textwidth]{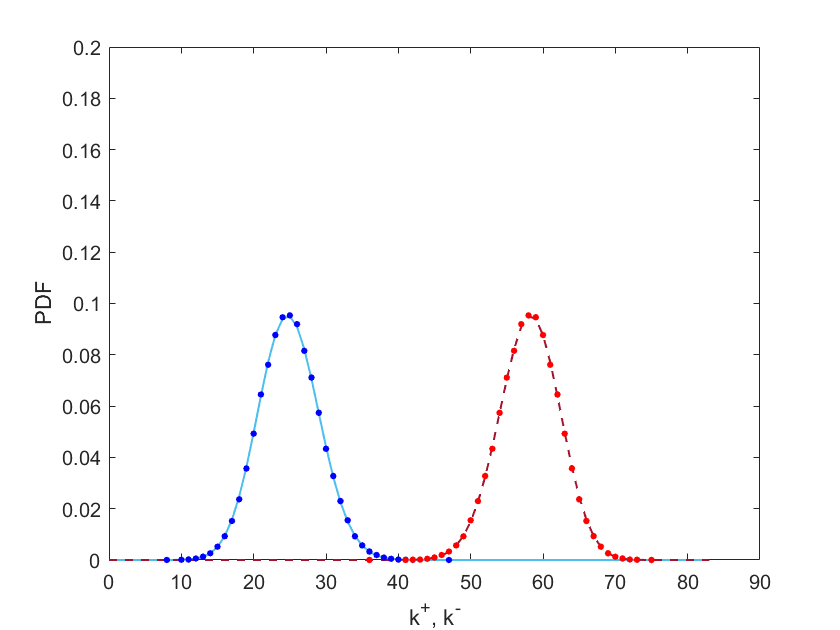}}
\subfigure[]{\includegraphics[width=0.49\textwidth]{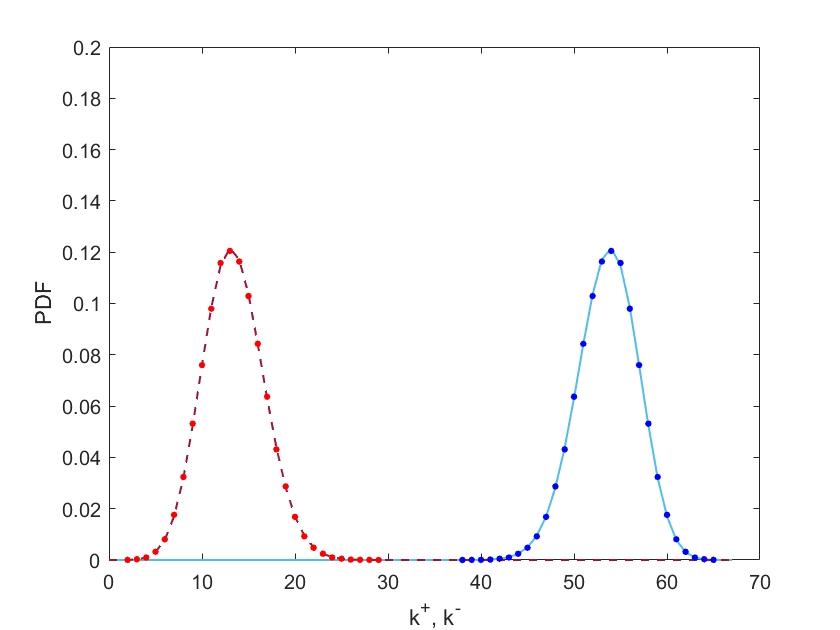}}
\caption{\textbf{Empirical distributions of $k^+$ and $k^-$ under the Signed Random Graph Model with Fixed Topology.} Distributions of $k^+$ (blue dots) and $k^-$ (red dots), for an arbitrarily chosen node, over an ensemble of $100.000$ configurations induced by the Signed Random Graph Model with Fixed Topology (SRGM-FT) whose parameters have been tuned to $N=100$, $p^+=0.3$, $p^-=0.7$ \textbf{(a)} and $N=100$, $p^+=0.8$, $p^-=0.2$ \textbf{(b)}. The fixed topologies have been chosen by sampling a `traditional' Random Graph Model with $p=0.8$ \textbf{(a)} and $p=0.6$ \textbf{(b)}. The red, dashed lines represent the binomial distributions $\text{Bin}\left(k,p^-\right)$ while the blue, solid lines represent the binomial distributions $\text{Bin}\left(k,p^+\right)$.}
\label{fig:4A}
\end{figure}

The probability, under the SRGM-FT, that a graph has exactly $L^+$ positive links reads

\begin{equation}
P(L^+)=\binom{L}{L^+}(p^-)^{L^-}(p^+)^{L^+}=\binom{L}{L^+}(p^+)^{L^+}(1-p^+)^{L-L^+}
\end{equation}
i.e. it is a binomial distribution, with $L$ indicating the total number of unsigned links. As a consequence, the total number of expected, positive links reads $\langle L^+\rangle=Lp^+$; analogously, $L^-\sim\text{Bin}(L,p^-)$. Similarly, the probability, under the SRGM-FT, that node $i$ establishes exactly $k_i^+$ positive links reads

\begin{equation}
P(k_i^+)=\binom{k_i}{k_i^+}(p^-)^{k_i^-}(p^+)^{k_i^+}=\binom{k_i}{k_i^+}(p^+)^{k_i^+}(1-p^-)^{k_i-k_i^+};
\end{equation}
again, it is a binomial distribution, with $k_i$ indicating the unsigned degree of node $i$. As a consequence, the expected, positive degree of node $i$ reads $\langle k_i^+\rangle=k_ip^+$; analogously, $k_i^-\sim\text{Bin}(k_i,p^-)$.\\

In order to determine the parameters that define the SRGM-FT, let us maximize the likelihood function

\begin{equation}
\mathcal{L}_\text{SRGM-FT}(x,y)\equiv\ln P_\text{SRGM-FT}(\mathbf{A}^*|x,y)=L^+(\mathbf{A}^*)\ln(x)+L^-(\mathbf{A}^*)\ln(y)-L(\mathbf{A}^*)\ln(x+y)
\end{equation}
with respect to $x$ and $y$. Upon doing so, we obtain the pair of equations

\begin{align}
\frac{\partial\mathcal{L}_\text{SRGM-FT}(x,y)}{\partial x}=\frac{L^+(\mathbf{A}^*)}{x}-\frac{L(\mathbf{A}^*)}{x+y},\quad\frac{\partial\mathcal{L}_\text{SRGM-FT}(x,y)}{\partial y}=\frac{L^-(\mathbf{A}^*)}{y}-\frac{L(\mathbf{A}^*)}{x+y};
\end{align}
equating them to zero leads us to find $L^+(\mathbf{A}^*)=L(\mathbf{A}^*)\frac{x}{x+y}=L(\mathbf{A}^*)p^+=\langle L^+\rangle$ and $L^-(\mathbf{A}^*)=L(\mathbf{A}^*)\frac{y}{x+y}=L(\mathbf{A}^*)p^-=\langle L^-\rangle$, i.e. 

\begin{align}
p^+=\frac{L^+(\mathbf{A}^*)}{L(\mathbf{A}^*)},\quad p^-=\frac{L^-(\mathbf{A}^*)}{L(\mathbf{A}^*)}.
\end{align}

For an illustrative example of the empirical distributions of $L^+$, $L^-$, $k^+$ and $k^-$ under our homogeneous network models see Supplementary Figures \ref{fig:1A}, \ref{fig:2A}, \ref{fig:3A} and \ref{fig:4A}.

\subsection{Signed Random Graph Model: free VS fixed topology}

In order to clarify the relationship between the SRGM and the SRGM-FT, let us write

\begin{align}
P_{\text{SRGM}}(\mathbf{A})&=P_{\text{RGM}}(\mathbf{A})\cdot\frac{ P_{\text{SRGM}}(\mathbf{A})}{P_{\text{RGM}}(\mathbf{A})}\nonumber\\
&=p^L(1-p)^{\binom{N}{2}-L}\cdot\frac{(p^-)^{L^-}(p^+)^{L^+}(1-p^--p^+)^{\binom{N}{2}-L^--L^+}}{p^L(1-p)^{\binom{N}{2}-L}}\nonumber\\
&=p^L(1-p)^{\binom{N}{2}-L}\cdot\frac{(p^-)^{L^-}(p^+)^{L^+}(1-p^--p^+)^{\binom{N}{2}-L^--L^+}}{p^{L^-}p^{L^+}(1-p)^{\binom{N}{2}-L^--L^+}}\nonumber\\
&=p^L(1-p)^{\binom{N}{2}-L}\cdot
\left(\frac{p^-}{p}\right)^{L^-}\left(\frac{p^+}{p}\right)^{L^+}\left(\frac{1-p^--p^+}{1-p}\right)^{\binom{N}{2}-L^--L^+};
\end{align}
since the RGM induced by the SRGM satisfies the relationship $p\equiv p^-+p^+$, one has that

\begin{align}
P_{\text{SRGM}}(\mathbf{A})&=p^L(1-p)^{\binom{N}{2}-L}\cdot
\left(\frac{p^-}{p}\right)^{L^-}\left(\frac{p^+}{p}\right)^{L^+}\nonumber\\
&=P_{\text{RGM}}(\mathbf{A})\cdot P_{\text{SRGM-FT}}(\mathbf{A})
\end{align}
the parameters defining the SRGM-FT being, now, $p^-/p$ and $p^+/p$. Beside having have an intuitive meaning, i.e.

\begin{align}
\frac{p^-}{p}&=\frac{p^-}{p^-+p^+}=\frac{P(\text{`link'}\:\cap\:\text{`link $-$'})}{P(\text{`link'})}=P(\text{`link $-$'}\:|\:\text{`link'}),\\
\frac{p^+}{p}&=\frac{p^+}{p^-+p^+}=\frac{P(\text{`link'}\:\cap\:\text{`link $+$'})}{P(\text{`link'})}=P(\text{`link $+$'}\:|\:\text{`link'})
\end{align}
these expressions are also consistent with the estimations of the parameters obtained via the likelihood maximization: in fact,

\begin{align}
\frac{p^-}{p}&=\frac{p^-}{p^-+p^+}=\frac{2L^-/N(N-1)}{2L^-/N(N-1)+2L^+/N(N-1)}=\frac{2L^-/N(N-1)}{2L/N(N-1)}=\frac{L^-}{L},\\
\frac{p^+}{p}&=\frac{p^+}{p^-+p^+}=\frac{2L^+/N(N-1)}{2L^-/N(N-1)+2L^+/N(N-1)}=\frac{2L^+/N(N-1)}{2L/N(N-1)}=\frac{L^+}{L}.
\end{align}

\subsection*{Signed Configuration Model}

The second set of constraints we consider is represented by the properties $\{k_i^+(\mathbf{A})\}_{i=1}^N$ and $\{k_i^-(\mathbf{A})\}_{i=1}^N$. The Hamiltonian describing such a problem reads

\begin{align}
H(\mathbf{A})=\sum_{i=1}^N[\alpha_i k_i^+(\mathbf{A})+\beta_i k_i^-(\mathbf{A})];
\end{align}
as a consequence, the partition function reads

\begin{align}
Z&=\sum_{\mathbf{A}\in\mathbb A}e^{-H(\mathbf A)}=\sum_{\mathbf{A}\in\mathbb A}e^{-\sum_{i=1}^N[\alpha_i k_i^+(\mathbf{A})+\beta_i k_i^-(\mathbf{A})]}=\sum_{\mathbf A\in\mathbb A}e^{-\sum_{i=1}^N\sum_{j(>i)=1}^N[(\alpha_i+\alpha_j)a_{ij}^++(\beta_i+\beta_j)a_{ij}^-]}\nonumber\\
&=\sum_{\mathbf A\in\mathbb A}\prod_{i=1}^N\prod_{\substack{j=1\\(j>i)}}^Ne^{-(\alpha_i+\alpha_j)a_{ij}^+-(\beta_i+\beta_j)a_{ij}^-}=\prod_{i=1}^N\prod_{\substack{j=1\\(j>i)}}^N\sum_{a_{ij}=-1,0,1}e^{-(\alpha_i+\alpha_j)a_{ij}^+-(\beta_i+\beta_j)a_{ij}^-}=\prod_{i=1}^N\prod_{\substack{j=1\\(j>i)}}^N(1+e^{-(\alpha_i+\alpha_j)}+e^{-(\beta_i+\beta_j)})
\end{align}
and induces the expression

\begin{align}
P_\text{SCM}(\mathbf{A})=\frac{e^{-\sum_{i=1}^N[\alpha_i k_i^+(\mathbf{A})+\beta_i k_i^-(\mathbf{A})]}}{\prod_{i=1}^N\prod_{\substack{j=1\\(j>i)}}^N(1+e^{-(\alpha_i+\alpha_j)}+e^{-(\beta_i+\beta_j)})}\equiv\frac{\prod_{i=1}^Nx_i^{k_i^+(\mathbf{A})}y_i^{k_i^-(\mathbf{A})}}{\prod_{i=1}^N\prod_{\substack{j=1\\(j>i)}}^N(1+x_ix_j+y_iy_j)}\equiv\prod_{i=1}^N\prod_{\substack{j=1\\(j>i)}}^N(p_{ij}^-)^{a_{ij}^-}(p_{ij}^0)^{a_{ij}^0}(p_{ij}^+)^{a_{ij}^+}
\end{align}
having posed $p_{ij}^-\equiv\frac{e^{-(\beta_i+\beta_j)}}{1+e^{-(\alpha_i+\alpha_j)}+e^{-(\beta_i+\beta_j)}}\equiv\frac{y_iy_j}{1+x_ix_j+y_iy_j}$, $p_{ij}^0\equiv\frac{1}{1+e^{-(\alpha_i+\alpha_j)}+e^{-(\beta_i+\beta_j)}}\equiv\frac{1}{1+x_ix_j+y_iy_j}$ and $p_{ij}^+\equiv\frac{e^{-(\alpha_i+\alpha_j)}}{1+e^{-(\alpha_i+\alpha_j)}+e^{-(\beta_i+\beta_j)}}\equiv\frac{x_ix_j}{1+x_ix_j+y_iy_j}$, where $p_{ij}^+$ is the probability that nodes $i$ and $j$ are linked by a positive edge, $p_{ij}^-$ is the probability that nodes $i$ and $j$ are linked by a negative edge and $p_{ij}^0$ is the probability that nodes $i$ and $j$ are no linked at all. Hence, according to the SCM, the generic entry of a signed network is a random variable following a generalized Bernoulli distribution, i.e. obeying the finite scheme

\begin{equation}
a_{ij}\sim
\begin{pmatrix}
-1 & 0 & +1\\
p_{ij}^- & p_{ij}^0 & p_{ij}^+
\end{pmatrix}\quad\forall\:i<j;
\end{equation}
as a consequence, any network belonging to $\mathbb{A}$ is a collection of independent random variables, each one obeying the finite scheme

\begin{equation}
\mathbf{A}\sim\bigotimes
\begin{pmatrix}
-1 & 0 & +1\\
p_{ij}^- & p_{ij}^0 & p_{ij}^+
\end{pmatrix}
\end{equation}
i.e. the direct product of the $\frac{N(N-1)}{2}=\binom{N}{2}$ finite schemes above.\\

In the case of the SCM, $L^+$ is a random variable obeying the Poisson-Binomial distribution that we indicate as $\text{PoissBin}\left(\binom{N}{2},\{p_{ij}^+\}_{i,j=1}^N\right)$; analogously, $L^-\sim\text{PoissBin}\left(\binom{N}{2},\{p_{ij}^-\}_{i,j=1}^N\right)$. Similarly, $k_i^+$ is a random variable obeying the Poisson-Binomial distribution that we indicate as $\text{PoissBin}\left(N-1,\{p_{ij}^+\}_{j=1}^N\right)$; analogously, $k_i^-\sim\text{PoissBin}\left(N-1,\{p_{ij}^-\}_{j=1}^N\right)$. Hence, the total number of expected, positive links reads $\langle L^+\rangle=\sum_{i=1}^N\sum_{j(>i)=1}^Np_{ij}^+$ while the total number of expected, negative links reads $\langle L^-\rangle=\sum_{i=1}^N\sum_{j(>i)=1}^Np_{ij}^-$; analogously, $\langle k_i^+\rangle=\sum_{j(\neq i)=1}^Np_{ij}^+$ and $\langle k_i^-\rangle=\sum_{j(\neq i)=1}^Np_{ij}^-$.\\

In order to determine the parameters that define the SCM, let us maximize the likelihood function

\begin{align}
\mathcal{L}_\text{SCM}(\{x_i\}_{i=1}^N,\{y_i\}_{i=1}^N)&\equiv\ln P_\text{SCM}(\mathbf{A}^*|\{x_i\}_{i=1}^N,\{y_i\}_{i=1}^N)\nonumber\\
&=\sum_{i=1}^Nk_i^+(\mathbf{A}^*)\ln(x_i)+\sum_{i=1}^Nk_i^-(\mathbf{A}^*)\ln(y_i)-\sum_{i=1}^N\sum_{\substack{j=1\\(j>i)}}^N\ln(1+x_ix_j+y_iy_j)
\end{align}
with respect to $x_i$ and $y_i$, $\forall\:i$. Upon doing so, we obtain the system of equations

\begin{align}
\frac{\partial\mathcal{L}_\text{SCM}(\{x_i\}_{i=1}^N,\{y_i\}_{i=1}^N)}{\partial x_i}&=\frac{k_i^+(\mathbf{A}^*)}{x_i}-\sum_{\substack{j=1\\(j\neq i)}}^N\frac{x_j}{1+x_ix_j+y_iy_j}\quad\forall\:i,\\
\frac{\partial\mathcal{L}_\text{SCM}(\{x_i\}_{i=1}^N,\{y_i\}_{i=1}^N)}{\partial y_i}&=\frac{k_i^-(\mathbf{A}^*)}{y_i}-\sum_{\substack{j=1\\(j\neq i)}}^N\frac{y_j}{1+x_ix_j+y_iy_j}\quad\forall\:i;
\end{align}
equating them to zero leads us to find

\begin{align}
k_i^+(\mathbf{A}^*)&=\sum_{\substack{j=1\\(j\neq i)}}^N\frac{x_ix_j}{1+x_ix_j+y_iy_j}=\sum_{\substack{j=1\\(j\neq i)}}^Np_{ij}^+=\langle k_i^+\rangle\quad\forall\:i,\\
k_i^-(\mathbf{A}^*)&=\sum_{\substack{j=1\\(j\neq i)}}^N\frac{y_iy_j}{1+x_ix_j+y_iy_j}=\sum_{\substack{j=1\\(j\neq i)}}^Np_{ij}^-=\langle k_i^-\rangle\quad\forall\:i.
\end{align}

Although the system above can be solved only numerically, particular conditions exist under which the equations constituting it can be approximated and solved explicitly. They are collectively named `sparse-case' approximation of the SCM and hold true whenever $x_i\ll1$ and $y_i\ll1$, $\forall\:i$. In this case, one can pose $p_{ij}^+\simeq x_ix_j$ and $p_{ij}^-\simeq y_iy_j$, $\forall\:i<j$, which allow the equations above to be simplified as follows

\begin{align}
k_i^+(\mathbf{A}^*)&\simeq\sum_{\substack{j=1\\(j\neq i)}}^Nx_ix_j\quad\forall\:i,\\
k_i^-(\mathbf{A}^*)&\simeq\sum_{\substack{j=1\\(j\neq i)}}^Ny_iy_j\quad\forall\:i;
\end{align}
the latter ones induce the expressions $x_i=\frac{k_i^+(\mathbf{A}^*)}{\sum_{j=1}^Nx_j}=\frac{k_i^+(\mathbf{A}^*)}{\sqrt{2L^+(\mathbf{A}^*)}}$ and $y_i=\frac{k_i^-(\mathbf{A}^*)}{\sum_{j=1}^Ny_j}=\frac{k_i^-(\mathbf{A}^*)}{\sqrt{2L^-(\mathbf{A}^*)}}$, $\forall\:i$, allowing us to find

\begin{align}
p_{ij}^+\simeq\frac{k_i^+(\mathbf{A}^*)k_j^+(\mathbf{A}^*)}{2L^+(\mathbf{A}^*)},\\
p_{ij}^-\simeq\frac{k_i^-(\mathbf{A}^*)k_j^-(\mathbf{A}^*)}{2L^-(\mathbf{A}^*)}.
\end{align}

The system of equations above is also known with the name of Signed Chung-Lu Model (SCLM).

\subsection*{Signed Configuration Model with fixed topology}

Let us, again, consider the properties $\{k_i^+(\mathbf{A})\}_{i=1}^N$ and $\{k_i^-(\mathbf{A})\}_{i=1}^N$, to be satisfied by keeping a network topology fixed. As usual, we will indicate the adopted topology as the one induced by the matrix $\mathbf{A}^*$. The Hamiltonian describing such a problem still reads

\begin{align}
H(\mathbf{A})=\sum_{i=1}^N[\alpha_i k_i^+(\mathbf{A})+\beta_i k_i^-(\mathbf{A})]
\end{align}
but induces a partition function reading 

\begin{align}
Z&=\sum_{\substack{\mathbf{A}\in\mathbb A\\(|\mathbf{A}|=|\mathbf{A}^*|)}}e^{-H(\mathbf A)}=\sum_{\substack{\mathbf{A}\in\mathbb A\\(|\mathbf{A}|=|\mathbf{A}^*|)}}e^{-\sum_{i=1}^N[\alpha_i k_i^+(\mathbf{A})+\beta_i k_i^-(\mathbf{A})]}=\sum_{\substack{\mathbf{A}\in\mathbb A\\(|\mathbf{A}|=|\mathbf{A}^*|)}}e^{-\sum_{i=1}^N\sum_{j(>i)=1}^N[(\alpha_i+\alpha_j)a_{ij}^++(\beta_i+\beta_j)a_{ij}^-]}\nonumber\\
&=\sum_{\substack{\mathbf{A}\in\mathbb A\\(|\mathbf{A}|=|\mathbf{A}^*|)}}\prod_{i=1}^N\prod_{\substack{j=1\\(j>i)}}^Ne^{-(\alpha_i+\alpha_j)a_{ij}^+-(\beta_i+\beta_j)a_{ij}^-}=\prod_{i=1}^N\prod_{\substack{j=1\\(j>i)}}^N\left(\sum_{a_{ij}=-1,1}e^{-(\alpha_i+\alpha_j)a_{ij}^+-(\beta_i+\beta_j)a_{ij}^-}\right)^{\left|a_{ij}^*\right|}\nonumber\\
&=\prod_{i=1}^N\prod_{\substack{j=1\\(j>i)}}
^N\left(e^{-(\alpha_i+\alpha_j)}+e^{-(\beta_i+\beta_j)}\right)^{\left|a_{ij}^*\right|}
\end{align}
which, in turn, induces the expression

\begin{align}
P_\text{SCM-FT}(\mathbf{A})&=\frac{e^{-\sum_{i=1}^N[\alpha_i k_i^+(\mathbf{A})+\beta_i k_i^-(\mathbf{A})]}}{\prod_{i=1}^N\prod_{\substack{j=1\\(j>i)}}^N\left(e^{-(\alpha_i+\alpha_j)}+e^{-(\beta_i+\beta_j)}\right)^{\left|a_{ij}^*\right|}}\equiv\frac{\prod_{i=1}^Nx_i^{k_i^+(\mathbf{A})}y_i^{k_i^-(\mathbf{A})}}{\prod_{i=1}^N\prod_{\substack{j=1\\(j>i)}}^N(x_ix_j+y_iy_j)^{\left|a_{ij}^*\right|}}\nonumber\\
&\equiv\prod_{i=1}^N\prod_{\substack{j=1\\(j>i)}}^N\left[(p_{ij}^-)^{a_{ij}^-}(p_{ij}^+)^{a_{ij}^+}\right]^{\left|a_{ij}^*\right|}=\prod_{i=1}^N\prod_{\substack{j=1\\(j>i)}}^N(p_{ij}^-)^{a_{ij}^-}(p_{ij}^+)^{a_{ij}^+}
\end{align}
having posed $p_{ij}^-\equiv\frac{e^{-(\beta_i+\beta_j)}}{e^{-(\alpha_i+\alpha_j)}+e^{-(\beta_i+\beta_j)}}\equiv\frac{y_iy_j}{x_ix_j+y_iy_j}$ and $p_{ij}^+\equiv\frac{e^{-(\alpha_i+\alpha_j)}}{e^{-(\alpha_i+\alpha_j)}+e^{-(\beta_i+\beta_j)}}\equiv\frac{x_ix_j}{x_ix_j+y_iy_j}$ where $p_{ij}^+$ is the probability that nodes $i$ and $j$ are linked by a positive edge and $p_{ij}^-$ is the probability that nodes $i$ and $j$ are linked by a negative edge. Hence, according to the SCM-FT, the generic entry of a signed network satisfying $|a_{ij}|=|a_{ij}^*|=1$ is a random variable following a Bernoulli distribution, i.e. obeying the finite scheme

\begin{equation}
a_{ij}\sim
\begin{pmatrix}
-1 & +1\\
p_{ij}^- & p_{ij}^+
\end{pmatrix}\quad\forall\:i<j\:|\:|a_{ij}^*|=1.
\end{equation}

In the case of the SCM-FT, $L^+$ is a random variable obeying the Poisson-Binomial distribution that we indicate as $\text{PoissBin}\left(L,\{p_{ij}^+\}_{i,j=1}^N\right)$; analogously, $L^-\sim\text{PoissBin}\left(L,\{p_{ij}^-\}_{i,j=1}^N\right)$. Similarly, $k_i^+$ is a random variable obeying the Poisson-Binomial distribution that we indicate as $\text{PoissBin}\left(k_i,\{p_{ij}^+\}_{j=1}^N\right)$; analogously, $k_i^-\sim\text{PoissBin}\left(k_i,\{p_{ij}^-\}_{j=1}^N\right)$. Hence, the total number of expected, positive links reads $\langle L^+\rangle=\sum_{i=1}^N\sum_{j(>i)=1}^N|a_{ij}^*|p_{ij}^+$ while the total number of expected, negative links reads $\langle L^-\rangle=\sum_{i=1}^N\sum_{j(>i)=1}^N|a_{ij}^*|p_{ij}^-$; analogously, $\langle k_i^+\rangle=\sum_{j(\neq i)=1}^N|a_{ij}^*|p_{ij}^+$ and $\langle k_i^-\rangle=\sum_{j(\neq i)=1}^N|a_{ij}^*|p_{ij}^-$.\\

In order to determine the parameters that define the SCM-FT, let us maximize the likelihood function

\begin{align}
\mathcal{L}_\text{SCM-FT}(\{x_i\}_{i=1}^N,\{y_i\}_{i=1}^N)&\equiv\ln P_\text{SCM-FT}(\mathbf{A}^*|\{x_i\}_{i=1}^N,\{y_i\}_{i=1}^N)\nonumber\\
&=\sum_{i=1}^Nk_i^+(\mathbf{A}^*)\ln(x_i)+\sum_{i=1}^Nk_i^-(\mathbf{A}^*)\ln(y_i)-\sum_{i=1}^N\sum_{\substack{j=1\\(j>i)}}^N|a_{ij}^*|\ln(x_ix_j+y_iy_j)
\end{align}
with respect to $x_i$ and $y_i$, $\forall\:i$. Upon doing so, we obtain the system of equations

\begin{align}
\frac{\partial\mathcal{L}_\text{SCM-FT}(\{x_i\}_{i=1}^N,\{y_i\}_{i=1}^N)}{\partial x_i}&=\frac{k_i^+(\mathbf{A}^*)}{x_i}-\sum_{\substack{j=1\\(j\neq i)}}^N|a_{ij}^*|\frac{x_j}{x_ix_j+y_iy_j}\quad\forall\:i,\\
\frac{\partial\mathcal{L}_\text{SCM-FT}(\{x_i\}_{i=1}^N,\{y_i\}_{i=1}^N)}{\partial y_i}&=\frac{k_i^-(\mathbf{A}^*)}{y_i}-\sum_{\substack{j=1\\(j\neq i)}}^N|a_{ij}^*|\frac{y_j}{x_ix_j+y_iy_j}\quad\forall\:i;
\end{align}
equating them to zero leads us to find

\begin{align}
k_i^+(\mathbf{A}^*)&=\sum_{\substack{j=1\\(j\neq i)}}^N|a_{ij}^*|\frac{x_ix_j}{x_ix_j+y_iy_j}=\sum_{\substack{j=1\\(j\neq i)}}^N|a_{ij}^*|p_{ij}^+=\langle k_i^+\rangle\quad\forall\:i,\\
k_i^-(\mathbf{A}^*)&=\sum_{\substack{j=1\\(j\neq i)}}^N|a_{ij}^*|\frac{y_iy_j}{x_ix_j+y_iy_j}=\sum_{\substack{j=1\\(j\neq i)}}^N|a_{ij}^*|p_{ij}^-=\langle k_i^-\rangle\quad\forall\:i.
\end{align}

The system above can be solved only numerically.

\subsection{Signed Configuration Model: free VS fixed topology}

In order to clarify the relationship between the SCM and the SCM-FT, let us write

\begin{align}
P_{\text{SCM}}(\mathbf{A})&=P_{\text{FM}}(\mathbf{A})\cdot\frac{ P_{\text{SCM}}(\mathbf{A})}{P_{\text{FM}}(\mathbf{A})}\nonumber\\
&=\prod_{i=1}^N\prod_{\substack{j=1\\(j>i)}}^Np_{ij}^{a_{ij}}(1-p_{ij})^{1-a_{ij}}\cdot\frac{\prod_{i=1}^N\prod_{\substack{j=1\\(j>i)}}^N(p_{ij}^-)^{a_{ij}^-}(p_{ij}^+)^{a_{ij}^+}(1-p_{ij}^--p_{ij}^+)^{1-a_{ij}^--a_{ij}^+}}{\prod_{i=1}^N\prod_{\substack{j=1\\(j>i)}}^Np_{ij}^{a_{ij}}(1-p_{ij})^{1-a_{ij}}}\nonumber\\
&=\prod_{i=1}^N\prod_{\substack{j=1\\(j>i)}}^Np_{ij}^{a_{ij}}(1-p_{ij})^{1-a_{ij}}\cdot\frac{\prod_{i=1}^N\prod_{\substack{j=1\\(j>i)}}^N(p_{ij}^-)^{a_{ij}^-}(p_{ij}^+)^{a_{ij}^+}(1-p_{ij}^--p_{ij}^+)^{1-a_{ij}^--a_{ij}^+}}{\prod_{i=1}^N\prod_{\substack{j=1\\(j>i)}}^Np_{ij}^{a_{ij}^-+a_{ij}^+}(1-p_{ij})^{1-a_{ij}^--a_{ij}^+}}\nonumber\\
&=\prod_{i=1}^N\prod_{\substack{j=1\\(j>i)}}^Np_{ij}^{a_{ij}}(1-p_{ij})^{1-a_{ij}}\cdot\prod_{i=1}^N\prod_{\substack{j=1\\(j>i)}}^N\left(\frac{p_{ij}^-}{p_{ij}}\right)^{a_{ij}^-}\left(\frac{p_{ij}^+}{p_{ij}}\right)^{a_{ij}^+}\left(\frac{1-p_{ij}^--p_{ij}^+}{1-p_{ij}}\right)^{1-a_{ij}^--a_{ij}^+}
\end{align}
where $P_{\text{FM}}(\mathbf{A})$ indicates the probability distribution of a generic, factorizable (null) model. Upon requiring $p_{ij}\equiv p_{ij}^-+p_{ij}^+$, we turn the FM into a Configuration Model (CM) whose coefficients are `induced' by the ones of the SCM (in fact, $k_i(\mathbf{A}^*)=k_i^-(\mathbf{A}^*)+k_i^+(\mathbf{A}^*)=\sum_{j(\neq i)=1}^Np_{ij}^-+\sum_{j(\neq i)=1}^Np_{ij}^+=\sum_{j(\neq i)=1}^N[p_{ij}^-+p_{ij}^+]=\sum_{j(\neq i)=1}^Np_{ij}$) and obtain

\begin{align}
P_{\text{SCM}}(\mathbf{A})=\prod_{i=1}^N\prod_{\substack{j=1\\(j>i)}}^Np_{ij}^{a_{ij}}(1-p_{ij})^{1-a_{ij}}\cdot\prod_{i=1}^N\prod_{\substack{j=1\\(j>i)}}^N\left(\frac{p_{ij}^-}{p_{ij}}\right)^{a_{ij}^-}\left(\frac{p_{ij}^+}{p_{ij}}\right)^{a_{ij}^+}
\end{align}
i.e. an expression that is the product of two probability distributions. Let us start from the second one, whose parameters read

\begin{align}
\frac{p_{ij}^-}{p_{ij}}&=\frac{p_{ij}^-}{p_{ij}^-+p_{ij}^+}=\frac{y_iy_j/(1+x_ix_j+y_iy_j)}{x_ix_j/(1+x_ix_j+y_iy_j)+y_iy_j/(1+x_ix_j+y_iy_j)}=\frac{y_iy_j}{x_ix_j+y_iy_j},\\
\frac{p_{ij}^+}{p_{ij}}&=\frac{p_{ij}^+}{p_{ij}^-+p_{ij}^+}=\frac{x_ix_j/(1+x_ix_j+y_iy_j)}{x_ix_j/(1+x_ix_j+y_iy_j)+y_iy_j/(1+x_ix_j+y_iy_j)}=\frac{x_ix_j}{x_ix_j+y_iy_j}
\end{align}
hence inducing the probability distribution of the SCM-FT, beside keeping the intuitive meaning made explicit by the expressions $p_{ij}^-/p_{ij}=P(\text{`link $-$'}\:|\:\text{`link'})$ and $p_{ij}^+/p_{ij}=P(\text{`link $+$'}\:|\:\text{`link'})$. The first one, on the other hand, can be identified with the probability distribution of the `induced' CM:

\begin{align}
p_{ij}=p_{ij}^-+p_{ij}^+&=\frac{y_iy_j}{1+x_ix_j+y_iy_j}+\frac{x_ix_j}{1+x_ix_j+y_iy_j}=\frac{y_iy_j+x_ix_j}{1+x_ix_j+y_iy_j}=\nonumber\\
&=\frac{\mathbf{z}_i\cdot\mathbf{z}_j}{1+\mathbf{z}_i\cdot\mathbf{z}_j}=\frac{|\mathbf{z}_i||\mathbf{z}_j|\cos\phi_{ij}}{1+|\mathbf{z}_i||\mathbf{z}_j|\cos\phi_{ij}}=\frac{\sqrt{(x_i^2+y_i^2)\cos\phi_{ij}}\cdot\sqrt{(x_j^2+y_j^2)\cos\phi_{ij}}}{1+\sqrt{(x_i^2+y_i^2)\cos\phi_{ij}}\cdot\sqrt{(x_j^2+y_j^2)\cos\phi_{ij}}}\equiv\frac{z_iz_j\cos\phi_{ij}}{1+z_iz_j\cos\phi_{ij}}
\end{align}
where $\mathbf{z}_i\equiv(x_i,y_i)$ is the vector of fitnesses of node $i$, $z_i\equiv|\mathbf{z}_i|=\sqrt{(x_i^2+y_i^2)}$ is its modulus and $\cos\phi_{ij}$ is the cosinus of the angle between vectors $\mathbf{z}_i$ and $\mathbf{z}_j$. As a consequence, we can write

\begin{equation}
P_{\text{SCM}}(\mathbf{A})=P_{\text{ICM}}(\mathbf{A})\cdot P_{\text{SCM-FT}}(\mathbf{A}).
\end{equation}

Notice that when $z_i=(x_i,0)$ and $z_j=(x_j,0)$, $\cos\phi_{ij}=1$ and $p_{ij}=p_{ij}^+=\frac{x_ix_j}{1+x_ix_j}$, i.e. the `induced' CM reduces to the proper CM: in this case, in fact, the information about signs is `redundant' as $k_i(\mathbf{A}^*)=k_i^+(\mathbf{A}^*)$ and $k_j(\mathbf{A}^*)=k_j^+(\mathbf{A}^*)$. On the other hand, when $z_i=(x_i,0)$ and $z_j=(0,y_j)$, $\cos\phi_{ij}=0$ and $p_{ij}=0$, i.e. nodes $i$ and $j$ cannot be linked: in this case, in fact, $k_i(\mathbf{A}^*)=k_i^+(\mathbf{A}^*)$ but $k_j(\mathbf{A}^*)=k_j^-(\mathbf{A}^*)$, whence the impossibility of (consistently) attributing a sign to the edge between $i$ and $j$.

\clearpage

\section*{Supplementary Note 4\\Numerical optimization of likelihood functions}

In order to numerically solve the systems of equations defining the SCM and the SCM-FT, we can follow the guidelines provided in Supplementary Reference \cite{vallarano2021}: more specifically, we will adapt the iterative recipe provided there to our (binary, undirected, signed) setting. First, let us consider the SCM whose system of equations can be rewritten as

\begin{align}
\label{eq:iterD1}
x_i&=\frac{k_i^+(\mathbf A^*)}{\sum_{\substack{j=1\\(j\neq i)}}^N\frac{x_j}{1+x_ix_j+y_iy_j}}\Longrightarrow x_i^{(n)}=\frac{k_i^+(\mathbf A^*)}{\sum_{\substack{j=1\\(j\neq i)}}^N\frac{x_j^{(n-1)}}{1+x_i^{(n-1)}x_j^{(n-1)}+y_i^{(n-1)}y_j^{(n-1)}}}\quad\forall\:i,\\
\label{eq:iterD2}
y_i&=\frac{k_i^-(\mathbf A^*)}{\sum_{\substack{j=1\\(j\neq i)}}^N\frac{y_j}{1+x_ix_j+y_iy_j}}\Longrightarrow y_i^{(n)}=\frac{k_i^-(\mathbf A^*)}{\sum_{\substack{j=1\\(j\neq i)}}^N\frac{y_j^{(n-1)}}{1+x_i^{(n-1)}x_j^{(n-1)}+y_i^{(n-1)}y_j^{(n-1)}}}\quad\forall\:i;
\end{align}
analogously, the system of equations defining the SCM-FT can be rewritten as 

\begin{align}
\label{eq:iterD3}
x_i&=\frac{k_i^+(\mathbf A^*)}{\sum_{\substack{j=1\\(j\neq i)}}^N|a_{ij}^*|\frac{x_j}{x_ix_j+y_iy_j}}\Longrightarrow x_i^{(n)}=\frac{k_i^+(\mathbf A^*)}{\sum_{\substack{j=1\\(j\neq i)}}^N|a_{ij}^*|\frac{x_j^{(n-1)}}{x_i^{(n-1)}x_j^{(n-1)}+y_i^{(n-1)}y_j^{(n-1)}}}\quad\forall\:i,\\
\label{eq:iterD4}
y_i&=\frac{k_i^-(\mathbf A^*)}{\sum_{\substack{j=1\\(j\neq i)}}^N|a_{ij}^*|\frac{y_j}{x_ix_j+y_iy_j}}\Longrightarrow y_i^{(n)}=\frac{k_i^-(\mathbf A^*)}{\sum_{\substack{j=1\\(j\neq i)}}^N|a_{ij}^*|\frac{y_j^{(n-1)}}{x_i^{(n-1)}x_j^{(n-1)}+y_i^{(n-1)}y_j^{(n-1)}}}\quad\forall\:i.
\end{align}

In order for each iterative recipe to converge, an appropriate vector of initial conditions need to be chosen; here, we have opted the following ones: $x_i=k_i^+(\mathbf{A}^*)/\sqrt{2L^+(\mathbf{A}^*)}$, $\forall\:i$ and $y_i=k_i^-(\mathbf{A}^*)/\sqrt{2L^-(\mathbf{A}^*)}$, $\forall\:i$. Besides, we have adopted two, different stopping criteria: the first one is a condition on the Euclidean norm of the vector of differences between the values of the parameters at subsequent iterations, i.e. $||\Delta\vec\theta||_2=\sqrt{\sum_{i=1}^N(\Delta\theta_i)^2}\le10^{-8}$; the second one is a condition on the maximum number of iterations of our iterative algorithm, set to $10^3$.

The accuracy of our method in estimating the constraints has been evaluated by computing the \textit{maximum absolute error} (MAE), defined as 

\begin{align}
\text{MAE}=\max_i\left\{|k_i^+(\mathbf A^*)-\langle k_i^+\rangle|,\:|k_i^-(\mathbf A^*)-\langle k_i^-\rangle|\right\}
\end{align}
(i.e. as the infinite norm of the difference between the vector of the empirical values of the constraints and the vector of their expected values) and the \textit{maximum relative error} (MRE), defined as

\begin{align}
\text{MRE}=\max_i\left\{\frac{|k_i^+(\mathbf A^*)-\langle k_i^+\rangle|}{k_i^+(\mathbf A^*)},\: \frac{|k_i^-(\mathbf A^*)-\langle k_i^-\rangle|}{k_i^-(\mathbf A^*)}\right\}
\end{align}
(i.e. as the infinite norm of the relative difference between the vector of the empirical values of the constraints and the vector of their expected values).

The Supplementary Tables \ref{tab:1A}, \ref{tab:2A} and \ref{tab:3A} sum up the time employed by our algorithm to converge as well as its accuracy in reproducing the constraints defining the SCM and the SCM-FT on each network considered in the present contribution. Overall, our method is fast and accurate: the numerical errors never exceed $O(10^{-1})$ and the time employed to achieve such an accuracy never exceeds minutes. To be noticed that the time required by our algorithm to solve the SCM is usually smaller than that required to solve the SCM-FT - although such a difference rises with the size of the considered configuration.

\clearpage

\begin{table}[t!]
\centering
\begin{tabular}{l|c|c|c|c|c|c|c|c|c|c|c}
\hline
\multicolumn{6}{c|}{} & \multicolumn{3}{c|}{\textbf{SCM}} & \multicolumn{3}{c}{\textbf{SCM-FT}}\\ \hline
\hline
& $N$ & $L$ & $L^+$ & $L^-$ & $c$ & MAE & MRE & Time (s) & MAE & MRE & Time (s)\\
\hline
\hline
CoW, 1946-49 & 60 & 360 & 319 & 41 & $\simeq$ 2.0$\cdot10^{-3}$ & $\simeq$ 6.4$\cdot10^{-2}$ & $\simeq$ 1.4$\cdot10^{-2}$ & $\simeq$ 0.02 & $\simeq$ 7.1$\cdot10^{-2}$ & $\simeq$ 3.9$\cdot10^{-3}$ & $\simeq$ 0.01\\
\hline
CoW, 1950-53 & 72 & 437 & 361 & 76 & $\simeq$ 1.7$\cdot10^{-3}$ & $\simeq$ 7.6$\cdot10^{-2}$ & $\simeq$ 2.1$\cdot10^{-2}$ & $\simeq$ 0.02 & $\simeq$ 6.0$\cdot10^{-2}$ & $\simeq$ 3.2$\cdot10^{-2}$ & $\simeq$ 0.01 \\
\hline
CoW, 1954-57 & 80 & 492 & 418 & 74 & $\simeq$ 1.5$\cdot10^{-3}$  & $\simeq$ 7.6$\cdot10^{-2}$ & $\simeq$ 2.2$\cdot10^{-2}$ & $\simeq$ 0.03 & $\simeq$ 8.3$\cdot10^{-2}$ & $\simeq$ 2.8$\cdot10^{-2}$ & $\simeq$ 0.02 \\
\hline
CoW, 1958-61 & 101 & 613 & 506 & 107 & $\simeq$ 1.2$\cdot10^{-3}$ & $\simeq$ 8.5$\cdot10^{-2}$ & $\simeq$ 2.2$\cdot10^{-2}$ & $\simeq$ 0.04 & $\simeq$ 9.7$\cdot10^{-2}$ & $\simeq$ 2.9$\cdot10^{-2}$ & $\simeq$ 0.03 \\
\hline
CoW, 1962-65 & 109 & 642 & 540 & 102 & $\simeq$ 1.1$\cdot10^{-3}$ & $\simeq$ 7.9$\cdot10^{-2}$ & $\simeq$ 1.9$\cdot10^{-2}$ & $\simeq$ 0.06 & $\simeq$ 9.8$\cdot10^{-2}$ & $\simeq$ 5.2$\cdot10^{-2}$ & $\simeq$ 0.03 \\
\hline
CoW, 1966-69 & 111 & 607 & 504 & 103 & $\simeq$ 9.9$\cdot10^{-4}$ & $\simeq$ 6.9$\cdot10^{-2}$ & $\simeq$ 2.2$\cdot10^{-2}$ & $\simeq$ 0.05 & $\simeq$ 5.8$\cdot10^{-2}$ & $\simeq$ 5.8$\cdot10^{-2}$ & $\simeq$ 0.03 \\
\hline
CoW, 1970-73 & 111 & 677 & 595 & 82 & $\simeq$ 1.1$\cdot10^{-3}$  & $\simeq$ 7.6$\cdot10^{-2}$ & $\simeq$ 2.0$\cdot10^{-2}$ & $\simeq$ 0.05 & $\simeq$ 7.2$\cdot10^{-2}$ & $\simeq$ 7.2$\cdot10^{-2}$ & $\simeq$ 0.03 \\
\hline
CoW, 1974-77 & 123 & 813 & 699 & 114 & $\simeq$ 1.1$\cdot10^{-3}$ & $\simeq$ 8.8$\cdot10^{-2}$ & $\simeq$ 2.3$\cdot10^{-2}$ & $\simeq$ 0.06 & $\simeq$ 7.2$\cdot10^{-2}$ & $\simeq$ 3.1$\cdot10^{-2}$ & $\simeq$ 0.04 \\
\hline
CoW, 1978-81 & 134 & 999 & 907 & 92 & $\simeq$ 1.1$\cdot10^{-3}$ & $\simeq$ 8.4$\cdot10^{-2}$ & $\simeq$ 2.3$\cdot10^{-2}$ & $\simeq$ 0.08 & $\simeq$ 6.5$\cdot10^{-2}$ & $\simeq$ 3.5$\cdot10^{-2}$ & $\simeq$ 0.06 \\
\hline
CoW, 1982-85 & 134 & 1042 & 935 & 107 & $\simeq$ 1.1$\cdot10^{-4}$ & $\simeq$ 9.8$\cdot10^{-2}$ & $\simeq$ 2.3$\cdot10^{-2}$ & $\simeq$ 0.07 & $\simeq$ 8.2$\cdot10^{-2}$ & $\simeq$ 2.8$\cdot10^{-2}$ & $\simeq$ 0.05 \\
\hline
CoW, 1986-89 & 139 & 1079 & 989 & 90 & $\simeq$ 1.1$\cdot10^{-4}$ & $\simeq$ 9.6$\cdot10^{-2}$ & $\simeq$ 2.6$\cdot10^{-2}$ & $\simeq$ 0.08 & $\simeq$ 8.9$\cdot10^{-2}$ & $\simeq$ 3.4$\cdot10^{-2}$ & $\simeq$ 0.06 \\
\hline
CoW, 1990-93 & 151 & 1286 & 1160 & 126 & $\simeq$ 1.1$\cdot10^{-4}$ & $\simeq$ 9.8$\cdot10^{-2}$ & $\simeq$ 2.7$\cdot10^{-2}$ & $\simeq$ 0.12 & $\simeq$ 9.0$\cdot10^{-2}$ & $\simeq$ 3.1$\cdot10^{-2}$ & $\simeq$ 0.07 \\
\hline
CoW, 1994-97 & 143 & 1220 & 1099 & 121 & $\simeq$ 1.2$\cdot10^{-4}$ & $\simeq$ 9.8$\cdot10^{-2}$ & $\simeq$ 2.8$\cdot10^{-2}$ & $\simeq$ 0.08 & $\simeq$ 5.3$\cdot10^{-2}$ & $\simeq$ 2.1$\cdot10^{-2}$ & $\simeq$ 0.06 \\
\hline
\end{tabular}
\caption{Performance of the fixed-point algorithm to solve the systems of equations defining the Signed Configuration Model (SCM) and the Signed Configuration Model (SCM-FT) on the snapshots of the Correlates of Wars dataset.}
\label{tab:1A}
\end{table}

\begin{table}[t!]
\centering
\begin{tabular}{l|c|c|c|c|c|c|c|c|c|c|c}
\hline
\multicolumn{6}{c|}{} & \multicolumn{3}{c|}{\textbf{SCM}} & \multicolumn{3}{c}{\textbf{SCM-FT}}\\ \hline
\hline
& $N$ & $L$ & $L^+$ & $L^-$ & $c$ & MAE & MRE & Time (s) & MAE & MRE & Time (s)\\
\hline
\hline
MMOG, Day 10 & 1312 & 3791 & 3725 & 66 & $\simeq$ 4.4$\cdot10^{-3}$ & $\simeq$ 1.2$\cdot10^{-1}$ & $\simeq$ 2.1$\cdot10^{-2}$ & $\simeq$ 7 & $\simeq$ 1.7$\cdot10^{-2}$ & $\simeq$ 1.0$\cdot10^{-2}$ & $\simeq$ 5 \\
\hline
MMOG, Day 20 & 1924 & 7032 & 6050 & 982 & $\simeq$ 3.8$\cdot10^{-3}$ & $\simeq$ 1.4$\cdot10^{-1}$ & $\simeq$ 2.6$\cdot10^{-2}$ & $\simeq$ 9 & $\simeq$ 3.1$\cdot10^{-2}$ & $\simeq$ 1.5$\cdot10^{-2}$ & $\simeq$ 21 \\
\hline
MMOG, Day 30 & 2261 & 9046 & 7371 & 1675 & $\simeq$ 3.5$\cdot10^{-3}$ & $\simeq$ 2.0$\cdot10^{-1}$ & $\simeq$ 3.1$\cdot10^{-2}$ & $\simeq$ 10 & $\simeq$ 6.5$\cdot10^{-2}$ & $\simeq$ 2.3$\cdot10^{-2}$ & $\simeq$ 34 \\
\hline
MMOG, Day 40 & 2544 & 11372 & 8403 & 2969 & $\simeq$ 3.5$\cdot10^{-3}$ & $\simeq$ 2.0$\cdot10^{-1}$ & $\simeq$ 3.2$\cdot10^{-2}$ & $\simeq$ 12 & $\simeq$ 8.4$\cdot10^{-2}$ & $\simeq$ 2.2$\cdot10^{-2}$ & $\simeq$ 35 \\
\hline
MMOG, Day 50 & 2714 & 13228 & 9110 & 4118 & $\simeq$ 3.6$\cdot10^{-3}$ & $\simeq$ 1.9$\cdot10^{-1}$ & $\simeq$ 3.3$\cdot10^{-2}$ & $\simeq$ 14 & $\simeq$ 1.2$\cdot10^{-1}$ & $\simeq$ 4.2$\cdot10^{-2}$ & $\simeq$ 54 \\
\hline
MMOG, Day 60 & 2923 & 13909 & 9711 & 4198 & $\simeq$ 3.3$\cdot10^{-3}$ & $\simeq$ 2.0$\cdot10^{-1}$ & $\simeq$ 3.0$\cdot10^{-2}$ & $\simeq$ 18 & $\simeq$ 1.1$\cdot10^{-1}$ & $\simeq$ 3.8$\cdot10^{-2}$ & $\simeq$ 80 \\
\hline
MMOG, Day 70 & 3068 & 14963 & 10151 & 4812 & $\simeq$ 3.2$\cdot10^{-3}$ & $\simeq$ 2.2$\cdot10^{-1}$ & $\simeq$ 3.2$\cdot10^{-2}$ & $\simeq$ 19 & $\simeq$ 9.1$\cdot10^{-2}$ & $\simeq$ 2.8$\cdot10^{-2}$ & $\simeq$ 85 \\
\hline
MMOG, Day 80 & 3221 & 16318 & 10745 & 5573 & $\simeq$ 3.2$\cdot10^{-3}$ & $\simeq$ 1.9$\cdot10^{-1}$ & $\simeq$ 3.0$\cdot10^{-2}$ & $\simeq$ 22 & $\simeq$ 1.1$\cdot10^{-1}$ & $\simeq$ 5.2$\cdot10^{-2}$  & $\simeq$ 109 \\
\hline
MMOG, Day 90 & 3363 & 17664 & 11342 & 6322 & $\simeq$ 3.1$\cdot10^{-3}$ & $\simeq$ 2.0$\cdot10^{-1}$ & $\simeq$ 3.4$\cdot10^{-2}$ & $\simeq$ 25 & $\simeq$ 1.3$\cdot10^{-1}$ & $\simeq$ 4.5$\cdot10^{-2}$ & $\simeq$ 120 \\
\hline
MMOG, Day 100 & 3523 & 19008 & 11915 & 7093 & $\simeq$ 3.1$\cdot10^{-3}$ & $\simeq$ 2.1$\cdot10^{-1}$ & $\simeq$ 3.2$\cdot10^{-2}$ & $\simeq$ 27  & $\simeq$ 1.1$\cdot10^{-1}$ & $\simeq$ 4.7$\cdot10^{-2}$ & $\simeq$ 138 \\
\hline
\end{tabular}
\caption{Performance of the fixed-point algorithm to solve the systems of equations defining the Signed Configuration Model (SCM) and the Signed Configuration Model with Fixed Topology (SCM-FT) on the snapshots of the MMOG dataset.}
\label{tab:2A}
\end{table}

\begin{table}[t!]
\centering
\begin{tabular}{l|c|c|c|c|c|c|c|c|c|c|c}
\hline
\multicolumn{6}{c|}{} & \multicolumn{3}{c|}{\textbf{SCM}} & \multicolumn{3}{c}{\textbf{SCM-FT}}\\ \hline
\hline
& $N$ & $L$ & $L^+$ & $L^-$ & $c$ & MAE & MRE & Time (s) & MAE & MRE & Time (s)\\
\hline
\hline
N.G.H. Tribes & 16 & 58 & 29 & 29 & $\simeq$ 0.48 & $\simeq$ 3.2$\cdot10^{-2}$ & $\simeq$ 1.1$\cdot10^{-2}$ & $\simeq0.005$ & $\simeq$ 2.7$\cdot10^{-2}$ & $\simeq$ 1.1$\cdot10^{-2}$ & $\simeq0.0004$\\
\hline
Monastery & 18 & 49 & 37 & 12 & $\simeq$ 3.2$\cdot10^{-1}$ & $\simeq$ 2.4$\cdot10^{-2}$ & $\simeq$ 1.4$\cdot10^{-2}$ & $\simeq$ 0.006 & $\simeq$ 1.2$\cdot10^{-2}$ & $\simeq$ 9.3$\cdot10^{-3}$ & $\simeq0.006$\\
\hline
Senate US & 100 & 2461 & 1414 & 1047 & $\simeq$ 4.9$\cdot10^{-1}$ & $\simeq$ 1.0$\cdot10^{-1}$ & $\simeq$ 1.1$\cdot10^{-2}$ & $\simeq$ 0.06 & $\simeq$ 5.9$\cdot10^{-2}$ & $\simeq$ 1.0$\cdot10^{-2}$ & $\simeq0.12$\\
\hline
EGFR & 313 & 755 & 499 & 256 & $\simeq$ 1.5$\cdot10^{-2}$ & $\simeq$ 5.8$\cdot10^{-2}$ & $\simeq$ 2.1$\cdot10^{-2}$ & $\simeq$ 0.54 & $\simeq$ 3.8$\cdot10^{-2}$ & $\simeq$ 1.1$\cdot10^{-2}$ & $\simeq0.35$\\
\hline
Macrophage & 660 & 1397 & 931 & 466 & $\simeq$ 6.4$\cdot10^{-3}$ & $\simeq$ 8.1$\cdot10^{-2}$ & $\simeq$ 2.1$\cdot10^{-2}$ & $\simeq$ 2 & $\simeq$ 2.6$\cdot10^{-2}$ & $\simeq$ 1.5$\cdot10^{-2}$ & $\simeq1.8$\\
\hline
E. Coli & 1376 & 3150 & 1848 & 1302 & $\simeq$ 3.3$\cdot10^{-3}$ & $\simeq$ 1.9$\cdot10^{-1}$ & $\simeq$ 2.2$\cdot10^{-2}$ & $\simeq$ 13 & $\simeq$ 1.3$\cdot10^{-2}$ & $\simeq$ 1.2$\cdot10^{-2}$ & $\simeq10$\\
\hline
Bitcoin Alpha & 3775 & 14120 & 12721 & 1399 & $\simeq$ 1.9$\cdot10^{-3}$ & $\simeq$ 2.1$\cdot10^{-1}$ & $\simeq$ 3.1$\cdot10^{-2}$ & $\simeq$ 76 & $\simeq$ 5.5$\cdot10^{-2}$ & $\simeq$ 2.2$\cdot10^{-2}$ & $\simeq138$\\
\hline
Bitcoin OTC & 5875 & 21489 & 18230 & 3259 & $\simeq$ 1.2$\cdot10^{-3}$ & $\simeq$ 2.5$\cdot10^{-1}$ & $\simeq$ 4.0$\cdot10^{-2}$ & $\simeq$ 24 & $\simeq$ 1.1$\cdot10^{-1}$ & $\simeq$ 3.5$\cdot10^{-2}$ & $\simeq267$\\
\hline
\end{tabular}
\caption{Performance of the fixed-point algorithm to solve the systems of equations defining the Signed Configuration Model (SCM) and the Signed Configuration Model with Fixed Topology (SCM-FT) on a bunch of real-world networks.}
\label{tab:3A}
\end{table}

\clearpage

\section*{Supplementary Note 5\\Sampling ensembles}

As each of our null models treats links independently, the ensemble it induces can be sampled quite straightforwardly as follows.

\begin{algorithm}[ht!]
\caption{Pseudocode for sampling the Signed Random Graph Model}
\begin{algorithmic}
\item[\hspace{1.4pt} 1:] \textbf{A=0};
\item[\hspace{1.4pt} 2:] \textbf{for} $i=1\dots N$ \textbf{do}
\item[\hspace{1.4pt} 3:] \hspace{15pt}\textbf{for} $j=i+1\dots N$ \textbf{do}
\item[\hspace{1.4pt} 4:] \hspace{15pt}$u=\text{RandomUniform}[0,1]$;
\item[\hspace{1.4pt} 5:] \hspace{30pt}\textbf{if} $u\le p^-$ \textbf{then}
\item[\hspace{1.4pt} 6:] \hspace{45pt}$a_{ij}=a_{ji}=-1$;
\item[\hspace{1.4pt} 7:] \hspace{30pt}\textbf{else if} $p^-<u\le (p^-+p^+)$ \textbf{then}
\item[\hspace{1.4pt} 8:] \hspace{45pt}$a_{ij}=a_{ji}=+1$;
\item[\hspace{1.4pt} 9:] \hspace{30pt}\textbf{end}
\item[10:] \hspace{15pt}\textbf{end}
\item[11:] \textbf{end}
\end{algorithmic} 
\label{alg:SRGM}
\end{algorithm}

\begin{algorithm}[ht!]
\caption{Pseudocode for sampling the Signed Random Graph Model with Fixed Topology}
\begin{algorithmic}
\item[\hspace{1.4pt} 1:] \textbf{A} $\leftarrow$ $N\times N$ matrix with $0,1$ entries;
\item[\hspace{1.4pt} 2:] \textbf{for} $i=1\dots N$ \textbf{do}
\item[\hspace{1.4pt} 3:] \hspace{15pt}\textbf{for} $j=i+1\dots N$ \textbf{do}
\item[\hspace{1.4pt} 4:] \hspace{30pt}\textbf{if} $a_{ij}=1$ \textbf{then}
\item[\hspace{1.4pt} 5:] \hspace{45pt}$u=\text{RandomUniform}[0,1]$;
\item[\hspace{1.4pt} 6:] \hspace{45pt}\textbf{if} $u\le p^-$ \textbf{then}
\item[\hspace{1.4pt} 7:] \hspace{60pt}$a_{i j}=a_{ji}=-1$;
\item[\hspace{1.4pt} 8:] \hspace{45pt}\textbf{else if} $p^-<u\le (p^-+p^+)$ \textbf{then}
\item[\hspace{1.4pt} 9:] \hspace{60pt}$a_{ij}=a_{ji}=+1$;
\item[10:] \hspace{45pt}\textbf{end}
\item[11:] \hspace{30pt}\textbf{end}
\item[12:] \hspace{15pt}\textbf{end}
\item[13:] \textbf{end}
\end{algorithmic} 
\label{alg:SRGMFT}
\end{algorithm}

\begin{algorithm}[ht!]
\caption{Pseudocode for sampling the Signed Configuration Model}
\begin{algorithmic}
\item[\hspace{1.4pt} 1:] \textbf{A=0};
\item[\hspace{1.4pt} 2:] \textbf{for} $i=1\dots N$ \textbf{do}
\item[\hspace{1.4pt} 3:] \hspace{15pt}\textbf{for} $j=i+1\dots N$ \textbf{do}
\item[\hspace{1.4pt} 4:] \hspace{15pt}$u=\text{RandomUniform}[0,1]$;
\item[\hspace{1.4pt} 5:] \hspace{30pt}\textbf{if} $u\le p_{ij}^-$ \textbf{then}
\item[\hspace{1.4pt} 6:] \hspace{45pt}$a_{ij}=a_{ji}=-1$;
\item[\hspace{1.4pt} 7:] \hspace{30pt}\textbf{else if} $p_{ij}^-<u\le (p_{ij}^-+p_{ij}^+)$ \textbf{then}
\item[\hspace{1.4pt} 8:] \hspace{45pt}$a_{ij}=a_{ji}=+1$;
\item[\hspace{1.4pt} 9:] \hspace{30pt}\textbf{end}
\item[10:] \hspace{15pt}\textbf{end}
\item[11:] \textbf{end}
\end{algorithmic} 
\label{alg:SCM}
\end{algorithm}

\begin{algorithm}[t!]
\caption{Pseudocode for sampling the Signed Configuration Model with Fixed Topology}
\begin{algorithmic}
\item[\hspace{1.4pt} 1:] \textbf{A} $\leftarrow$ $N\times N$ matrix with $0,1$ entries;
\item[\hspace{1.4pt} 2:] \textbf{for} $i=1\dots N$ \textbf{do}
\item[\hspace{1.4pt} 3:] \hspace{15pt}\textbf{for} $j=i+1\dots N$ \textbf{do}
\item[\hspace{1.4pt} 4:] \hspace{30pt}\textbf{if} $a_{ij}=1$ \textbf{then}
\item[\hspace{1.4pt} 5:] \hspace{45pt}$u=\text{RandomUniform}[0,1]$;
\item[\hspace{1.4pt} 6:] \hspace{45pt}\textbf{if} $u\le p_{ij}^-$ \textbf{then}
\item[\hspace{1.4pt} 7:] \hspace{60pt}$a_{ij}=a_{ji}=-1$;
\item[\hspace{1.4pt} 8:] \hspace{45pt}\textbf{else if} $p_{ij}^-<u\le (p_{ij}^-+p_{ij}^+)$ \textbf{then}
\item[\hspace{1.4pt} 9:] \hspace{60pt}$a_{ij}=a_{ji}=+1$;
\item[10:] \hspace{45pt}\textbf{end}
\item[11:] \hspace{30pt}\textbf{end}
\item[12:] \hspace{15pt}\textbf{end}
\item[13:] \textbf{end}
\end{algorithmic} 
\label{alg:SCMFT}
\end{algorithm}

An estimation of the time needed to sample the ensemble induced by each of our models, for each of our datasets, is reported in the Supplementary Tables \ref{tab:4A}, \ref{tab:5A} and \ref{tab:6A}.

\begin{table}[h!]
\centering
\begin{tabular}{l|c|c|c|c|c|c|c|c}
\hline
\multicolumn{1}{c|}{} & \multicolumn{8}{c}{Time (s)}\\ \hline
\hline
& \multicolumn{2}{c|}{SRGM} & \multicolumn{2}{c|}{SRGM-FT} & \multicolumn{2}{c|}{SCM} & \multicolumn{2}{c}{SCM-FT}\\
\hline 
\hline
& 1 network & $10^3$ networks & 1 network & $10^3$ networks & 1 network & $10^3$ networks & 1 network & $10^3$ networks \\
\hline
\hline
CoW, 1946-49 & $\simeq$ 0.002 & $\simeq$ 0.09 & $\simeq$ 0.001 & $\simeq$ 0.05 & $\simeq$ 0.0005 & $\simeq$ 0.09 & $\simeq$ 0.0008 & $\simeq$ 0.04\\
\hline
CoW, 1950-53 & $\simeq$ 0.001 & $\simeq$ 0.14 & $\simeq$ 0.001 & $\simeq$ 0.05 & $\simeq$ 0.0005 & $\simeq$ 0.12 & $\simeq$ 0.0004 & $\simeq$ 0.05\\
\hline
CoW, 1954-57 & $\simeq$ 0.001 & $\simeq$ 0.16 & $\simeq$ 0.0008 & $\simeq$ 0.06 & $\simeq$ 0.001 & $\simeq$ 0.15 & $\simeq$ 0.0004 & $\simeq$ 0.06\\
\hline
CoW, 1958-61 & $\simeq$ 0.001 & $\simeq$ 0.26 & $\simeq$ 0.0006 & $\simeq$ 0.09 & $\simeq$ 0.001 & $\simeq$ 0.24 & $\simeq$ 0.001 & $\simeq$ 0.09\\
\hline
CoW, 1962-65 & $\simeq$ 0.002 & $\simeq$ 0.29 & $\simeq$ 0.001 & $\simeq$ 0.10 & $\simeq$ 0.001 & $\simeq$ 0.27 & $\simeq$ 0.001 & $\simeq$ 0.10\\
\hline
CoW, 1966-69 & $\simeq$ 0.003 & $\simeq$ 0.30 & $\simeq$ 0.0007 & $\simeq$ 0.10 & $\simeq$ 0.001 & $\simeq$ 0.28 & $\simeq$ 0.001 & $\simeq$ 0.10\\
\hline
CoW, 1970-73 & $\simeq$ 0.002 & $\simeq$ 0.29 & $\simeq$ 0.0007 & $\simeq$ 0.10 & $\simeq$ 0.001 & $\simeq$ 0.28 & $\simeq$ 0.001 & $\simeq$ 0.10\\
\hline
CoW, 1974-77 & $\simeq$ 0.002 & $\simeq$ 0.37 & $\simeq$ 0.0009 & $\simeq$ 0.13 & $\simeq$ 0.002 & $\simeq$ 0.34 & $\simeq$ 0.0006 & $\simeq$ 0.12\\
\hline
CoW, 1978-81 & $\simeq$ 0.003 & $\simeq$ 0.43 & $\simeq$ 0.0006 & $\simeq$ 0.19 & $\simeq$ 0.001 & $\simeq$ 0.43 & $\simeq$ 0.0007 & $\simeq$ 0.17\\
\hline
CoW, 1982-85 & $\simeq$ 0.003 & $\simeq$ 0.44 & $\simeq$ 0.0006 & $\simeq$ 0.20 & $\simeq$ 0.001 & $\simeq$ 0.43 & $\simeq$ 0.0007 & $\simeq$ 0.19\\
\hline
CoW, 1986-89 & $\simeq$ 0.002 & $\simeq$ 0.47 & $\simeq$ 0.0007 & $\simeq$ 0.19 & $\simeq$ 0.002 & $\simeq$ 0.46 & $\simeq$ 0.001 & $\simeq$ 0.23\\
\hline
CoW, 1990-93 & $\simeq$ 0.001 & $\simeq$ 0.55 & $\simeq$ 0.0005 & $\simeq$ 0.26 & $\simeq$ 0.001 & $\simeq$ 0.55 & $\simeq$ 0.0009 & $\simeq$ 0.25\\
\hline
CoW, 1994-97 & $\simeq$ 0.002 & $\simeq$ 0.51 & $\simeq$ 0.0006 & $\simeq$ 0.19 & $\simeq$ 0.001 & $\simeq$ 0.47 & $\simeq$ 0.0007 & $\simeq$ 0.24\\
\hline
\end{tabular}
\caption{Time required by the Signed Random Graph Model (SRGM), Signed Random Graph Model with Fixed Topology (SRGM-FT),  Signed Configuration Model (SCM) and the Signed Configuration Model with Fixed Topology (SCM-FT) to sample its ensemble - CoW dataset.}
\label{tab:4A}
\end{table}

\begin{table}[h!]
\centering
\begin{tabular}{l|c|c|c|c|c|c|c|c}
\hline
\multicolumn{1}{c|}{} & \multicolumn{8}{c}{Time (s)}\\ \hline
\hline
& \multicolumn{2}{c|}{SRGM} &  \multicolumn{2}{c|}{SRGM-FT} &  \multicolumn{2}{c|}{SCM} &  \multicolumn{2}{c}{SCM-FT}\\
\hline 
\hline
& 1 network & $10^3$ networks & 1 network & $10^3$ networks & 1 network & $10^3$ networks & 1 network & $10^3$ networks \\
\hline
\hline
MMOG, Day 10 & $\simeq$ 0.07 & $\simeq$ 40 & $\simeq$ 0.07 & $\simeq$ 22 & $\simeq$ 0.08 & $\simeq$ 55 & $\simeq$ 0.08 & $\simeq$ 34\\
\hline
MMOG, Day 20 & $\simeq$ 0.11 & $\simeq$ 0.88 & $\simeq$ 0.09 & $\simeq$ 57 & $\simeq$ 0.14 & $\simeq$ 115 & $\simeq$ 0.11 & $\simeq$ 85\\
\hline
MMOG, Day 30 & $\simeq$ 0.14 & $\simeq$ 125 & $\simeq$ 0.11 & $\simeq$ 113 & $\simeq$ 0.19 & $\simeq$ 179 & $\simeq$ 0.17 & $\simeq$ 159\\
\hline
MMOG, Day 40 & $\simeq$ 0.18 & $\simeq$ 180 & $\simeq$ 0.15 & $\simeq$ 146 & $\simeq$ 0.20 & $\simeq$ 210 & $\simeq$ 0.19 & $\simeq$ 212\\
\hline
MMOG, Day 50 & $\simeq$ 0.20 & $\simeq$ 188 & $\simeq$ 0.16 & $\simeq$ 156 & $\simeq$ 0.27 & $\simeq$ 261 & $\simeq$ 0.25 & $\simeq$ 237\\
\hline
MMOG, Day 60 & $\simeq$ 0.23 & $\simeq$ 214 & $\simeq$ 0.19 & $\simeq$ 190 & $\simeq$ 0.31 & $\simeq$ 315 & $\simeq$ 0.29 & $\simeq$ 279\\
\hline
MMOG, Day 70 & $\simeq$ 0.25 & $\simeq$ 235 & $\simeq$ 0.19 & $\simeq$ 196 & $\simeq$ 0.45 & $\simeq$ 338 & $\simeq$ 0.32 & $\simeq$ 306\\
\hline
MMOG, Day 80 & $\simeq$ 0.29 & $\simeq$ 261 & $\simeq$ 0.24 & $\simeq$ 233 & $\simeq$ 0.39 & $\simeq$ 379 & $\simeq$ 0.35 & $\simeq$ 342\\
\hline
MMOG, Day 90 & $\simeq$ 0.29 & $\simeq$ 283 & $\simeq$ 0.26 & $\simeq$ 254 & $\simeq$ 0.43 & $\simeq$ 422 & $\simeq$ 0.38 & $\simeq$ 376\\
\hline
MMOG, Day 100 & $\simeq$ 0.33 & $\simeq$ 317 & $\simeq$ 0.29 & $\simeq$ 286 & $\simeq$ 0.47 & $\simeq$ 459 & $\simeq$ 0.43 & $\simeq$ 427\\
\hline
\end{tabular}
\caption{Time required by the Signed Random Graph Model (SRGM), Signed Random Graph Model with Fixed Topology (SRGM-FT),  Signed Configuration Model (SCM) and the Signed Configuration Model with Fixed Topology (SCM-FT) to sample its ensemble - MMOG dataset.}
\label{tab:5A}
\end{table}

\begin{table}[h!]
\centering
\begin{tabular}{l|c|c|c|c|c|c|c|c}
\hline
\multicolumn{1}{c|}{} & \multicolumn{8}{c}{Time (s)}\\ \hline
\hline
 & \multicolumn{2}{c|}{SRGM} &  \multicolumn{2}{c|}{SRGM-FT} &  \multicolumn{2}{c|}{SCM} &  \multicolumn{2}{c}{SCM-FT}\\
\hline 
\hline
 & 1 network & $10^3$ networks & 1 network & $10^3$ networks & 1 network & $10^3$ networks & 1 network & $10^3$ networks \\
\hline
\hline
E. Coli & $\simeq$ 0.055 & $\simeq$ 42 & $\simeq$ 0.051 & $\simeq$ 24 & $\simeq$ 0.093 & $\simeq$ 60 & $\simeq$ 0.041 & $\simeq$ 36\\
\hline
Macrophage & $\simeq$ 0.021 & $\simeq$ 14 & $\simeq$ 0.016 & $\simeq$ 10 & $\simeq$ 0.017 & $\simeq$ 14 & $\simeq$ 0.015 & $\simeq$ 12\\
\hline
EGFR & $\simeq$ 0.007 & $\simeq$ 2.0 & $\simeq$ 0.008 & $\simeq$ 0.83 & $\simeq$ 0.009 & $\simeq$ 2.4 & $\simeq$ 0.005 & $\simeq$ 0.9\\
\hline
N.G.H. Tribes & $\simeq$ 0.035 & $\simeq$ 0.24 & $\simeq$ 0.028 & $\simeq$ 0.04 & $\simeq$ 0.009 & $\simeq$ 0.015 & $\simeq$ 0.006 & $\simeq$ 0.02\\
\hline
Senate US & $\simeq$ 0.006 & $\simeq$ 0.32 & $\simeq$ 0.004 & $\simeq$ 0.27 & $\simeq$ 0.001 & $\simeq$ 0.32 & $\simeq$ 0.003 & $\simeq$ 0.27\\
\hline
Monastery & $\simeq$ 0.001 & $\simeq$ 0.01 & $\simeq$ 0.002 & $\simeq$ 0.02 & $\simeq$ 0.001 & $\simeq$ 0.015 & $\simeq$ 0.001 & $\simeq$ 0.01\\
\hline
Bitcoin Alpha & $\simeq$ 0.39 & $\simeq$ 323 & $\simeq$ 0.29 & $\simeq$ 283 & $\simeq$ 0.51 & $\simeq$ 492 & $\simeq$ 0.44 & $\simeq$ 441\\
\hline
Bitcoin OTC & $\simeq$ 0.76 & $\simeq$ 757 & $\simeq$ 0.73 & $\simeq$ 728 & $\simeq$ 1.3 & $\simeq$ 1110 & $\simeq$ 1.0 & $\simeq$ 1230\\
\hline
\end{tabular}
\caption{Time required by the Signed Random Graph Model (SRGM), Signed Random Graph Model with Fixed Topology (SRGM-FT),  Signed Configuration Model (SCM) and the Signed Configuration Model with Fixed Topology (SCM-FT) to sample its ensemble - socio-political, biological, financial networks.}
\label{tab:6A}
\end{table}

\clearpage

\section*{Supplementary Note 6\\Inspecting local frustration on signed networks}

Let us provide the fully analytical expressions for the quantities entering into the definition of our $z$-scores. The empirical abundances of our motifs read

\begin{align}
T^{(+++)}&=\frac{1}{6}\sum_{i=1}^N\sum_{\substack{j=1\\(j\neq i)}}^N\sum_{\substack{k=1\\(k\neq i,j)}}^Na_{ij}^+a_{jk}^+a_{ki}^+\equiv\sum_{i<j<k}a_{ij}^+a_{jk}^+a_{ki}^+,\\
T^{(++-)}&=\frac{1}{2}\sum_{i=1}^N\sum_{\substack{j=1\\(j\neq i)}}^N\sum_{\substack{k=1\\(k\neq i,j)}}^Na_{ij}^+a_{jk}^+a_{ki}^-\equiv\sum_{i<j<k}[a_{ij}^+a_{jk}^+a_{ki}^-+a_{ij}^+a_{jk}^-a_{ki}^++a_{ij}^-a_{jk}^+a_{ki}^+],\\
T^{(+--)}&=\frac{1}{2}\sum_{i=1}^N\sum_{\substack{j=1\\(j\neq i)}}^N\sum_{\substack{k=1\\(k\neq i,j)}}^Na_{ij}^+a_{jk}^-a_{ki}^-\equiv\sum_{i<j<k}[a_{ij}^+a_{jk}^-a_{ki}^-+a_{ij}^-a_{jk}^+a_{ki}^-+a_{ij}^-a_{jk}^-a_{ki}^+],\\
T^{(---)}&=\frac{1}{6}\sum_{i=1}^N\sum_{\substack{j=1\\(j\neq i)}}^N\sum_{\substack{k=1\\(k\neq i,j)}}^Na_{ij}^-a_{jk}^-a_{ki}^-\equiv\sum_{i<j<k}a_{ij}^-a_{jk}^-a_{ki}^-
\end{align}
while their expected abundances read

\begin{align}
\langle T^{(+++)}\rangle&=\frac{1}{6}\sum_{i=1}^N\sum_{\substack{j=1\\(j\neq i)}}^N\sum_{\substack{k=1\\(k\neq i,j)}}^Np_{ij}^+p_{jk}^+p_{ki}^+\equiv\sum_{i<j<k}p_{ij}^+p_{jk}^+p_{ki}^+
,\\
\langle T^{(++-)}\rangle&=\frac{1}{2}\sum_{i=1}^N\sum_{\substack{j=1\\(j\neq i)}}^N\sum_{\substack{k=1\\(k\neq i,j)}}^Np_{ij}^+p_{jk}^+p_{ki}^-\equiv\sum_{i<j<k}[p_{ij}^+p_{jk}^+p_{ki}^-+p_{ij}^+p_{jk}^-p_{ki}^++p_{ij}^-p_{jk}^+p_{ki}^+],\\
\langle T^{(+--)}\rangle&=\frac{1}{2}\sum_{i=1}^N\sum_{\substack{j=1\\(j\neq i)}}^N\sum_{\substack{k=1\\(k\neq i,j)}}^Np_{ij}^+p_{jk}^-p_{ki}^-\equiv\sum_{i<j<k}[p_{ij}^+p_{jk}^-p_{ki}^-+p_{ij}^-p_{jk}^+p_{ki}^-+p_{ij}^-p_{jk}^-p_{ki}^+],\\
\langle T^{(---)}\rangle&=\frac{1}{6}\sum_{i=1}^N\sum_{\substack{j=1\\(j\neq i)}}^N\sum_{\substack{k=1\\(k\neq i,j)}}^Np_{ij}^-p_{jk}^-p_{ki}^-\equiv\sum_{i<j<k}p_{ij}^-p_{jk}^-p_{ki}^-.
\end{align}

The standard deviation of motif $T^{+++}$ reads

\begin{align}
\sigma[T^{(+++)}]&=\sqrt{
\sum_\mathbf{I}\text{Var}[a_\mathbf{I}]+2\cdot\sum_{\mathbf{I<\mathbf{J}}}\text{Cov}[a_\mathbf{I},a_\mathbf{J}]
}
\end{align}
where we have employed the multi-index notation, i.e. $\mathbf{I}\equiv(i,j,k)$ and $\mathbf{J}\equiv(l,m,n)$. Naturally,

\begin{align}
\sum_\mathbf{I}\text{Var}[a_\mathbf{I}]=\sum_{i<j<k}p_{ij}^+p_{jk}^+p_{ki}^+(1-p_{ij}^+p_{jk}^+p_{ki}^+);
\end{align}
let us, now, consider that

\begin{align}
\text{Cov}[a_\mathbf{I},a_\mathbf{J}]&=\langle a_{ij}^+a_{jk}^+a_{ki}^+\cdot a_{lm}^+a_{mn}^+a_{nl}^+\rangle-\langle a_{ij}^+a_{jk}^+a_{ki}^+\rangle\cdot\langle a_{lm}^+a_{mn}^+a_{nl}^+\rangle\nonumber\\
&=\langle a_{ij}^+a_{jk}^+a_{ki}^+\cdot a_{lm}^+a_{mn}^+a_{nl}^+\rangle-(p_{ij}^+p_{jk}^+p_{ki}^+)\cdot(p_{lm}^+p_{mn}^+p_{nl}^+)
\end{align}
is different from zero, i.e. any two triads co-variate, if they share an edge. In this case, they form a diamond whose vertices can be labelled as $i\equiv l$, $j\equiv m$, $k$, $n$ and induce the expression

\begin{align}
\text{Cov}[a_\mathbf{I},a_\mathbf{J}]=p_{ij}^+p_{jk}^+p_{ki}^+p_{jn}^+p_{ni}^+-(p_{ij}^+)^2p_{jk}^+p_{ki}^+p_{jn}^+p_{ni}^+=p_{ij}^+(1-p_{ij}^+)p_{jk}^+p_{ki}^+p_{jn}^+p_{ni}^+.
\end{align}

Let us, now, calculate the number of times such an expression appears, i.e. the number of triples sharing an edge: since we need to choose the pair of nodes individuating the common edge, first, and, then, the pair of nodes individuating the `free' vertices of the two triads, such a number amounts at $\binom{N}{2}\binom{N-2}{2}=N(N-1)(N-2)(N-3)/4$; in case $N=4$, it amounts at $3!=6$ - indeed, let us concretely focus on the triads $(1,2,3)$, $(1,2,4)$, $(1,3,4)$, $(2,3,4)$: $(1,2,3)$ co-variates with $(1,2,4)$, $(1,3,4)$, $(2,3,4)$; $(1,2,4)$ co-variates with $(1,3,4)$, $(2,3,4)$; $(1,3,4)$ co-variates with $(2,3,4)$. Overall, then,

\begin{align}
\sum_{\mathbf{I<\mathbf{J}}}\text{Cov}[a_\mathbf{I},a_\mathbf{J}]=3!\sum_{i<j<k<n}p_{ij}^+(1-p_{ij}^+)p_{jk}^+p_{ki}^+p_{jn}^+p_{ni}^+.
\end{align}

Analogously, the standard deviation of motif $T^{---}$ reads

\begin{align}
\sigma[T^{(---)}]&=\sqrt{\sum_\mathbf{I}\text{Var}[a_\mathbf{I}]+2\sum_{\mathbf{I<\mathbf{J}}}\text{Cov}[a_\mathbf{I},a_\mathbf{J}]}\nonumber\\
&=\sqrt{\sum_{i<j<k}p_{ij}^-p_{jk}^-p_{ki}^-(1-p_{ij}^-p_{jk}^-p_{ki}^-)+2\cdot3!\sum_{i<j<k<n}p_{ij}^-(1-p_{ij}^-)p_{jk}^-p_{ki}^-p_{jn}^-p_{ni}^-}.
\end{align}

For what concerns motif $T^{++-}$, its standard deviation reads

\begin{align}
\sigma[T^{(++-)}]&=\sqrt{\sum_\mathbf{I}\text{Var}[a_\mathbf{I}]+2\sum_{\mathbf{I<\mathbf{J}}}\text{Cov}[a_\mathbf{I},a_\mathbf{J}]}
\end{align}
where

\begin{align}
\sum_\mathbf{I}\text{Var}[a_\mathbf{I}]=\sum_{i<j<k}&\text{Var}[a_{ij}^+a_{jk}^+a_{ki}^-+a_{ij}^+a_{jk}^-a_{ki}^++a_{ij}^-a_{jk}^+a_{ki}^+]\nonumber\\
=\sum_{i<j<k}&\{p_{ij}^+p_{jk}^+p_{ki}^-(1-p_{ij}^+p_{jk}^+p_{ki}^-)+p_{ij}^+p_{jk}^-p_{ki}^+(1-p_{ij}^+p_{jk}^-p_{ki}^+)+p_{ij}^-p_{jk}^+p_{ki}^+(1-p_{ij}^-p_{jk}^+p_{ki}^+)\nonumber\\
&-2[(p_{ij}^+p_{jk}^+p_{ki}^-)(p_{ij}^+p_{jk}^-p_{ki}^+)+(p_{ij}^+p_{jk}^-p_{ki}^+)(p_{ij}^-p_{jk}^+p_{ki}^+)+(p_{ij}^-p_{jk}^+p_{ki}^+)(p_{ij}^+p_{jk}^+p_{ki}^-)]\}\nonumber\\
=\sum_{i<j<k}&\{p_{ij}^+p_{jk}^+p_{ki}^-(1-p_{ij}^+p_{jk}^+p_{ki}^--2p_{ij}^+p_{jk}^-p_{ki}^+)+p_{ij}^+p_{jk}^-p_{ki}^+(1-p_{ij}^+p_{jk}^-p_{ki}^+-2p_{ij}^-p_{jk}^+p_{ki}^+)\nonumber\\
&+p_{ij}^-p_{jk}^+p_{ki}^+(1-p_{ij}^-p_{jk}^+p_{ki}^+-2p_{ij}^+p_{jk}^+p_{ki}^-)\}.
\end{align}

The analysis of covariances requires a more detailed explanation. Let us consider that the generic addendum of $\sum_{\mathbf{I<\mathbf{J}}}\text{Cov}[a_\mathbf{I},a_\mathbf{J}]=\sum_{i<j<k<n}\text{Cov}[a_{ij}^+a_{jk}^+a_{ki}^-+a_{ij}^+a_{jk}^-a_{ki}^++a_{ij}^-a_{jk}^+a_{ki}^+, a_{lm}^+a_{mn}^+a_{nl}^-+a_{lm}^+a_{mn}^-a_{nl}^++a_{lm}^-a_{mn}^+a_{nl}^+]$ reads

\begin{align}
\text{Cov}[a_{ij}^+a_{jk}^+a_{ki}^-+a_{ij}^+a_{jk}^-a_{ki}^++a_{ij}^-a_{jk}^+a_{ki}^+, a_{lm}^+a_{mn}^+a_{nl}^-+a_{lm}^+a_{mn}^-a_{nl}^++a_{lm}^-a_{mn}^+a_{nl}^+]\equiv\text{Cov}[X+Y+Z,A+B+C]
\end{align}
and that it can be decomposed as

\begin{align}\label{covarianzestramaledette}
\text{Cov}[X,A]+\text{Cov}[X,B]+\text{Cov}[X,C]+\text{Cov}[Y,A]+\text{Cov}[Y,B]+\text{Cov}[Y,C]+\text{Cov}[Z,A]+\text{Cov}[Z,B]+\text{Cov}[Z,C].
\end{align}

Let us focus on $\text{Cov}[X,A]=\text{Cov}[a_{ij}^+a_{jk}^+a_{ki}^-,a_{lm}^+a_{mn}^+a_{nl}^-]$ and consider that the aforementioned $3!=6$ pairs of triads leads to the following events

\begin{align}
\text{Cov}[X,A]&=\text{Cov}[a_{12}^+a_{23}^+a_{31}^-,a_{12}^+a_{24}^+a_{41}^-]=p_{ij}^+(1-p_{ij}^+)p_{jk}^+p_{ki}^-p_{jn}^+p_{ni}^-,\\
\text{Cov}[X,A]&=\text{Cov}[a_{12}^+a_{23}^+a_{31}^-,a_{13}^+a_{34}^+a_{41}^-]=-(p_{ij}^+p_{jk}^+p_{ki}^-)(p_{ik}^+p_{kn}^+p_{ni}^-),\\
\text{Cov}[X,A]&=\text{Cov}[a_{12}^+a_{23}^+a_{31}^-,a_{23}^+a_{34}^+a_{42}^-]=p_{jk}^+(1-p_{jk}^+)p_{ij}^+p_{ki}^-p_{kn}^+p_{nj}^-,\\
\text{Cov}[X,A]&=\text{Cov}[a_{12}^+a_{24}^+a_{41}^-,a_{13}^+a_{34}^+a_{41}^-]=p_{ni}^-(1-p_{ni}^-)p_{ij}^+p_{jn}^+p_{ik}^+p_{kn}^+,\\
\text{Cov}[X,A]&=\text{Cov}[a_{12}^+a_{24}^+a_{41}^-,a_{23}^+a_{34}^+a_{42}^-]=-(p_{ij}^+p_{jn}^+p_{ni}^-)(p_{jk}^+p_{kn}^+p_{nj}^-),\\
\text{Cov}[X,A]&=\text{Cov}[a_{13}^+a_{34}^+a_{41}^-,a_{23}^+a_{34}^+a_{42}^-]=p_{kn}^+(1-p_{kn}^+)p_{ik}^+p_{ni}^-p_{jk}^+p_{nj}^-;
\end{align}
repeating these calculations for the remaining eight addenda of Supplementary Equation~\eqref{covarianzestramaledette} leads to the expression

\begin{align}
\text{Cov}[X+Y+Z,A+B+C]=&[p_{ij}^-(1-p_{ij}^-)+p_{kn}^-(1-p_{kn}^-)]p_{ik}^+p_{kj}^+p_{jn}^+p_{ni}^+\nonumber\\
&+[p_{jk}^-(1-p_{jk}^-)+p_{in}^-(1-p_{in}^-)]p_{ij}^+p_{jn}^+p_{nk}^+p_{ki}^+\nonumber\\
&+[p_{ik}^-(1-p_{ik}^-)+p_{jn}^-(1-p_{jn}^-)]p_{ij}^+p_{jk}^+p_{kn}^+p_{ni}^+\nonumber\\
&+p_{ij}^+(1-p_{ij}^+)[p_{jk}^+p_{ki}^-+p_{jk}^-p_{ki}^+][p_{jn}^+p_{ni}^-+p_{jn}^-p_{ni}^+]\nonumber\\
&+p_{jk}^+(1-p_{jk}^+)[p_{ij}^+p_{ki}^-+p_{ij}^-p_{ki}^+][p_{kn}^+p_{nj}^-+p_{kn}^-p_{nj}^+]\nonumber\\
&+p_{ik}^+(1-p_{ik}^+)[p_{ij}^+p_{jk}^-+p_{ij}^-p_{jk}^+][p_{kn}^+p_{ni}^-+p_{kn}^-p_{ni}^+]\nonumber\\
&+p_{in}^+(1-p_{in}^+)[p_{nk}^+p_{ki}^-+p_{nk}^-p_{ki}^+][p_{nj}^+p_{ji}^-+p_{nj}^-p_{ji}^+]\nonumber\\
&+p_{jn}^+(1-p_{jn}^+)[p_{ni}^+p_{ij}^-+p_{ni}^-p_{ij}^+][p_{nk}^+p_{kj}^-+p_{nk}^-p_{kj}^+]\nonumber\\
&+p_{kn}^+(1-p_{kn}^+)[p_{ni}^+p_{ik}^-+p_{ni}^-p_{ik}^-][p_{nj}^+p_{jk}^-+p_{nj}^-p_{jk}^+]\nonumber\\
&-(p_{ij}^+p_{jk}^+p_{ki}^-)(p_{ik}^+p_{kn}^+p_{ni}^-+p_{ik}^+p_{kn}^-p_{ni}^++p_{ij}^-p_{jn}^+p_{ni}^++p_{jk}^-p_{kn}^+p_{nj}^+)\nonumber\\
&-(p_{ij}^+p_{jk}^-p_{ki}^+)(p_{jk}^+p_{kn}^+p_{nj}^-+p_{jk}^+p_{kn}^-p_{nj}^++p_{ij}^-p_{jn}^+p_{ni}^++p_{jk}^-p_{kn}^+p_{ni}^+)\nonumber\\
&-(p_{ij}^-p_{jk}^+p_{ki}^+)(p_{ij}^+p_{jn}^+p_{ni}^-+p_{ij}^+p_{jn}^-p_{ni}^++p_{ik}^-p_{kn}^+p_{ni}^++p_{jk}^-p_{kn}^+p_{nj}^+)\nonumber\\
&-(p_{ij}^+p_{jn}^+p_{ni}^-)(p_{jk}^+p_{kn}^+p_{nj}^-+p_{ik}^+p_{kn}^-p_{ni}^++p_{ik}^-p_{kn}^+p_{ni}^+)\nonumber\\
&-(p_{ij}^+p_{jn}^-p_{ni}^+)(p_{ik}^+p_{kn}^+p_{ni}^-+p_{jk}^+p_{kn}^-p_{nj}^++p_{jk}^-p_{kn}^+p_{nj}^+)\nonumber\\
&-(p_{ij}^-p_{jn}^+p_{ni}^+)(p_{ik}^+p_{kn}^+p_{ni}^-+p_{jk}^+p_{kn}^+p_{nj}^-)\nonumber\\
&-(p_{ik}^+p_{kn}^+p_{ni}^-)(p_{jk}^+p_{kn}^-p_{nj}^+)-(p_{ik}^+p_{kn}^-p_{ni}^+)(p_{jk}^+p_{kn}^+p_{nj}^-+p_{jk}^-p_{kn}^+p_{nj}^+)-(p_{ik}^-p_{kn}^+p_{ni}^+)(p_{jk}^+p_{kn}^-p_{nj}^+)
\end{align}
accounting for the $\binom{4}{2}$ ways two triads can share the negative link, the $4\binom{4}{2}$ ways two triads can share a positive link and the $4\binom{4}{2}$ ways two triads represent incompatible events. Analogously, the standard deviation of motif $T^{+--}$ reads

\begin{align}
\sigma[T^{(+--)}]&=\sqrt{\sum_\mathbf{I}\text{Var}[a_\mathbf{I}]+2\sum_{\mathbf{I<\mathbf{J}}}\text{Cov}[a_\mathbf{I},a_\mathbf{J}]}
\end{align}
where

\begin{align}
\sum_\mathbf{I}\text{Var}[a_\mathbf{I}]=\sum_{i<j<k}&\text{Var}[a_{ij}^+a_{jk}^-a_{ki}^-+a_{ij}^-a_{jk}^+a_{ki}^-+a_{ij}^-a_{jk}^-a_{ki}^+]\nonumber\\
=\sum_{i<j<k}&\{p_{ij}^+p_{jk}^-p_{ki}^-(1-p_{ij}^+p_{jk}^-p_{ki}^-)+p_{ij}^-p_{jk}^+p_{ki}^-(1-p_{ij}^-p_{jk}^+p_{ki}^-)+p_{ij}^-p_{jk}^-p_{ki}^+(1-p_{ij}^-p_{jk}^-p_{ki}^+)\nonumber\\
&-2[(p_{ij}^+p_{jk}^-p_{ki}^-)(p_{ij}^-p_{jk}^+p_{ki}^-)+(p_{ij}^-p_{jk}^+p_{ki}^-)(p_{ij}^-p_{jk}^-p_{ki}^+)+(p_{ij}^-p_{jk}^-p_{ki}^+)(p_{ij}^+p_{jk}^-p_{ki}^-)]\}\nonumber\\
=\sum_{i<j<k}&\{p_{ij}^+p_{jk}^-p_{ki}^-(1-p_{ij}^+p_{jk}^-p_{ki}^--2p_{ij}^-p_{jk}^+p_{ki}^-)+p_{ij}^-p_{jk}^+p_{ki}^-(1-p_{ij}^-p_{jk}^+p_{ki}^--2p_{ij}^-p_{jk}^-p_{ki}^+)\nonumber\\
&+p_{ij}^-p_{jk}^-p_{ki}^+(1-p_{ij}^-p_{jk}^-p_{ki}^+-2p_{ij}^+p_{jk}^-p_{ki}^-)\}
\end{align}
and $\sum_{\mathbf{I<\mathbf{J}}}\text{Cov}[a_\mathbf{I},a_\mathbf{J}]=\sum_{i<j<k<n}\text{Cov}[a_{ij}^+a_{jk}^-a_{ki}^-+a_{ij}^-a_{jk}^+a_{ki}^-+a_{ij}^-a_{jk}^-a_{ki}^+, a_{lm}^+a_{mn}^-a_{nl}^-+a_{lm}^-a_{mn}^+a_{nl}^-+a_{lm}^-a_{mn}^-a_{nl}^+]$. Upon posing

\begin{align}
\text{Cov}[a_{ij}^+a_{jk}^-a_{ki}^-+a_{ij}^-a_{jk}^+a_{ki}^-+a_{ij}^-a_{jk}^-a_{ki}^+,a_{lm}^+a_{mn}^-a_{nl}^-+a_{lm}^-a_{mn}^+a_{nl}^-+a_{lm}^-a_{mn}^-a_{nl}^+]\equiv\text{Cov}[X+Y+Z,A+B+C]
\end{align}
and decomposing it as

\begin{align}
\text{Cov}[X,A]+\text{Cov}[X,B]+\text{Cov}[X,C]+\text{Cov}[Y,A]+\text{Cov}[Y,B]+\text{Cov}[Y,C]+\text{Cov}[Z,A]+\text{Cov}[Z,B]+\text{Cov}[Z,C]
\end{align}
one can write

\begin{align}
\text{Cov}[X+Y+Z,A+B+C]=&[p_{ij}^+(1-p_{ij}^+)+p_{kn}^+(1-p_{kn}^+)]p_{ik}^-p_{kj}^-p_{jn}^-p_{ni}^-
\nonumber\\
&+[p_{jk}^+(1-p_{jk}^+)+p_{in}^+(1-p_{in}^+)]p_{ij}^-p_{jn}^-p_{nk}^-p_{ki}^-
\nonumber\\
&+[p_{ik}^+(1-p_{ik}^+)+p_{jn}^+(1-p_{jn}^+)]p_{ij}^-p_{jk}^-p_{kn}^-p_{ni}^-
\nonumber\\
&+p_{ij}^-(1-p_{ij}^-)[p_{jk}^+p_{ki}^-+p_{jk}^-p_{ki}^+][p_{jn}^+p_{ni}^-+p_{jn}^-p_{ni}^+]\nonumber\\
&+p_{jk}^-(1-p_{jk}^-)[p_{ij}^+p_{ki}^-+p_{ij}^-p_{ki}^+][p_{kn}^+p_{nj}^-+p_{kn}^-p_{nj}^+]\nonumber\\
&+p_{ik}^-(1-p_{ik}^-)[p_{ij}^+p_{jk}^-+p_{ij}^-p_{jk}^+][p_{kn}^+p_{ni}^-+p_{kn}^-p_{ni}^+]\nonumber\\
&+p_{in}^-(1-p_{in}^-)[p_{nk}^+p_{ki}^-+p_{nk}^-p_{ki}^+][p_{nj}^+p_{ji}^-+p_{nj}^-p_{ji}^+]\nonumber\\
&+p_{jn}^-(1-p_{jn}^-)[p_{ni}^+p_{ij}^-+p_{ni}^-p_{ij}^+][p_{nk}^+p_{kj}^-+p_{nk}^-p_{kj}^+]\nonumber\\
&+p_{kn}^-(1-p_{kn}^-)[p_{ni}^+p_{ik}^-+p_{ni}^-p_{ik}^-][p_{nj}^+p_{jk}^-+p_{nj}^-p_{jk}^+]\nonumber\\
&-(p_{ij}^-p_{jk}^-p_{ki}^+)(p_{ik}^-p_{kn}^-p_{ni}^++p_{ik}^-p_{kn}^+p_{ni}^-+p_{ij}^+p_{jn}^-p_{ni}^-+p_{jk}^+p_{kn}^-p_{nj}^-)\nonumber\\
&-(p_{ij}^-p_{jk}^+p_{ki}^-)(p_{jk}^-p_{kn}^-p_{nj}^++p_{jk}^-p_{kn}^+p_{nj}^-+p_{ij}^+p_{jn}^-p_{ni}^-+p_{jk}^+p_{kn}^-p_{ni}^-)\nonumber\\
&-(p_{ij}^+p_{jk}^-p_{ki}^-)(p_{ij}^-p_{jn}^-p_{ni}^++p_{ij}^-p_{jn}^+p_{ni}^-+p_{ik}^+p_{kn}^-p_{ni}^-+p_{jk}^+p_{kn}^-p_{nj}^-)\nonumber\\
&-(p_{ij}^-p_{jn}^-p_{ni}^+)(p_{jk}^-p_{kn}^-p_{nj}^++p_{ik}^-p_{kn}^+p_{ni}^-+p_{ik}^+p_{kn}^-p_{ni}^-)\nonumber\\
&-(p_{ij}^-p_{jn}^+p_{ni}^-)(p_{ik}^-p_{kn}^-p_{ni}^++p_{jk}^-p_{kn}^+p_{nj}^-+p_{jk}^+p_{kn}^-p_{nj}^-)\nonumber\\
&-(p_{ij}^+p_{jn}^-p_{ni}^-)(p_{ik}^-p_{kn}^-p_{ni}^++p_{jk}^-p_{kn}^-p_{nj}^+)-(p_{ik}^-p_{kn}^-p_{ni}^+)(p_{jk}^-p_{kn}^+p_{nj}^-)\nonumber\\
&-(p_{ik}^-p_{kn}^+p_{ni}^-)(p_{jk}^-p_{kn}^-p_{nj}^++p_{jk}^+p_{kn}^-p_{nj}^-)-(p_{ik}^+p_{kn}^-p_{ni}^-)(p_{jk}^-p_{kn}^+p_{nj}^-).
\end{align}

For what concerns fixed-topology benchmarks, instead, the formulas above hold in a `conditional' fashion: in other words, the sums run over the connected sets of nodes (be they triples or quadruples).

\begin{figure}[t!]
\centering
\includegraphics[width=0.3\textwidth]{fig2.png}\\
\subfigure[]{\includegraphics[width=0.49\textwidth]{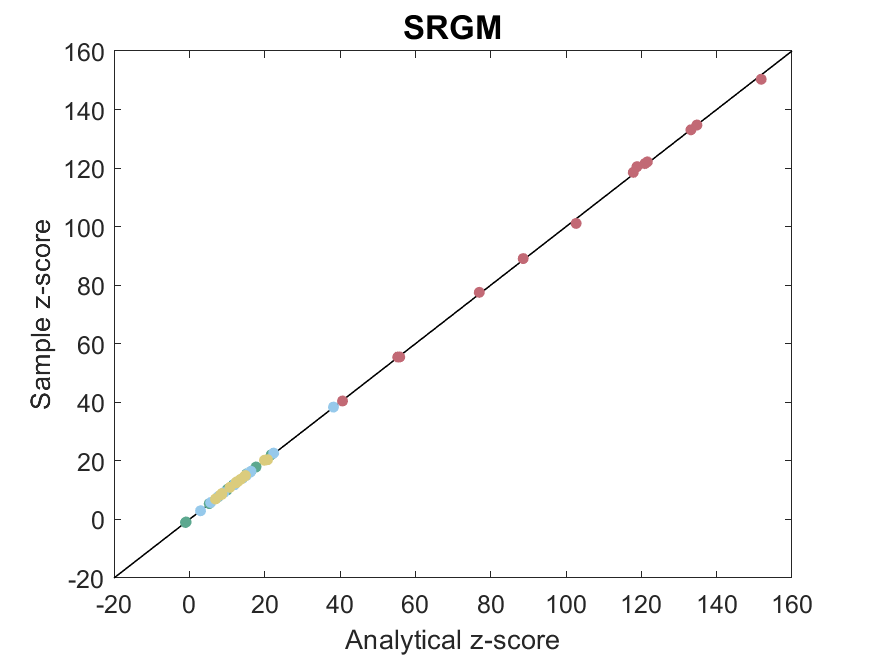}}
\subfigure[]{\includegraphics[width=0.49\textwidth]{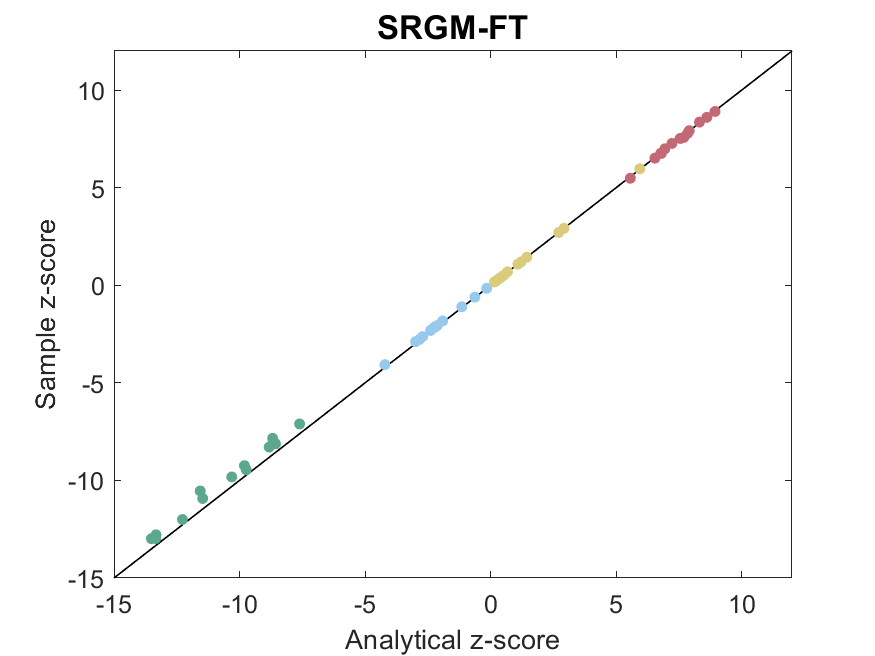}}\\
\subfigure[]{\includegraphics[width=0.49\textwidth]{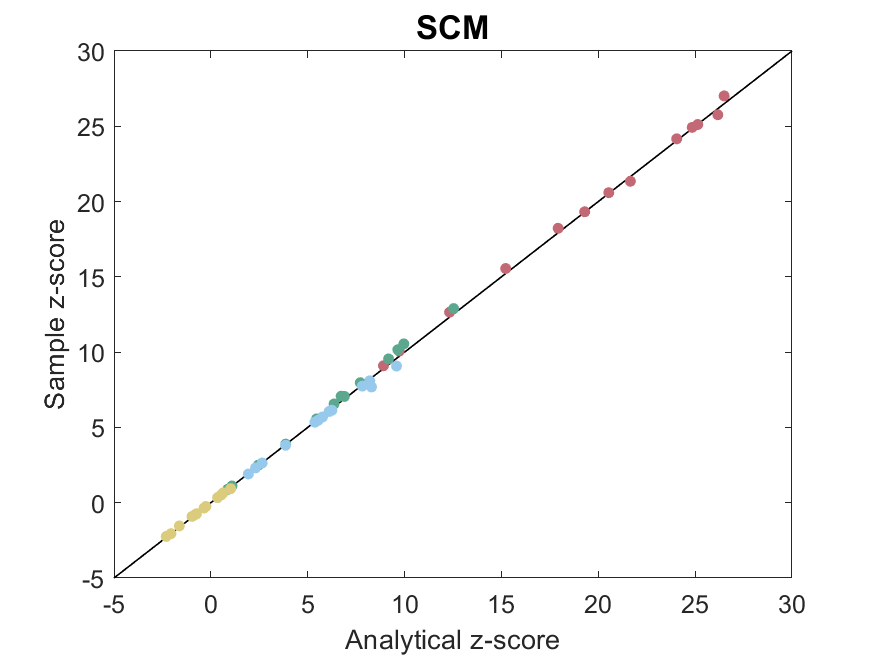}}
\subfigure[]{\includegraphics[width=0.49\textwidth]{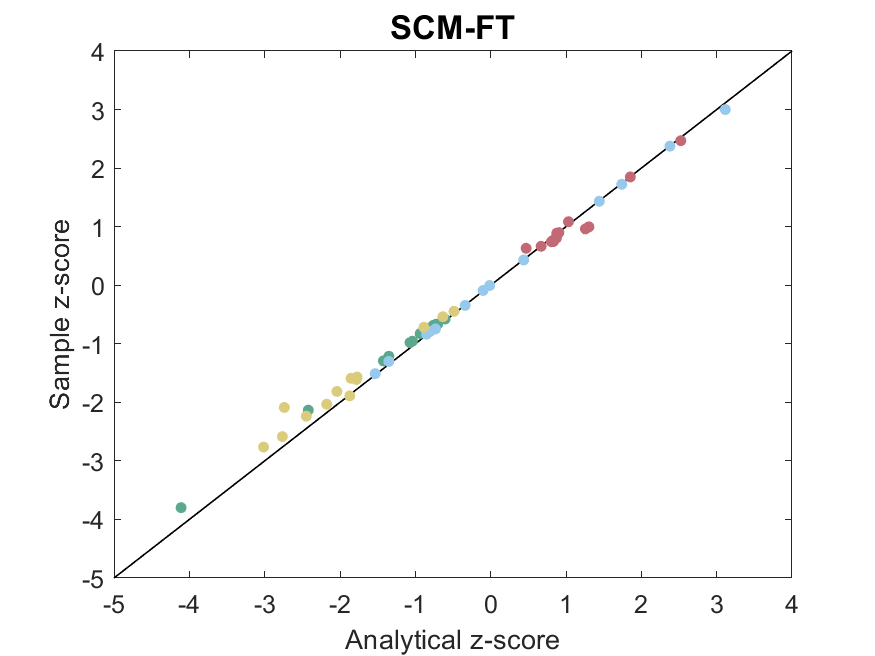}}
\caption{\textbf{Sample VS analytical $z$-scores of triadic motifs.} Sample VS analytical $z$-scores of our motifs, for each null model and each snapshot of the Correlates of Wars dataset. Each ensemble is constituted by 10.000 realizations.}
\label{fig:5A}
\end{figure}

Supplementary Figure \ref{fig:5A} shows the agreement between the analytical and the numerical estimations of our motifs $z$-scores, for each null model and each snapshot of the CoW dataset. The residual discrepancies may be due to undersampling of rare events, caused by the conditions $p^-\ll p^+$ and $p_{ij}^-\ll p_{ij}^+$, $\forall\:i<j$.

The Gaussianity of the ensemble distributions of the abundances of our motifs is shown in Supplementary Figure \ref{fig:6A}, for each null model and a couple of snapshots of the CoW dataset.

\begin{figure}[b!]
\centering
\subfigure[]{\includegraphics[width=0.24\textwidth]{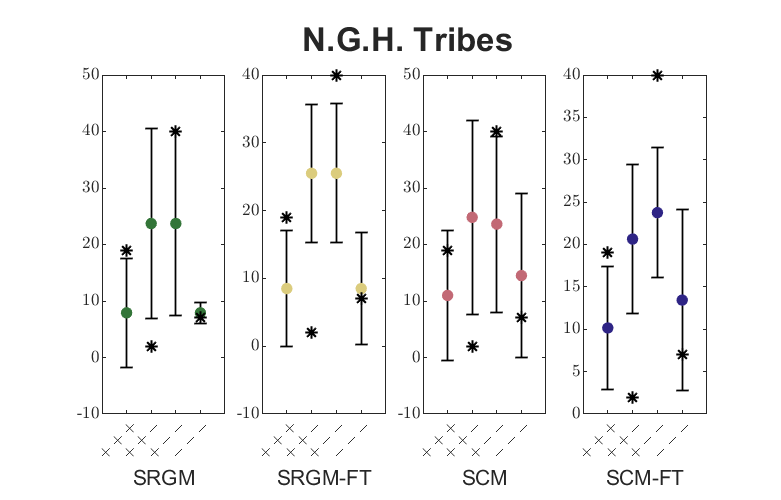}}
\subfigure[]{\includegraphics[width=0.24\textwidth]{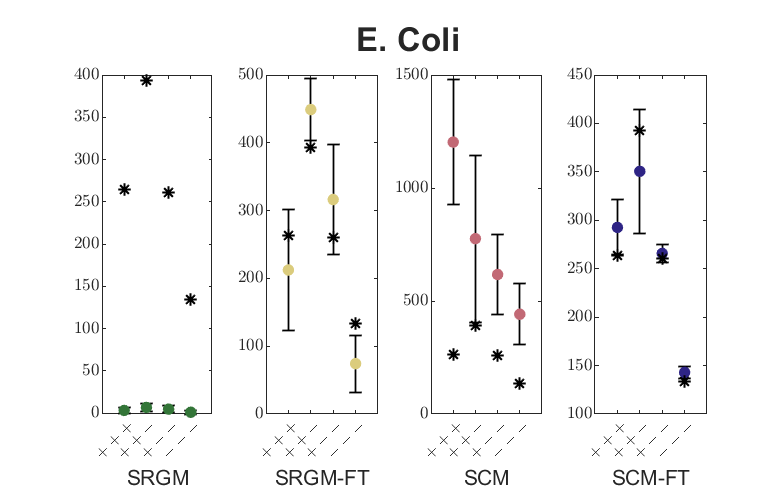}}
\subfigure[]{\includegraphics[width=0.24\textwidth]{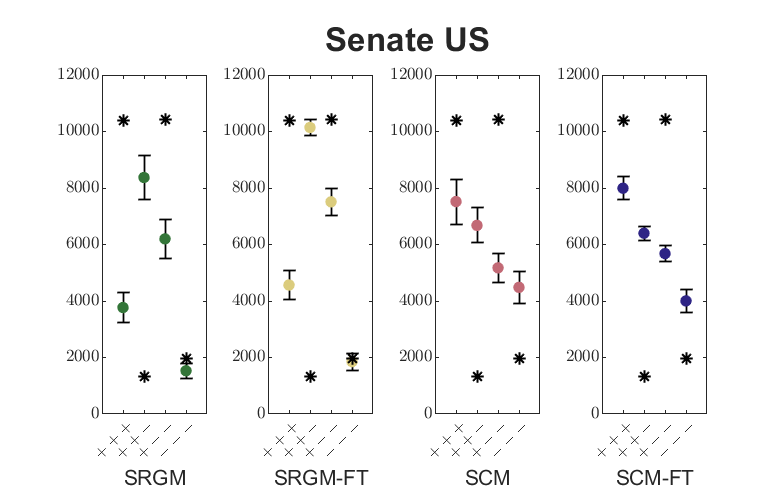}}
\subfigure[]{\includegraphics[width=0.24\textwidth]{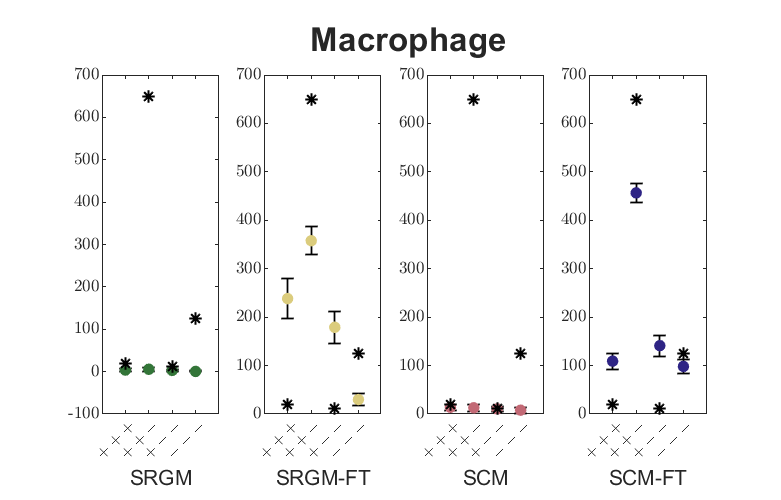}}\\
\subfigure[]{\includegraphics[width=0.24\textwidth]{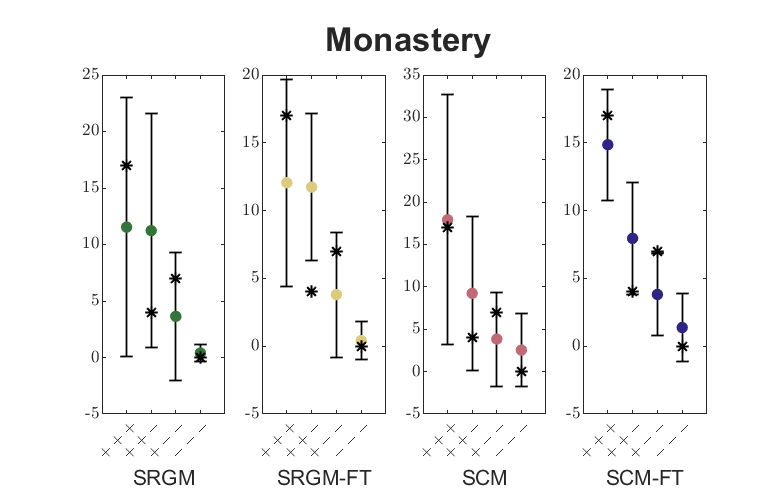}}
\subfigure[]{\includegraphics[width=0.24\textwidth]{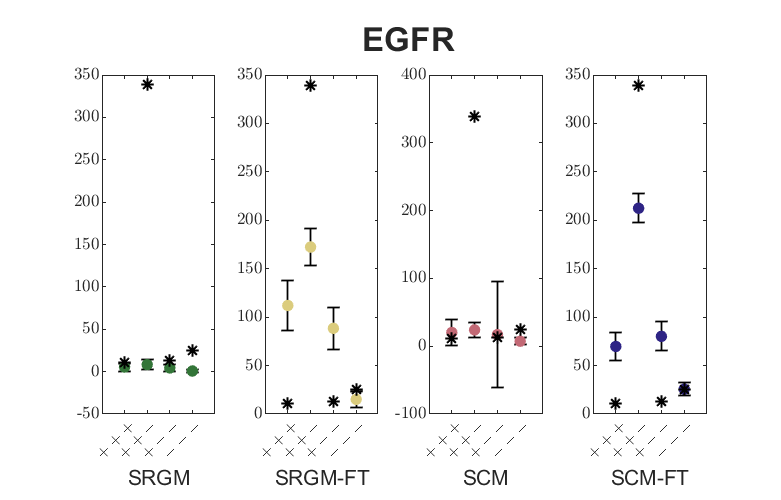}}
\subfigure[]{\includegraphics[width=0.24\textwidth]{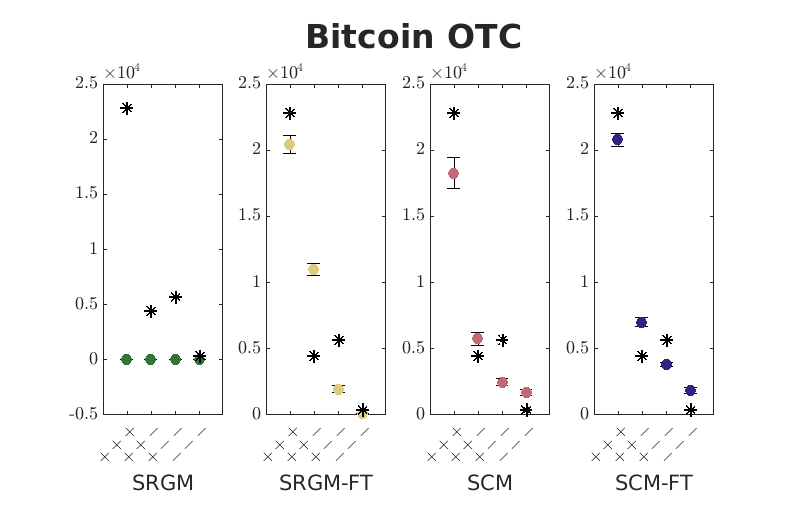}}
\subfigure[]{\includegraphics[width=0.24\textwidth]{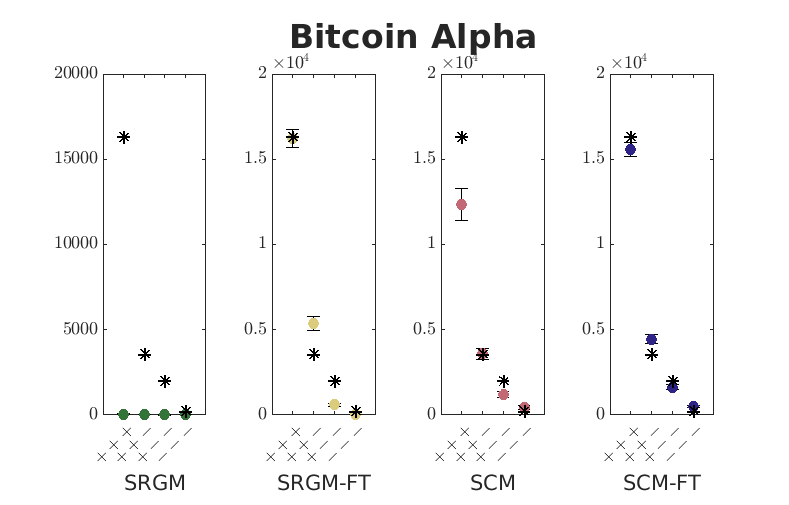}}
\caption{\textbf{Alternative representation of the $z$-scores of triadic motifs.} $\langle N_m\rangle\pm2\sigma[N_m]$ values, with $N_m$ indicating the abundance of motif $m$, for each null model and a bunch of networks. Colors refer to the Signed Random Graph Model (SRGM) (\textcolor{colSRGM}{$\bullet$}), the  Signed Random Graph Model with Fixed Topology (SRGM-FT) (\textcolor{colSRGMFT}{$\bullet$}), the Signed Configuration Model (SCM) (\textcolor{colSCM}{$\bullet$}) and the Signed Configuration Model with Fixed Topology (SCM-FT) (\textcolor{colSCMFT}{$\bullet$}).}
\label{fig:7A}
\end{figure}

Finally, Supplementary Figure \ref{fig:7A} depicts the $\langle N_m\rangle\pm2\sigma[N_m]$ values - with $N_m$ indicating the abundance of motif $m$ - for each null model and a bunch of networks. Although being less general than the representation provided by $z$-scores - it forces us to choose the width of the interval accompanying the expected value, e.g. $2\sigma[N_m]$ - it allows one to disentangle the contributions of the mean and the variance to the $z$-score value.

\begin{figure}[t!]
\centering
\subfigure[]{\includegraphics[width=0.495\textwidth]{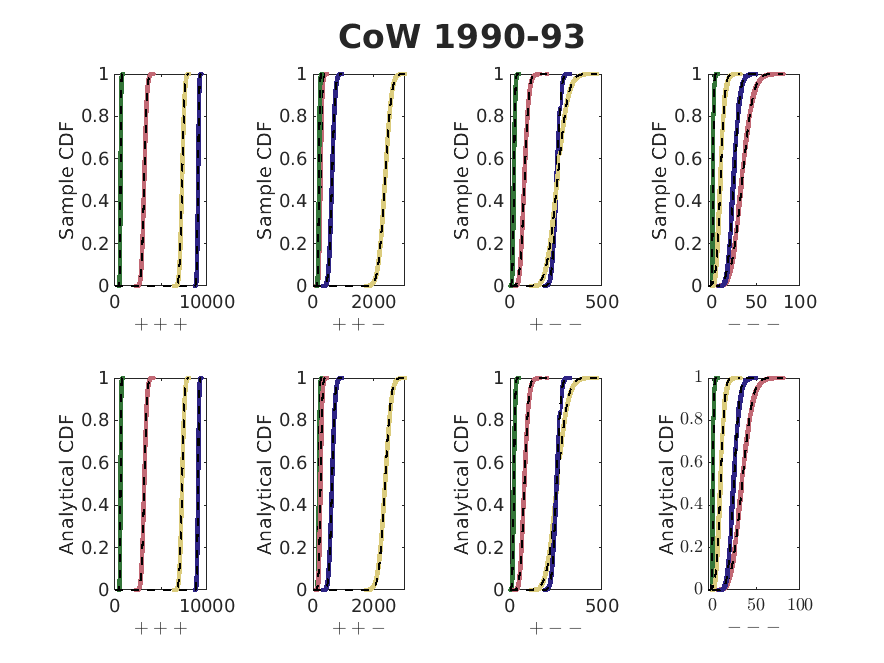}}
\subfigure[]{\includegraphics[width=0.495\textwidth]{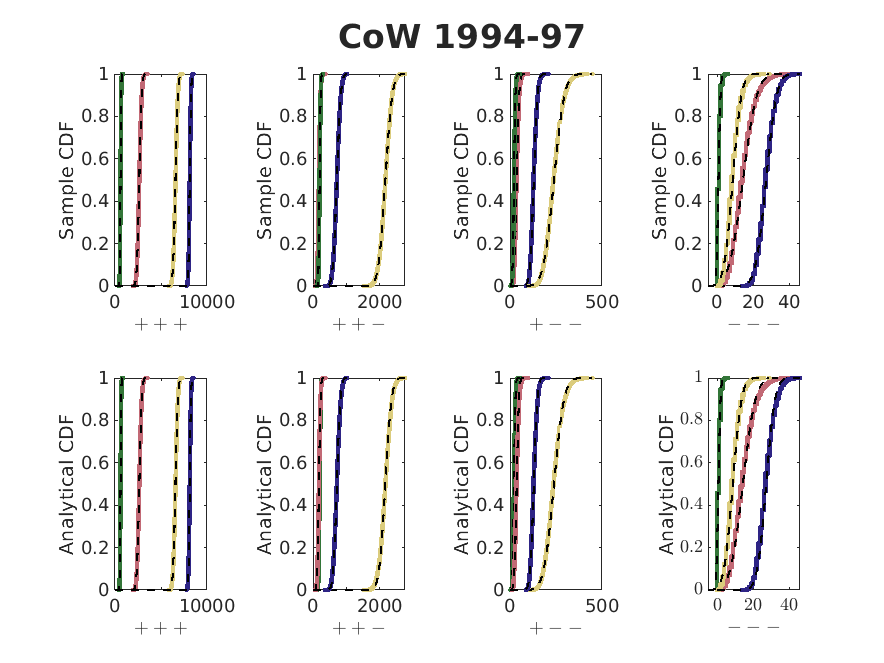}}
\subfigure[]{\includegraphics[width=0.495\textwidth]{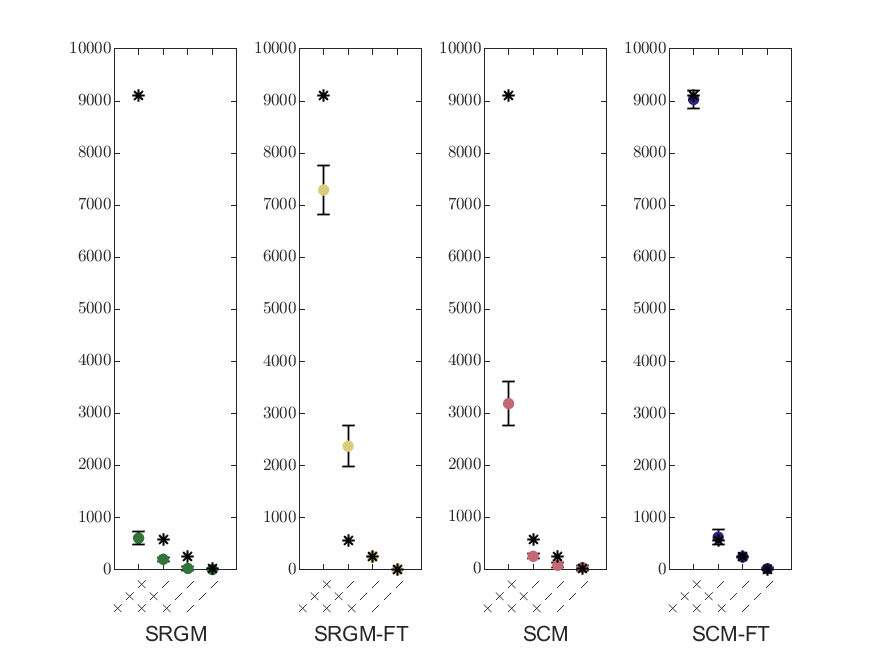}}
\subfigure[]{\includegraphics[width=0.495\textwidth]{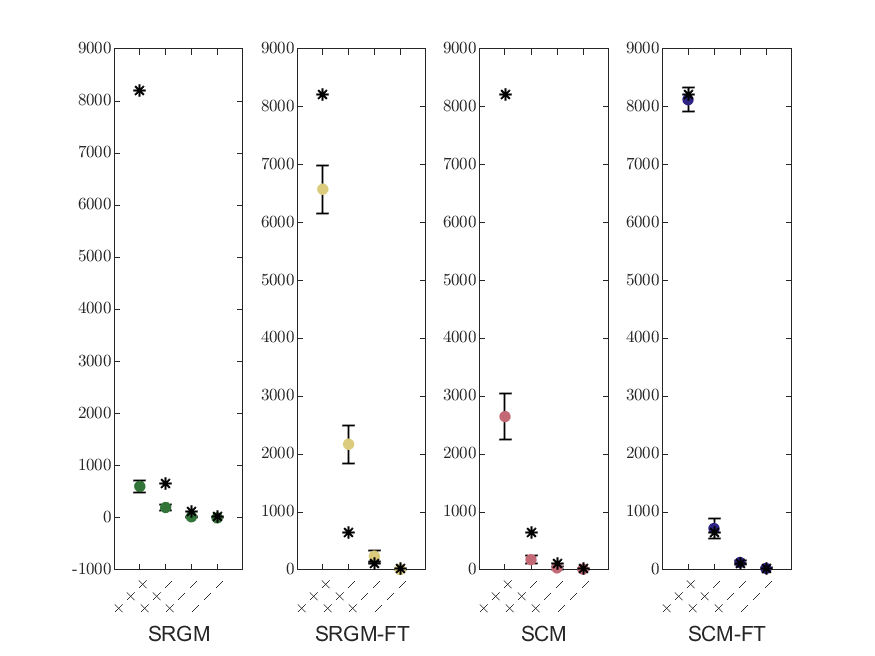}}
\caption{\textbf{Check of Gaussianity of our motifs ensemble distributions.} Check of the Gaussianity of the ensemble distributions of the abundances of our motifs, for each null model and the 1990-93 and 1994-97 snapshots of the Correlates of Wars dataset. The distribution are computed under the Signed Random Graph Model (SRGM) (\textcolor{colSRGM}{$\bullet$}), the  Signed Random Graph Model with Fixed Topology (SRGM-FT) (\textcolor{colSRGMFT}{$\bullet$}), the Signed Configuration Model (SCM) (\textcolor{colSCM}{$\bullet$}) and the Signed Configuration Model with Fixed Topology (SCM-FT) (\textcolor{colSCMFT}{$\bullet$}).}
\label{fig:6A}
\end{figure}

\clearpage

\section*{Supplementary Note 7\\Inspecting mesoscopic frustration on signed networks}

In order to inspect the behaviour of frustration at the mesoscopic level, we have considered the signed modularity, defined as

\begin{align}
Q&=\sum_{i=1}^N\sum_{\substack{j=1\\(j>i)}}^N[a_{ij}^*-\langle a_{ij}\rangle]\delta_{c_ic_j}\nonumber\\
&=\sum_{i=1}^N\sum_{\substack{j=1\\(j>i)}}^N[(a_{ij}^+)^*-(a_{ij}^-)^*-\langle a_{ij}^+\rangle+\langle a_{ij}^-\rangle]\delta_{c_ic_j}\nonumber\\
&=\sum_{i=1}^N\sum_{\substack{j=1\\(j>i)}}^N(a_{ij}^+)^*\delta_{c_ic_j}-\sum_{i=1}^N\sum_{\substack{j=1\\(j>i)}}^N(a_{ij}^-)^*\delta_{c_ic_j}-\sum_{i=1}^N\sum_{\substack{j=1\\(j>i)}}^N\langle a_{ij}^+\rangle\delta_{c_ic_j}+\sum_{i=1}^N\sum_{\substack{j=1\\(j>i)}}^N\langle a_{ij}^-\rangle\delta_{c_ic_j}\nonumber\\
&=\sum_{i=1}^N\sum_{\substack{j=1\\(j>i)}}^N(a_{ij}^+)^*\delta_{c_ic_j}-\sum_{i=1}^N\sum_{\substack{j=1\\(j>i)}}^N(a_{ij}^-)^*\delta_{c_ic_j}-\sum_{i=1}^N\sum_{\substack{j=1\\(j>i)}}^Np_{ij}^+\delta_{c_ic_j}+\sum_{i=1}^N\sum_{\substack{j=1\\(j>i)}}^Np_{ij}^-\delta_{c_ic_j}\nonumber\\
&=L_\bullet^+-L_\bullet^--\langle L_\bullet^+\rangle+\langle L_\bullet^-\rangle\nonumber\\
&=L^+-L_\circ^+-L_\bullet^--\langle L^+-L_\circ^+\rangle+\langle L_\bullet^-\rangle\nonumber\\
&=-(L_\circ^++L_\bullet^-)+\langle L_\circ^++L_\bullet^-\rangle+L^+-\langle L^+\rangle\nonumber\\
&=-[(L_\circ^++L_\bullet^-)-\langle L_\circ^++L_\bullet^-\rangle]+L^+-\langle L^+\rangle
\end{align}
where 

\begin{align}
L_\bullet^+&=\sum_{i=1}^N\sum_{\substack{j=1\\(j>i)}}^N(a_{ij}^+)^*\delta_{c_ic_j}=\sum_{i=1}^N\sum_{\substack{j=1\\(j>i)}}^N(a_{ij}^+)^*-\sum_{i=1}^N\sum_{\substack{j=1\\(j>i)}}^N(a_{ij}^+)^*(1-\delta_{c_ic_j})=L^+-L_\circ^+
\end{align}
and analogously for $L_\bullet^-$: since the total number of positive links is preserved under any null model considered here, we obtain $Q=-[(L_\circ^++L_\bullet^-)-\langle L_\circ^++L_\bullet^-\rangle]$; moreover, if we employ a null model that preserves a network topology, the stronger result $Q=-L\cdot(\text{FI}-\langle\text{FI}\rangle)$ holds true and, since $L>0$, maximizing $Q$ becomes equivalent at minimizing the FI (that coincides with the percentage of `misplaced' links, i.e. the total number of positive links between communities, $L^+_\circ$, plus the total number of negative links within communities, $L^-_\bullet$, divided by the total number of links, $L$).

Intuitively, maximizing modularity amounts at placing the nodes connected by a positive link within the same modules and the nodes connected by a negative link within different modules. Indeed, under the assumption $0\leq p_{ij}^-\leq p_{ij}^+\leq 1$, one has that

\begin{align}
a_{ij}^*-(p_{ij}^+-p_{ij}^-)=&
\begin{cases}
+[1-(p_{ij}^+-p_{ij}^-)]>0,&\text{if}\:a_{ij}^*=+1\\
-[1+(p_{ij}^+-p_{ij}^-)]<0,&\text{if}\:a_{ij}^*=-1
\end{cases}
\end{align}
i.e. $\text{sgn}[a_{ij}^*-(p^+_{ij}-p^-_{ij})]=\text{sgn}[a_{ij}^*]$; hence, $Q$ rises (decreases) if $\delta_{c_ic_j}=1$, i.e. $c_i=c_j$, and $a_{ij}^*=+1$ ($\delta_{c_ic_j}=1$ and $a_{ij}^*=-1$).

A frustration-based community detection algorithm where the number of blocks, say $k$, is fixed a priori, thus remains naturally defined (see below). Notice that our exercise is defined in such a way that the numerical value of the generic addendum $a_{ij}^*-(p_{ij}^+-p_{ij}^-)$ is fixed, once and for all, by the choice of the benchmark to be solved: in other words, the definition of modularity does not change with the level of aggregation, being just recomputed (as any other score function) as the partition changes.

\clearpage

\begin{algorithm}[t!]
\caption{Pseudocode to partition nodes in order to maximize the signed modularity $Q$}
\begin{algorithmic}
\item[\hspace{1.4pt} 1:
\textbf{function} \textit{ModularityBasedCommunityDetection}$(N,k,\mathbf A)$]
\item[\hspace{1.4pt} 2: $C$ $\leftarrow$ array of length $N$, randomly initialized with $k$ different integers $1\dots k$;]
\item[\hspace{1.4pt} 4: $Q \leftarrow$ \textit{UpdateSignedModularity}$(N,\mathbf A,C)$;]
\item[\hspace{1.4pt} 5: $E \leftarrow$ randomly sorted edges;] 
\item[\hspace{1.4pt} 6: \textbf{for} $(u,v)\in E$ \textbf{do}]
\item[\hspace{1.4pt} 7: \hspace{15pt} $C_0 \leftarrow C$; ]
\item[\hspace{1.4pt} 8: \hspace{15pt} $Q_0 \leftarrow Q$; ]
\item[\hspace{1.4pt} 9: \hspace{15pt} \textbf{if} $C(u)\neq C(v)$ \textbf{then}]
\item[\hspace{1.4pt} 10: \hspace{30pt} $C_1 \leftarrow C$; ]
\item[\hspace{1.4pt} 11: \hspace{30pt} $C_1(u) \leftarrow C(v)$; ]
\item[\hspace{1.4pt} 13: \hspace{30pt} $Q_1 \leftarrow$ \textit{UpdateSignedModularity}$(N,\mathbf A,C_1)$;]
\item[\hspace{1.4pt} 14: \hspace{30pt} $C_2 \leftarrow C$; ]
\item[\hspace{1.4pt} 15: \hspace{30pt} $C_2(v) \leftarrow C(u)$; ]
\item[\hspace{1.4pt} 16: \hspace{30pt} $Q_2 \leftarrow$ \textit{UpdateSignedModularity}$(N,\mathbf A,C_2)$;]
\item[\hspace{1.4pt} 17: \hspace{10pt}
\textbf{else if} $C(u)=C(v)$
\textbf{then}]
\item[\hspace{1.4pt} 18: \hspace{30pt} $C_1 \leftarrow C$;]
\item[\hspace{1.4pt} 19: \hspace{30pt} $C_1(u) \leftarrow$ randomly sorted community different from $C(v)$;]
\item[\hspace{1.4pt} 20: \hspace{30pt} $Q_1 \leftarrow$ \textit{UpdateSignedModularity}$(N,\mathbf A,C_1)$;]
\item[\hspace{1.4pt} 21: \hspace{30pt} $C_2 \leftarrow C$;]
\item[\hspace{1.4pt} 22: \hspace{30pt} $C_2(v) \leftarrow$ randomly sorted community different from $C(u)$;]
\item[\hspace{1.4pt} 23: \hspace{30pt} $Q_2 \leftarrow$ \textit{UpdateSignedModularity}$(N,\mathbf A,C_2)$;]
\item[\hspace{1.4pt} 24: \hspace{10pt} \textbf{end if}]
\item[\hspace{1.4pt} 25: \hspace{10pt} $i \leftarrow \text{argmax}\{Q_0,Q_1,Q_2\}$; ]
\item[\hspace{1.4pt} 26: \hspace{10pt} $C \leftarrow C_i$; ]
\item[\hspace{1.4pt} 27: \hspace{10pt} $Q \leftarrow Q_i$;]
\item[\hspace{1.4pt} 28: \textbf{end for}]
\end{algorithmic} 
\label{alg:minfru}
\end{algorithm}

\begin{algorithm}[t!]
\caption{Pseudocode to update the signed modularity $Q$}
\begin{algorithmic}
\item[\hspace{1.4pt} 1: \textbf{function} \textit{UpdateSignedModularity}$(N,\mathbf A,C)$]
\item[\hspace{1.4pt} 2: $\mathbf{P}^+\leftarrow$ matrix whose generic element is the probability that the corresponding nodes are linked by a $+1$;]
\item[\hspace{1.4pt} 3: $\mathbf{P}^-\leftarrow$ matrix whose generic element is the probability that the corresponding nodes are linked by a $-1$;]
\item[\hspace{1.4pt} 4: $Q\leftarrow 0$;]
\item[\hspace{1.4pt} 5: \textbf{for} $(u,v)\in E$ \textbf{do}]
\item[\hspace{1.4pt} 6: \hspace{15pt} \textbf{if} $C(u)=C(v)$ \textbf{do}]
\item[\hspace{1.4pt} 7: \hspace{30pt} $Q=Q+\mathbf{A}(u,v)-\mathbf{P}^+(u,v)+\mathbf{P}^-(u,v)$;]
\item[\hspace{1.4pt} 8: \hspace{15pt} \textbf{end}]
\item[\hspace{1.4pt} 9: \textbf{end}]
\end{algorithmic} 
\label{alg:upfru}
\end{algorithm}

\end{document}